\documentclass[
reprint,
showpacs,
nofootinbib,
amsmath,amssymb,prd]{revtex4-1}

\usepackage{graphicx}
\usepackage{pstricks}
\usepackage{dcolumn}
\usepackage{bm}
\usepackage{esdiff}
\usepackage{hyperref}
\usepackage[mathlines]{lineno}
\usepackage{subfigure}

\begin{document}

\title{Definition of the relativistic geoid in terms of isochronometric surfaces}

\author{Dennis Philipp, Volker Perlick, Dirk Puetzfeld, Eva Hackmann, and Claus L\"ammerzahl}

\affiliation{ZARM, University of Bremen, 28359 Bremen, Germany}

\begin{abstract}
We present a definition of the geoid that is based on the formalism of general relativity without approximations; i.e.\ it allows for arbitrarily strong gravitational fields. For this reason, it applies not only to the Earth and other planets but also to compact objects such as neutron stars. We define the geoid as a level surface of a time-independent redshift potential. Such a redshift potential exists in any stationary spacetime. Therefore, our geoid is well defined for any rigidly rotating object with constant angular velocity and a fixed rotation axis that is not subject to external forces. Our definition is operational because the level surfaces of a redshift potential can be realized with the help of standard clocks, which may be connected by optical fibers. Therefore, these surfaces are also called ``isochronometric surfaces.'' We deliberately base our definition of a relativistic geoid on the use of clocks since we believe that clock geodesy offers the best methods for probing gravitational fields with highest precision in the future. However, we also point out that our definition of the geoid is mathematically equivalent to a definition in terms of an acceleration potential, i.e.\ that our geoid may also be viewed as a level surface orthogonal to plumb lines. Moreover, we demonstrate that our definition reduces to the known Newtonian and post-Newtonian notions in the appropriate limits. As an illustration, we determine the isochronometric surfaces for rotating observers in axisymmetric static and axisymmetric stationary solutions to Einstein's vacuum field equation, with the Schwarzschild metric, the Erez-Rosen metric, the $q$-metric and the Kerr metric as particular examples. 

\end{abstract}

\maketitle



\section{Introduction}

One of the fundamental tasks of geodesy is to determine the Earth's geoid from gravity field measurements. Within a Newtonian framework, the definition of the geoid combines the Newtonian gravitational potential and the potential related to centrifugal forces that act on the rotating Earth. Therefore, the gradient of the total potential describes the free fall of particles in the corotating frame. From acceleration measurements, and the knowledge of the Earth's state of rotation, one can deduce the pure Newtonian potential. Afterward, via geodetic modeling schemes, information about the change of mass distributions and mass transport can be obtained. These temporal variations and long time trends are usually translated into water height equivalent mass changes on the Earth's surface for visualization. The geoid itself is also commonly used as a reference surface for height measurements \cite{TorgeMueller:2012}.

Within the last years, the accuracy of measurements of the gravitational field has improved considerably, and it is expected to improve even more in the near future. For example, such an improvement is expected from the upcoming geodetic space mission GRACE-FO, which consists of two spacecraft in a polar orbit around the Earth. The influence of the varying gravitational field along the orbit causes a variation in the separation of the two satellites. With the onboard Laser Ranging Interferometer (LRI), it is expected that such variations can be measured to within an accuracy of $10\,$nm \cite{Loomis:2012, Flechtner:2016}. Another important improvement is expected from the use of clocks in the context of chronometric geodesy. The basic idea is to surround the Earth with a network of clocks and to measure their mutual redshifts (or their redshifts with respect to a master clock). As clocks now approach a stability of $10^{-18}$ \cite{Bloom:2014}, it will soon be possible to measure gravitational redshifts that correspond to height differences of about $1\,$cm. 

Both examples show that for a correct evaluation of present or near-future measurements of the gravitational field of the Earth it is mandatory to take general relativity into account. Of course, the geodetic community is well aware of this fact. The usual way to consider relativistic effects is by starting with the Newtonian theory and applying post-Newtonian (PN) corrections. In particular, the notion of the geoid was already discussed in such a PN setting in 1988 by Soffel \emph{et al.}\ \cite{Soffel:1988}. They defined a so-called a-geoid, which is based on acceleration measurements, and a so-called u-geoid which is based on using clocks. The authors showed that, within their setting, the two definitions are equivalent. For a more recent discussion of the Earth's geoid in terms of PN calculations, we refer to the work by Kopeikin \textit{et al.}\ \cite{Kopeikin:2015}. Although the PN approach is certainly sufficient for calculating all relevant effects with the desired accuracy in the vicinity of the Earth, from a methodological point of view, it is more satisfactory to start out from a fully relativistic setting and then to apply approximations where appropriate. This makes it necessary to provide fully relativistic definitions of all the basic concepts, in particular of the Earth's geoid.

It is the purpose of this paper to present and discuss such a fully relativistic definition of the geoid. As we allow the gravitational field to be arbitrarily strong, our definition applies not only to the Earth and to other planets but also to compact objects such as neutron stars. For lack of a better word, we always speak of the ``geoid,'' for all kinds of gravitating bodies. Our definition is operational, using clocks as measuring devices. That is to say, in the terminology of the above-mentioned paper by Soffel \textit{et al.}, we define a fully relativistic u-geoid. However, we also discuss the notion of an a-geoid and we show that, also in the relativistic theory without approximations, the two notions are equivalent. We believe that high-precision geodesy will be mainly based on the use of clocks in the future; therefore, we consider the u-geoid as the primary notion and the fact that it coincides with the a-geoid as convenient but of secondary importance only. 

Our definition assumes a central body that rotates rigidly with constant angular velocity, where we have to recall that in general relativity a ``rigid motion'' is defined by vanishing shear and vanishing expansion for a timelike congruence of worldlines. (This is often called ``Born rigidity.'') Of course, the motion of the Earth (or of neutron stars) is not perfectly rigid. However, rigidity may be viewed as a reasonable first approximation, and the effect of deformations may be considered in terms of small perturbations afterward. Our definition is based on the mathematical fact that the gravitational field of a body that rotates rigidly with constant angular velocity admits a time-independent redshift potential. We define the geoid as a surface of constant redshift potential, which is also called an \emph{isochronometric} surface. The equivalence of our (u-)geoid with an appropriately defined a-geoid follows from the fact that the redshift potential is also an acceleration potential.

As we will outline below, our definition of a relativistic geoid may be viewed as a translation into mathematical language of a definition that was given, just in words, already in 1985 by Bjerhammar \cite{Bjerhammar:1985, Bjerhammar:1986}. More recently, inspired by Bjerhammar's wording, Kopeikin \textit{et al.}\ \cite{Kopeikin:2016} discussed a relativistic notion of the u-geoid assuming a particular fluid model for the Earth. Also, Oltean \textit{et al.}\ \cite{Oltean:2015} gave another fully relativistic definition of the geoid, which is mathematically quite satisfactory. However, we believe that our definition is more operational. A major difference is in the fact that, in the above-mentioned terminology, Oltean \textit{et al.}\ defined an a-geoid. In contrast to our work, Bjerhammar's, and Kopeikin's, they do not make any reference to the use of clocks. We see the advantage of our framework in the exploration of the use of clocks and their description in terms of an isometric timelike congruence. We ask for the redshift of any pair of clocks within such a congruence and use the redshift potential as the basis for the definition of the relativistic geoid.

For a general review of relativistic geodesy and related problems, see, e.g.\ Refs.\ \cite{Mueller:2008} and \cite{Kopeikin:Book:2011}. Reference \cite{Delva:2017} contains a comprehensive summary of theoretical methods in relativistic gravimetry, chronometric geodesy, and related fields as well as applications to a parametrized post-Newtonian metric. Our notational conventions and a list of symbols can be found in Appendix \ref{app_conventions}.
\section{Nonrelativistic geoid}
\label{Sec_NonRelGeoid}

The field equation that Newtonian gravity is based upon is the Poisson equation
\begin{align}
	\label{Eq_NewtonianFieldEquation}
	\Delta U = 4\pi G \rho \, ,
\end{align}
where $U$ is the Newtonian gravitational potential, $G$ is Newton's gravitational constant, and $\rho$ is the mass density of the gravitating source. In the region outside the source, i.e.\ in vacuum, the field equation reduces to the Laplace equation $\Delta U = 0$. 

On the rotating Earth, the centrifugal effects give an additional contribution to the acceleration of a freely falling particle that is dropped from rest. This total acceleration can be derived from the potential
\begin{align}
	\label{Eq_NewtonianTotalPotential}
	W = U + V = U - \dfrac{1}{2} \Omega^2 d_z^2 \, .
\end{align}
Here, $V$ is the centrifugal potential, $\Omega$ is the angular velocity of the Earth, and $d_z$ is the distance to the rotation axis, which is defined as the $z$-axis. Whereas the attractive gravitational potential is a harmonic function in empty space, the centrifugal part is not.

The shape of the Earth as well as its gravity field shows an enormous complexity. The idea of using an equipotential surface for defining an idealized ``mathematical figure of the Earth'' was brought forward by C.\ F.\ Gauss in 1828. The name \emph{geoid} was coined by J.\ F.\ Listing in 1873. In modern terminology, here quoted from the U.S.\ National Geodetic Survey \cite{NGS}, the geoid is defined as ``the equipotential surface of the Earth's gravity field which best fits, in a least squares sense, global mean sea level.'' Here, the term ``equipotential surface'' refers to the potential $W$ in Eq.\ \eqref{Eq_NewtonianTotalPotential}. The question of which equipotential surface is chosen as the geoid is largely a matter of convention; for the Earth, it is convenient to choose a best fit to the sea level, while for celestial bodies without a water surface, such as Mars or the Moon, one could choose a best fit to the surface. 

In a strict sense, the geoid is not time independent because the Earth undergoes various kinds of deformations and its angular velocity is not strictly constant. However, all temporal variabilities  may be treated as perturbations of a time-independent geoid. For having such a time-independent geoid, one makes the following idealizing assumptions:
\begin{itemize}
\item[(A1)]   
The Earth is in rigid motion.
\item[(A2)]   
The Earth rotates with constant angular velocity about a fixed rotation axis.
\item[(A3)]   
There are no external forces acting on the Earth.
\end{itemize}
Note that assumption (A3) also excludes time-independent deformations caused by other gravitating bodies such as the so-called ``permanent tides;'' see, e.g.\ Ref.\ \cite{TorgeMueller:2012}. Just as the time-dependent variations mentioned above, they may be considered as perturbations at a later stage.
Physical effects that must be treated in that way include, among others, the intrinsic time dependence of the mass multipoles, tidal effects, anelastic deformations, friction, ocean loading, atmospheric effects, mass variations in the hydrosphere and cryonosphere, and postglacial mass variations.

In geodesy, different notions of the geoid are commonly used. See, e.g.\ the standard textbook on geodesy \cite{TorgeMueller:2012} for the definitions of the mean geoid, the non-tidal geoid, and the zero-geoid. In this work, since we exclude the influence of external forces by assumption (A3), we refer to the concept of the non-tidal geoid.
 
The assumptions (A1), (A2), and (A3) guarantee the existence of the time-independent potential $W$ as given in Eq.\ \eqref{Eq_NewtonianTotalPotential}; the geoid is then defined as the time-independent surface
\begin{align}
	\label{Eq_NewtonianGeoid}
	W = W_0 \, ,
\end{align}
with the constant $W_0$ chosen by an appropriate convention, as indicated above. By definition, the geoid is perpendicular to the acceleration 
\begin{align}
 \nabla W = \nabla U + \nabla V \, .
\end{align}
The magnitude $|\nabla W|$ is called gravity in the geodetic community. The gravitational part of the potential is usually expanded into spherical harmonics, cf., e.g.\ Refs.\ \cite{TorgeMueller:2012, Barthelmes:2013},
\begin{multline}
	\label{Eq_NewtonianPotenialDecomposition}
	U = -\dfrac{GM}{r} \sum_{l=0}^\infty \sum_{m=0}^l \left( \dfrac{R_E}{r} \right)^{l} P_{lm}(\cos \vartheta) \left[ C_{lm} \cos(m \varphi) \right. \\ + \left. S_{lm} \sin(m \varphi) \right] \, .
\end{multline}
An additional assumption of axial symmetry reduces the decomposition \eqref{Eq_NewtonianPotenialDecomposition} to
\begin{align}
	\label{Eq_NewtonianPotenialAxialSymDecomposition}
	U = - G \sum_{l=0}^\infty N_l \dfrac{P_{l}(\cos \vartheta)}{r^{l+1}} \, .
\end{align}
Here, $M$ is the mass of the Earth, $R_E$ is some reference radius (e.g.\ the equatorial radius of the Earth), $(r, \vartheta, \varphi)$ are geocentric spherical coordinates, $P_{l}\, (P_{lm})$ are the (associated) Legendre polynomials, and $C_{lm}, S_{lm}, N_l$ are the multipole coefficients. In geodesy, Eq.\ \eqref{Eq_NewtonianPotenialAxialSymDecomposition} is often rewritten as
\begin{align}
	\label{Eq_NewtonianPotenialAxialSymDecomposition2}
  	U = -\dfrac{GM}{r} \sum_{l=0}^\infty \left( \dfrac{R_E}{r} \right)^{l} J_l \, P_{l}(\cos \vartheta) \, ,
\end{align}
where the relation between the dimensionless quantities $J_l$ and the multipole moments $N_l$ is given by ${N_l = J_l R_E^l M}$.

The multipole coefficients $C_{lm}, S_{lm}$ (or $N_l$ in an axisymmetric model) can be determined by different measurements. Among others, satellite missions such as GOCE and GRACE as well as ground-based gravimetry and leveling observations on the surface of the Earth contribute to the knowledge of the gravitational field and the derivation of precise models of the geoid \cite{TorgeMueller:2012}. Modern space missions use laser ranging (LAGEOS), laser interferometry (GRACE-FO), and GPS tracking for providing such precise models.

We end this section by rewriting the three assumptions (A1), (A2), and (A3), which guarantee the existence of a time-independent geoid, in a way that facilitates comparison with the relativistic version to be discussed below. We start out from the well-known transformation formula from an inertial system $\Sigma$ to a reference system $\Sigma '$ attached to a rigidly moving body,
\begin{align}
	\label{Eq_rigid}
  	\vec{x} = \vec{x}{}_0 (t) + \boldsymbol{R} (t) \, \vec{x}{\,'} \, .
\end{align}
Here, $\vec{x}{}_0 (t)$ is the position vector in $\Sigma$ of the center of mass of the central body and $\boldsymbol{R} (t)$ is an orthogonal matrix that describes the momentary rotation of the central body about an axis through its center of mass. The orthogonality condition $\boldsymbol{R} (t)^{-1} = \boldsymbol{R} (t)^T$ implies that the matrix
\begin{align}
	\label{Eq_omega}
  	\boldsymbol{\omega} (t) = {\dot{\boldsymbol{R}}}{}(t) \, \boldsymbol{R}{}(t)^{-1}
\end{align}
is antisymmetric. From Eq.\ \eqref{Eq_rigid}, we find that
\begin{align}
	\label{Eq_v}
  	\vec{v} = \dot{\vec{x}} = \dot{\vec{x}}{}_0 + \boldsymbol{\omega}  \, (\vec{x} - \vec{x}{}_0) \, ,
\end{align}
where the dot means a derivative with respect to $t$, keeping $\vec{x}{\, '}$ fixed. Successive differentiation results in
\begin{align}
	\label{Eq_a}
  	\vec{a} = \dot{\vec{v}} = \ddot{\vec{x}}{}_0 + \dot{\boldsymbol{\omega}} \, (\vec{x} - \vec{x}{}_0) + \boldsymbol{\omega} \, (\vec{v} - \dot{\vec{x}}{}_0) \, , \\
	\label{Eq_dota}
  	\dot{\vec{a}} = \dddot{\vec{x}}{}_0 + \ddot{\boldsymbol{\omega}} \, ( \vec{x} - \vec{x}{}_0 ) + 2 \, \dot{\boldsymbol{\omega}} \, ( \vec{v} - \dot{\vec{x}}{}_0 ) + \boldsymbol{\omega} \, ( \vec{a} - \ddot{\vec{x}}{}_0 ) \, . 
\end{align}
We will now verify that the three assumptions (A1), (A2), and (A3) imply the following:
\begin{itemize}
\item[(A1')]   
The velocity gradient $\nabla \otimes \vec{v}$ is antisymmetric.
\item[(A2')]   
$\dot{\boldsymbol{\omega}} = 0$.
\item[(A3')]   
$\dot{\vec{a}} = \boldsymbol{\omega} \, \vec{a} $.
\end{itemize}
Clearly, from Eq.\ \eqref{Eq_v}, we read that the assumption of rigid motion implies (A1'). Moreover, (A2) obviously requires (A2'). Finally, (A3) implies that $\ddot{\vec{x}}{}_0 (t) = \vec{0}$ (which means that we may choose the inertial system such that $\vec{x}{}_0= \vec{0}$); this result inserted into \eqref{Eq_dota}, together with (A2'), gives indeed (A3'). The three conditions (A1'), (A2'), and (A3'), which are necessary for defining a time-independent geoid in the Newtonian theory, have natural analogs in the relativistic theory as we will demonstrate below.  


\section{Relativistic geoid}

Since clocks are the most precise measurement devices that modern technology offers, a relativistic definition of the geoid that is based on time and frequency measurements might be most convenient and operationally realizable with high accuracy. In one of the first articles on a relativistic treatment of geodetic concepts Bjerhammar \cite{Bjerhammar:1985}, see also Ref.\ \cite{Bjerhammar:1986}, proposed the following definition:

\begin{quote}
\textit{The relativistic geoid is the surface nearest to mean sea level on which precise clocks run with the same speed.}
\end{quote}

\subsection{Redshift potential}\label{subsec:redpot}

If one wants to translate Bjerhammar's definition into the language of mathematics, one has to specify what ``precise clocks'' are and what is meant by saying that clocks ``run at the same speed''. Presupposing the formalism of general relativity, without approximations, we suggest the following: ``precise clocks'' are standard clocks, i.e.\ clocks that measure proper time along their respective worldlines. The notion of standard clocks is mathematically well defined in the formalism of general relativity by the condition that for a worldline parametrized by proper time the tangent vector is normalized; moreover, standard clocks can be equivalently characterized by an operational definition with the help of light rays and freely falling particles, using the notions of radar time and radar distance; see Perlick \cite{Perlick:1987}. When comparing predictions from general relativity with observations one always assumes that atomic clocks are standard clocks. This hypothesis is in agreement with all experiments to date.

\begin{figure}
	\centering
	\includegraphics[width=0.38\textwidth]{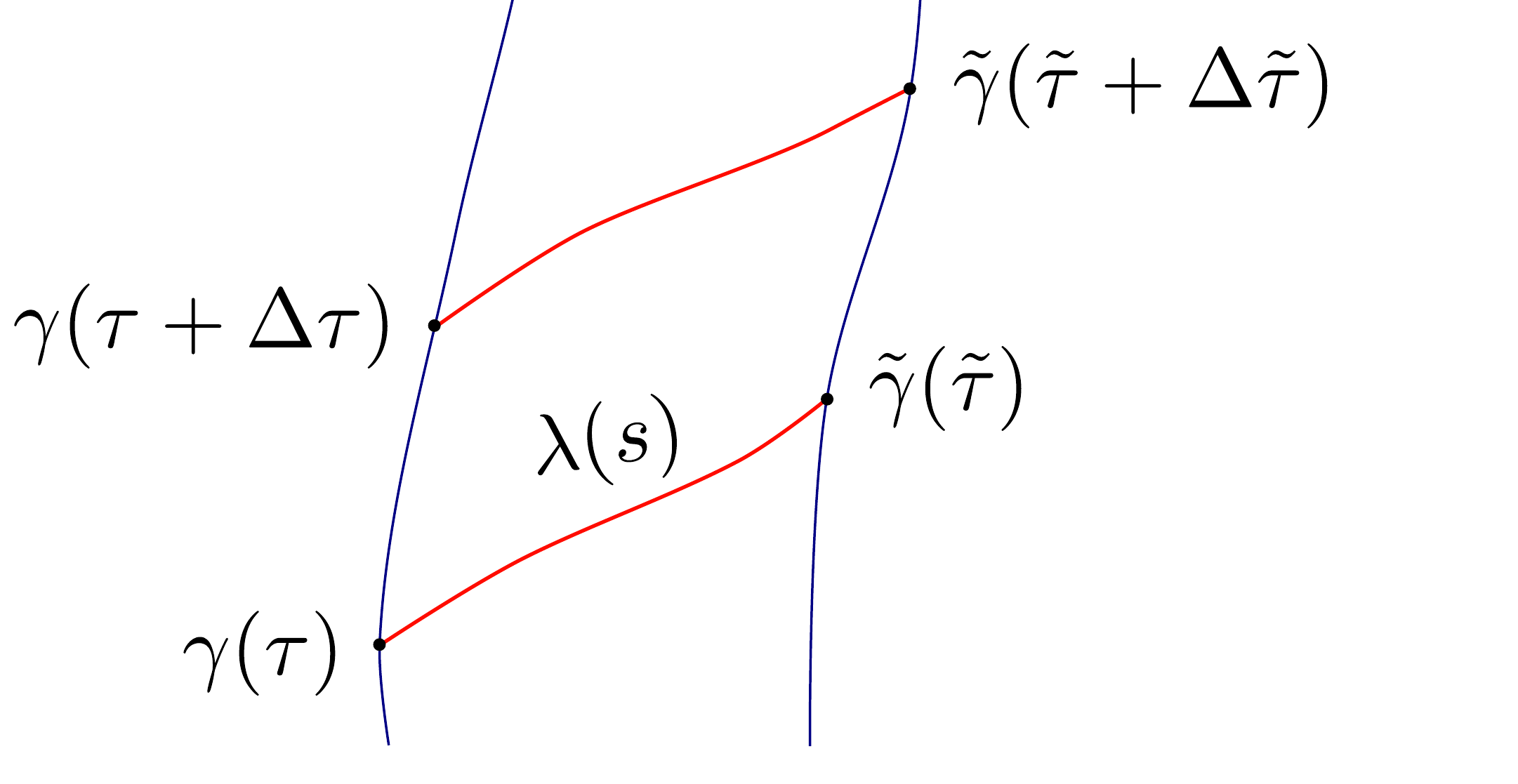}
	\caption{Definition of the redshift in general relativity: exchanging light signals between two worldlines $\gamma$ and $\tilde{\gamma}$.}
	\label{Fig_redshift}
\end{figure}

Knowing what is meant by ``precise clocks,'' we still have to explain what we mean by saying that two clocks ``run at the same speed''. For comparing two clocks, it is obviously necessary to send signals from one clock to the other. In a general relativistic setting, it is natural to use light signals which, in the mathematical formalism, are given by lightlike geodesics. This gives rise to the following well-known definition of the general-relativistic redshift: let $\gamma$ and $\tilde{\gamma}$ be the worldlines of two standard clocks that measure proper times $\tau$ and $\tilde{\tau}$, respectively. Assume that a light ray $\lambda$ is emitted at $\gamma(\tau)$ and received at $\tilde{\gamma}({\tilde{\tau}})$ while a second light ray is emitted at $\gamma ( \tau + \Delta \tau )$ and received at $\tilde{\gamma}({\tilde{\tau}} + \Delta \tilde{\tau})$, see Fig.\ \ref{Fig_redshift}. 
One defines the redshift $z$ by 
\begin{align}
	\label{Eq_RedshiftDefinition}
	z+1 = \dfrac{\nu}{\tilde{\nu}} = \dfrac{d \tilde{\tau}}{d \tau} = \underset{{\Delta \tau \to 0}}{\mathrm{lim}}  \dfrac{\Delta \tilde{\tau}}{\Delta \tau} \, ,
\end{align}
where $\nu$ and $\tilde{\nu}$ are the frequencies measured by the emitter $\gamma$ and by the receiver $\tilde{\gamma}$, respectively. In general relativity there is a universal formula for the redshift of standard clocks \cite{Kermack:1934},
\begin{align}
	\label{Eq_RedshiftGR}
	z+1 = \dfrac{\nu}{\tilde{\nu}} = \dfrac{\left( \left. g_{\mu\nu} \dfrac{d\lambda^\mu}{ds} \, \dfrac{d\gamma^\nu}{d\tau} \right) \right|_{\gamma(\tau)} }{\left( \left. g_{\rho\sigma} \dfrac{d \lambda^\rho}{ds} \, \dfrac{d \tilde{\gamma}^\sigma}{d \tilde{\tau}} \right) \right|_{\tilde{\gamma}(\tilde{\tau})}} \, .
\end{align}
Here, $s$ is an affine parameter for the lightlike geodesic $\lambda$. A simple derivation of the redshift formula was given by Brill \cite{Brill:1972}; this derivation can also be found in the book by Straumann \cite{Straumann:1984}. We are now ready to explain how we interpret the statement that $\gamma$ and $\tilde{\gamma}$ run at the same speed: it is supposed to mean that $z=0$.

In this interpretation, Bjerhammar's definition requires pairwise vanishing redshift for an entire family of clocks. Therefore, we now consider a congruence of worldlines and we ask for the redshift of any pair of worldlines in this congruence. The congruence is defined by a four-velocity field $u$, which is normalized according to $g_{\mu\nu}u^\mu u^\nu=-c^2$, i.e.\ such that its integral curves are parametrized by proper time. We say that $\phi$ is a \emph{redshift potential} for $u$ if 
\begin{align}
	\label{Eq_RedPot}
	\log (z+1) = \phi \big( \tilde{\gamma} (\tilde{\tau} ) \big) - \phi \big( \gamma ( \tau ) \big)
\end{align}
for any two integral curves $\gamma$ and $\tilde{\gamma}$ of $u$. According to Ref.\ \cite{Hasse:1988}, $\phi$ is a redshift potential if and only if ${\exp(\phi)u =: \xi}$ is a conformal Killing vector field of the spacetime. The redshift potential is time independent (i.e.\ constant along the integral curves of $\xi$) if and only if $\xi$ is a Killing vector field. The integral curves of $u$ are then called \emph{Killing observers}. The existence of a time-independent redshift potential is, thus, guaranteed if and only if the spacetime is stationary. In this case, we may introduce coordinates $(t, x^1,x^2,x^3)$ with $\xi = \partial_t$ such that the metric reads
\begin{align}
	\label{Eq_StationaryMetric}
    g = e^{2 \phi (x) } \left[ -(c \, dt + \alpha_a(x) dx^a)^2 + \alpha_{ab}(x) dx^a dx^b \right] \, ,
\end{align}
where the metric functions $\phi$, $\alpha _a$, and $\alpha _{ab}$ depend on $x=(x^1,x^2,x^3)$ but not on $t$.

The redshift potential $\phi (x)$ foliates the three-dimensional space into surfaces which we call \emph{isochronometric surfaces}. According to Eq.\ \eqref{Eq_RedPot}, any two standard clocks, mathematically described by integral curves of the vector field $u = \exp (- \phi ) \xi$, that are on the same isochronometric surface $\phi = \phi_0 = \text{constant}$ show zero redshift with respect to each other. We are thus led to the conclusion that Bjerhammar's definition (with our interpretation of his wording) makes sense in any stationary spacetime, and that the geoid is an isochronometric surface.

One might ask if the assumption of stationarity is really necessary for this definition to make sense. As a matter of fact, it can be shown that a four-velocity field $u$ must be proportional to a Killing vector field if any two clocks on integral curves of $u$ see each other with temporally constant redshift and if these integral curves are complete; see Theorem 10 in Ref.\ \cite{Perlick:1990}. This demonstrates that, based on redshift measurements, a time-independent geoid can be defined only in the case of stationarity.

We end this subsection by briefly discussing the notion of a redshift potential in the Newtonian limit. Given a stationary spacetime with a metric in the form above, the redshift potential $\phi$ is given by the equation 
\begin{align}
	\label{Eq_gtt}
  	c^2 e^{2\phi} = -g_{\mu\nu} \xi^\mu \xi^\nu = -g_{tt} \, .
\end{align}
Clearly, the redshift between any two stationary standard clocks (i.e.\ standard clocks of which the worldlines are integral curves of the vector field $u = \exp(- \phi ) \xi$) is
\begin{align}
	\label{Eq_RedshiftByRedshiftPotential}
	z+1 = \dfrac{\nu}{\tilde{\nu}} &= e^{\phi|_{\tilde{\gamma}} - \phi|_{\gamma}} = \dfrac{ e^{\phi}|_{\tilde{\gamma}}}{e^{\phi}|_{\gamma}} = \dfrac{\sqrt{-g_{tt}}|_{\tilde{\gamma}}}{\sqrt{-g_{tt}}|_{\gamma}} \, .
\end{align}
For the Newtonian limit of general relativity, we know that in a suitable coordinate system $-g_{tt} \to c^2(1+2U/c^2)$; hence,
\begin{align}
	\label{Eq_RedshiftPotentialLimit}
	e^{\phi} \approx 1+U/c^2 \, .	
\end{align}
This demonstrates that in the Newtonian approximation the level sets of the redshift potential $\phi$ correspond to equipotential surfaces of the Newtonian gravitational potential $U$. In the same approximation, the redshift is determined by the potential difference between the emitter and receiver,
\begin{align}
	\label{Eq_RedshiftPotentialLimit2}
	\dfrac{\nu}{\tilde{\nu}} \approx 1+\dfrac{U_2-U_1}{c^2} =: 1 + \dfrac{\Delta U}{c^2} \, .
\end{align}
Near the surface of the Earth, such a potential difference corresponds to a height difference. From Eq.\ \eqref{Eq_RedshiftPotentialLimit2}, one concludes that the relative frequency change, i.e.\ the redshift, is about $10^{-16}$ per meter near the Earth's surface. Hence, modern clocks with a stability in the $10^{-18}$ regime can be used to measure height differences at the centimeter level.

Fig.\ 2 shows a sketch of the level sets of the redshift potential and fibers connecting these surfaces. The redshifts measured using fibers I and II are identical, whereas the redshift measured using fiber III vanishes.


\subsection{Clock comparison through optical fibers}\label{subsec:fiber}
\begin{figure}
	\centering
	\includegraphics[width=0.4\textwidth]{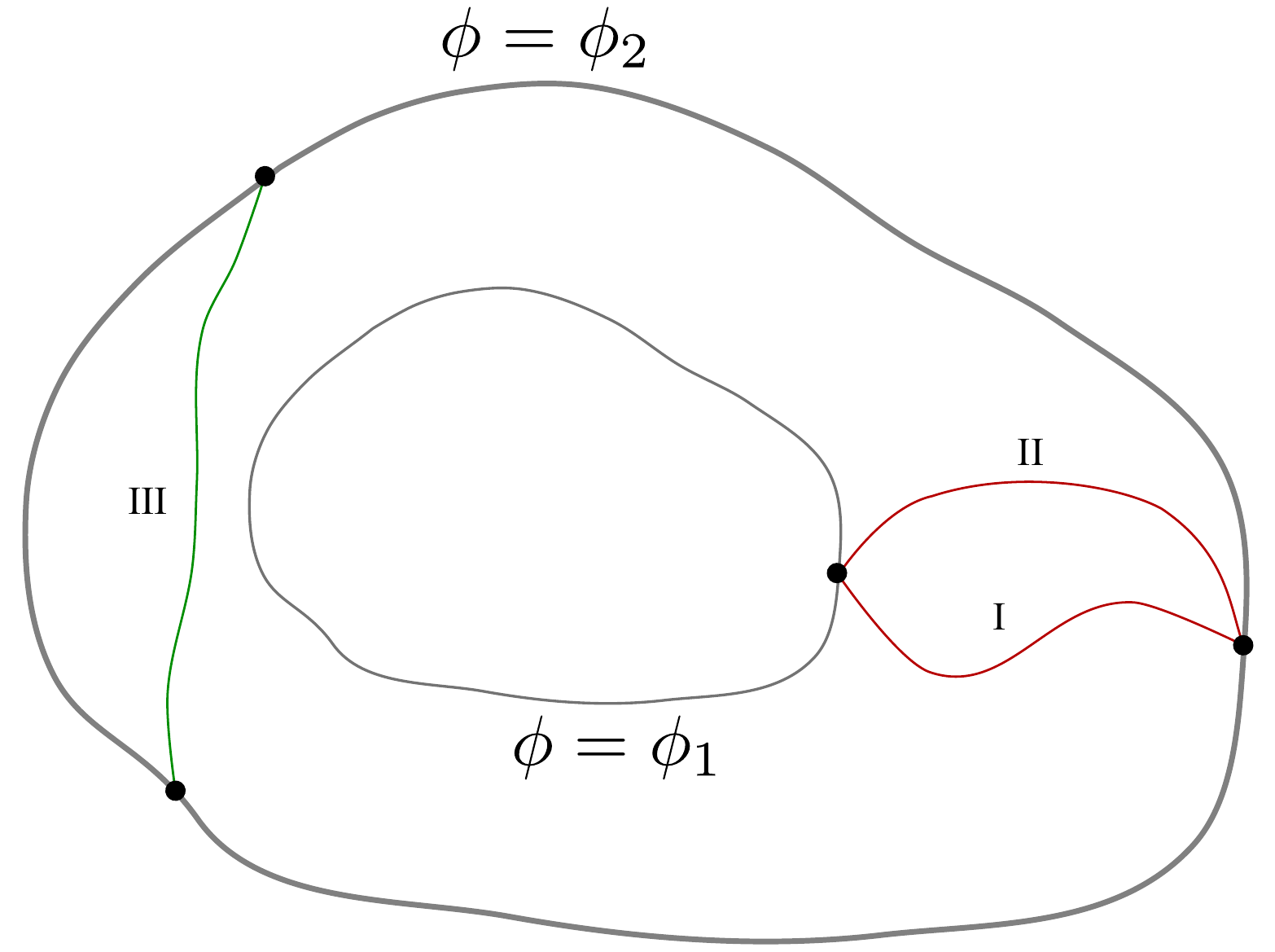}
	\caption{Sketch of surfaces of constant redshift potential $\phi$ and optical fibers connecting them. The redshift is independent of the spatial shape of the chosen fiber as long as the fibers are at rest with respect to the Killing observers. The redshifts measured using fiber I and fiber II will be identical, whereas the redshift measured using fiber III is zero.}
	\label{Fig_fiber}
\end{figure}
The general redshift formula \eqref{Eq_RedshiftGR} is valid only if the comparison between the two clocks is made with the help of freely propagating light rays, i.e.\ with the help of lightlike geodesics. We will now show that, by contrast, in the case of a stationary spacetime, the formula \eqref{Eq_RedPot} is valid whenever the comparison between the two clocks is made with signals that move at the speed of light, even if they are not freely propagating (i.e.\ nongeodesic). This has the important consequence that this formula may be used if the signals are transmitted through an optical fiber. We have to assume that the fiber is at rest with respect to the Killing observers, i.e.\ that it establishes a time-independent path in the coordinate representation \eqref{Eq_StationaryMetric} of the metric. A signal that propagates along this fiber with the speed of light has to satisfy the condition
\begin{align}
	g_{\mu\nu} \dot{x}^\mu \dot{x}^\nu = 0 \, ,
\end{align}
where the dot denotes the derivative with respect to a curve parameter $s$. As the signal is future oriented, this is equivalent to
\begin{align}
	c \, dt + \alpha_a dx^a = \sqrt{\alpha_{ab}dx^a dx^b} \, .
\end{align}
As a consequence, the coordinate travel time 
\begin{align}
	\label{Eq_FiberCoordinateTravelTime}
	& \Delta t := t_2 - t_1 = \int _{t_1} ^{t_2} dt 
	\notag \\
	& = \dfrac{1}{c} \int _{s_1} ^{s_2} \left( \sqrt{\alpha_{ab}\dfrac{dx^a}{ds} 
	\dfrac{dx^b}{ds}} - \alpha_c \dfrac{dx^c}{ds} \right) ds 
\end{align}
of the signal through the fiber is independent of the emission time since $\partial_t \alpha_a = 0$ and $\partial_t \alpha_{ab} = 0$. This implies that two signals that are emitted with a time difference $\Delta t$ will be received with the same time difference $\Delta t$. Together with the fact that, for observers with four-velocity $u = \exp(-\phi) \partial_t$, proper time and coordinate time are related by
\begin{align}
	\dfrac{d \tau}{dt} = e^{\phi} \, ;
\end{align}
this shows that the redshift of signals sent through the fiber is 
\begin{align}
	z + 1 = \dfrac{\nu}{\tilde{\nu}} =  \dfrac{d\tilde{\tau}}{d\tau} = \dfrac{d\tilde{\tau}}{dt} \dfrac{dt}{d\tau} = \dfrac{e^{\phi}|_{\tilde{\gamma}}}{e^{\phi}|_{\gamma}} \, .
\end{align}
Hence, the redshift potential also gives the correct frequency ratio $\nu / \tilde{\nu}$ for clock comparison by signal transmission through an arbitrarily shaped optical fiber, provided that the fiber is at rest with respect to the Killing observers.

Using the framework of optical metrics, see for instance Ref.\ \cite{Perlick:2000}, we can also consider fiber links with an index of refraction $n$ in which the signal does not propagate with the vacuum speed of light as assumed above. Instead of Eq.\ \eqref{Eq_StationaryMetric}, the metric now reads
\begin{align}
	\label{Eq_opticalMetric}
	g = e^{2 \phi (x) } \left[ -n(x)^{-2}(c \, dt + \alpha_a(x) dx^a)^2 + \alpha_{ab}(x) dx^a dx^b \right] \, .
\end{align}
We again assume that the fiber is at rest w.r.t.\ the Killing observers, i.e.\ w.r.t.\ the emitter and observer of the signal. The redshift between the two ends of the fiber now results in
\begin{align}
	z + 1 = \dfrac{\nu}{\tilde{\nu}} = \dfrac{e^{\phi}|_{\tilde{\gamma}}}{e^{\phi}|_{\gamma}} \dfrac{n|_{\gamma}}{n|_{\tilde{\gamma}}}  \, ,
\end{align}
such that, again, the redshift potential $\phi$ gives the correct result for frequency comparison if the index of refraction is constant. As can be seen by the equation above, the vacuum redshift potential $\phi$ can also be deduced from redshift measurements using optical fibers when the position-dependent index of refraction of the fiber is known.


\subsection{Definition of the relativistic geoid}\label{subsec:defgeoid}

Based on our deliberations in Sec.\ \ref{subsec:redpot}, we suggest the following definition of the relativistic geoid:

\begin{quote}
\textit{The relativistic geoid is the level surface of the redshift potential $\phi$ that is closest to mean sea level.} 
\end{quote}

In the case of celestial bodies without a water surface, one has to single out one particular level surface of the redshift potential by some other convention. This definition of the relativistic geoid makes sense for any celestial body that is associated with a stationary spacetime, i.e.\ with a family of Killing observers. In the next section, we will show that the assumption of stationarity is tantamount to three conditions that are analogous to the three conditions (A1'), (A2'), and (A3'), which are necessary for defining a time-independent geoid in the Newtonian theory; recall Sec.\ \ref{Sec_NonRelGeoid}.

Our definition is operational in the sense that standard clocks and fiber links can be used to determine the relativistic geoid. A clock network may be built such that all clocks show pairwise zero redshift, and one of them is positioned at mean sea level. The spatial grid of clocks then determines the shape of the Earth's geoid. 

We emphasize that our definition of the geoid allows for arbitrarily strong gravitational fields. For weak fields, we may use the Newtonian limit for which the redshift potential can be expressed in terms of the Newtonian potential; see Sec.\ \ref{subsec:redpot}. In this limit, our definition of the geoid becomes the usual Newtonian one. At the PN level, our geoid reduces to the u-geoid of Soffel \emph{et al.}\ \cite{Soffel:1988}. 

Our definition of the geoid should be compared with the one by Oltean \emph{et al.}\ \cite{Oltean:2015}, which is also fully relativistic. A major difference is in the fact that we give an operational definition in terms of clocks that are connected by fiber links while their mathematical construction is not immediately related with an operational prescription. In particular, they do not make any reference to clocks. 


\section{General relativistic model of the solid earth}

Our definition of the geoid requires stationarity, i.e.\ the existence of a timelike Killing vector field. In this section, we will recall some known facts about timelike congruences. They will demonstrate that the stationarity assumption is equivalent to a relativistic version of the three conditions (A1'), (A2'), and (A3') we have discussed in Sec.\ \ref{Sec_NonRelGeoid}.

\subsection{Rigid and isometric congruences}

We consider a timelike congruence of worldlines (see, e.g.\ Refs.\ \cite{Ehlers:1961, Ehlers:1993}), i.e.\ a family of timelike curves which do not intersect and fill a certain region of the four-dimensional spacetime. The tangents to the worldlines are given by a timelike vector field $u=u^\mu \partial_\mu$, which we assume to be normalized, $g_{\mu\nu}u^\mu u^\nu=-c^2$. We interpret $u$ as the four-velocity field of a gravitating body. On the surface of the body, $u$ may be interpreted as the four-velocity of observers with standard clocks that are attached to the surface. Moreover, we may extend $u$ into the exterior region where it may be interpreted as the four-velocity of observers hovering above the surface, e.g.\ in satellites. We will characterize the case that $u$ is proportional to a Killing vector field; in this case, the congruence is called \emph{isometric}. 

The projection onto the local rest space of the congruence is given by the projection operator
\begin{align}
	\label{Eq_ProjectionOperator}
	P^{\mu}_{\nu} = \delta^\mu_\nu + \dfrac{1}{c^2} \, u^\mu u_\nu \, .
\end{align}
The acceleration $a = a^\mu \partial_\mu$ of the congruence is defined by
\begin{align}
	a^\mu := \dot{u}^\mu = u^\nu D_\nu u^\mu \, .
\end{align}
The acceleration vanishes along a particular integral curve of $u$ if and only if this curve is a geodesic.

As in nonrelativistic physics, a congruence can be characterized by the kinematic quantities rotation $\omega_{\mu\nu}$, shear $\sigma_{\mu\nu}$, and expansion $\theta$,
\begin{subequations}
\begin{align}
	\label{Eq_KinematicQuantities}
	\omega_{\mu\nu} &:= P^{\rho}_{\mu} \, P^{\sigma}_{\nu} \, D_{[\sigma} u_{\rho]} = 
	D_{[\nu}u_{\mu]} + \dfrac{1}{c^2} \, \dot{u}_{[\mu} u_{\nu]} \, , \\ 
	\sigma_{\mu\nu} &:= P^{\rho}_{\mu} \, P^{\sigma}_{\nu} \, D_{(\sigma} u_{\rho)} - \dfrac{1}{3} \theta P_{\mu\nu} \notag \\
					&= D_{(\nu}u_{\mu)} + \dfrac{1}{c^2} \, \dot{u}_{(\mu} u_{\nu)} - \dfrac{1}{3} \theta P_{\mu\nu} \, , \\
	\theta 			&:= D_\mu u^\mu \, .
\end{align}
\end{subequations}
The rotation is antisymmetric, while the shear is symmetric and traceless. The motion of neighboring worldlines with respect to a chosen worldline with tangent $u$ is determined by
\begin{equation}
	\label{Eq_CongruenceMotion}
	D_\nu u_\mu = \omega_{\mu\nu} + \sigma_{\mu\nu} + \dfrac{1}{3} \theta P_{\mu\nu} - \dfrac{1}{c^2} \, u_\nu a_\mu \, .
\end{equation}

A congruence with vanishing expansion, $\theta = 0$, is isochoric, i.e.\ the volume of a comoving spatial region does not change over time \cite{Ehlers:1961, Ehlers:1993}. If the shear vanishes as well, $\sigma_{\mu\nu} = 0$, the congruence is called \emph{Born rigid}. This is true if and only if the spatial distance between any two infinitesimally neighboring integral curves of $u$ remains constant over time. In this case, Eq.\ \eqref{Eq_CongruenceMotion} reduces to
\begin{align}
	\label{Eq_CongruenceMotionBornRigid}
	D_\nu u_\mu	= \omega_{\mu\nu} - \dfrac{1}{c^2} \, u_\nu a_\mu \, .
\end{align}
In analogy to the Newtonian condition (A1'), we require the congruence to be Born rigid, i.e.:
\\[0.2cm]
(A1'') 
$P^{\rho}_{\mu} \, P^{\sigma}_{\nu} \, D_{(\sigma} u_{\rho)}
= 0 \, $.
\\[0.2cm]
For defining the analogs of the Newtonian conditions (A2') and (A3'), we introduce the rotation four-vector $\omega^\mu$ by
\begin{align}
	\omega^\mu := \dfrac{1}{2c} \eta^{\mu\nu\sigma\lambda} u_\nu \omega_{\sigma\lambda} = 
	\dfrac{1}{c} \, \eta^{\mu\nu\sigma\lambda} u_\nu \partial_\lambda u_{\sigma} \, .
\end{align}
As $\omega^\mu u_\mu = 0$, the vector $\omega ^{\mu}$ is spacelike. If we write it in the form $\omega ^{\mu} = \omega \, e^{\mu}$ with $e^{\mu}e_{\mu} =1$, the unit vector $e^{\mu}$ gives the direction of the momentary rotation axis, and the scalar $\omega$ gives the modulus of the momentary angular velocity. The Newtonian requirements (A2') and (A3') now translate into the following conditions: 
\\[0.2cm]
(A2'')
$P^{\mu}_{\nu} \dot{\omega}^\nu = 0$.
\\[0.2cm]
(A3'')
$P^{\mu}_{\nu} \dot{a}^\nu = \omega^{\mu}{}_{\nu} a^\nu$.
\\[0.2cm]
Condition (A2'') states that the unit vector $e^{\mu}$ is Fermi-Walker transported and that the scalar $\omega$ is constant along each worldline of the congruence; in other words, it states that the rotation axis and the angular velocity are time independent. Condition (A3'') states that the change of the acceleration along the congruence is only due to the rotation and that the acceleration vector always points to the same neighboring worldline. 


\subsection{Acceleration potential}
Ehlers \cite{Ehlers:1961} has shown that for a rigid congruence the two requirements (A2'') and (A3'') together are equivalent to
\begin{align}
	D_{[\nu} a_{\mu]} = 0 \, .
\end{align}
The latter condition means that there exists a potential $\phi$ for the acceleration,
\begin{equation}
	\label{Eq_CongrucenceAcceleration}
	a_\mu = c^2 \partial_\mu \phi \, .
\end{equation}
This, in turn, is true for a rigid congruence if and only if $u$ is proportional to a timelike Killing vector field $\xi$ \cite{Salzmann:1954}, where the proportionality is given by
\begin{align}
	\label{Eq_CongruenceProportionality}
	\xi = e^\phi u \, .
\end{align}
Clearly, $\phi$ is equal to the redshift potential considered above. We have now seen that at the same time it plays the role of an acceleration potential. Moreover, we have seen that stationarity is equivalent to the three conditions (A1''), (A2''), and (A3''). A congruence with these properties is called isometric. The existence of a time-independent redshift potential is thus based on assumptions that are quite analogous to the assumptions (A1'), (A2'), and (A3') we have discussed in the Newtonian theory. 

The Killing vector field $\xi$ corresponds to a corotating family of observers. Note that $\xi$ is defined and timelike on a cylindrical neighborhood of the body. This neighborhood extends to infinity for a nonrotating (isolated) body but for a rotating body it is finite. If extended outside of this neighborhood, the Killing vector field becomes spacelike.


\subsection{General relativistic geoid revisited}

We summarize our observations in the following way. We have seen that a natural generalization of the classical assumptions (A1'), (A2'), and (A3') requires the congruence associated with the Earth to be isometric, i.e.\ the spacetime to be stationary. The assumption of stationarity gives rise to a time-independent potential $\phi$ with two properties. First, $\phi$ is a redshift potential, which means that the surfaces $\phi = \mathrm{constant}$ in 3-space are isochronometric. Second, $\phi$ is an acceleration potential, which means that the acceleration $a^{\mu}$ (which is a spatial vector field) is the gradient of the surfaces $\phi = \mathrm{constant}$ in 3-space. Note that freely falling particles undergo the acceleration $- a^{\mu}$ relative to comoving observers. Therefore, the acceleration of freely falling bodies on the Earth, e.g.\ in falling corner-cube devices, is governed by the potential $\phi$. By the same token, plumb lines are perpendicular to the surfaces $\phi = \mathrm{constant}$.

As a consequence, we could rewrite our definition of the relativistic geoid, as it is given in Sec.\  \ref{subsec:defgeoid}, by replacing the words ``redshift potential'' with the words ``acceleration potential.'' The geoid may be determined by a family of Killing observers with standard clocks. Once a reference point defining the mean sea level has been chosen, the geoid may be realized either by clock comparison or by measuring the gravitational acceleration in falling corner cubes as shown by Eqs.\ \eqref{Eq_CongrucenceAcceleration} and \eqref{Eq_RedshiftByRedshiftPotential}. In this sense, one may say that also in the full relativistic theory the notions of the u-geoid and a-geoid are equivalent; it was already mentioned that a similar result was proven by Soffel \emph{et al.}\ \cite{Soffel:1988} in a PN setting. This fact is very convenient because it implies that the geoid may be determined with two independent types of measurements that complement each other. As the notions of redshift potential and acceleration potential coincide, we will speak just of the \emph{relativistic potential} in the following.

Our definition of the geoid is based on the assumption of stationarity. Of course, this is only an approximation. Just as in the Newtonian theory, temporal variations may be taken into account by modifying the time-independent (rigid) geoid by time-dependent perturbations, i.e.\ by considering a nonstationary metric $\Sigma_{\mu\nu}$ of the form 
\begin{align}
	\Sigma_{\mu\nu} = g_{\mu\nu} + h_{\mu\nu} 
\end{align}
where $g_{\mu\nu}$ is stationary. In practical geodesy, the stationary part is defined as the mean value over a sufficiently long time interval. Thus, this part also contains the permanent tide effects from the external gravitational field of celestial bodies like the Moon or the Sun. For the stationary part $g_{\mu\nu}$, we may still use our definition of the geoid in terms of a relativistic potential $\phi$. In this paper, we will not work out a theory for such time-dependent perturbations of the relativistic geoid. For examples of such effects, we refer to the list given in Sec.\  \ref{Sec_NonRelGeoid}.

However, as our formalism also applies, e.g.\ to rapidly rotating neutron stars with ``mountains'' and other non-axisymmetric stationary objects, we should mention that our assumption of stationarity ignores the fact that an irregularly shaped rotating body emits gravitational radiation, so its angular velocity will actually not be constant over time. Of course, this is a small effect; for the Earth and other planets, it is completely negligible.

For rigid motion inside the gravitating body, the four-velocity field $u$ and, consequently, the Killing vector field $\xi$ are defined within the interior as well. The extension of equipotential surfaces (i.e.\ of the geoid) to regions inside the body is also well defined. An interior solution should be considered, and the corresponding isochronometric surfaces need to be calculated. The particular interior solution must be matched, at the surface, to the vacuum solution. The level surface that defines the geoid by the condition of pairwise vanishing redshift for any two clocks on this particular surface will then be continuous but in general not differentiable.

In the following two sections, we consider axisymmetric static and axisymmetric stationary spacetimes, respectively, and we determine the isochronometric surfaces for various examples of such spacetimes. Of course, axisymmetric models are highly overidealized in view of applications to the Earth; see e.g.\ the analysis in Ref.\ \cite{Soffel:2016}. However, we believe that these examples are instructive because they illustrate the general idea behind our definition and its applicability to compact objects. We emphasize that our general definition of the geoid does of course not assume axisymmetry or any other kind of spatial symmetry. However, the axisymmetric stationary case is mathematically distinguished by the fact that then we have two linearly independent Killing vector fields; one of them is timelike and hypersurface orthogonal near spatial infinity. This allows the use of asymptotically defined time-independent multipole moments; see below. The only other case where a Killing vector field exists that is timelike up to spatial infinity and hypersurface orthogonal (near spatial infinity) is the case of a static (i.e.\ nonrotating) gravitating body. In the exterior of an irregularly shaped rotating body, we have only one Killing vector field, which becomes spacelike at a certain distance from the rotation axis; in this case, the asymptotic definition of time-independent multipole moments is not applicable.

All our examples are vacuum solutions of Einstein's field equation. For modeling a gravitating body they have to be matched to an interior matter solution. Correspondingly, the isochronometric surfaces we are calculating are valid only outside of the gravitating body.


\section{Axisymmetric static spacetimes}


\subsection{Axisymmetric static solutions to Einstein's vacuum field equation}
\label{Sec_staticSpacetimes}

Any axisymmetric and static spacetime that satisfies Einstein's vacuum field equation is given by the Weyl metric \cite{Weyl:1917}
\begin{multline}
	\label{Eq_WeylMetric}
	g _{\mu \nu} dx^{\mu} dx^{\nu} = - e^{2\psi} c^2 dt^2  + e^{- 2\psi} \rho^2 d\varphi^2 
	\\
	+ e^{-2\psi} e^{2\gamma} (d\rho^2 + dz^2)\, ,
\end{multline}
where $(t,\rho,z,\varphi)$ are Weyl's canonical coordinates. The metric functions $\psi$ and $\gamma$ depend only on the coordinates $\rho$ and $z$. The coordinates $t$ and $\varphi$ are associated with the two Killing vector fields $\partial_t$ and $\partial_\varphi$. Some important examples are the Schwarzschild metric, the Erez-Rosen metric \cite{Erez:1959}, and the $q$-metric \cite{Quevedo:2013} (Zipoy-Voorhees metric \cite{Zipoy:1966, Voorhees:1970}). Using the metric \eqref{Eq_WeylMetric}, the vacuum field equations reduce to, see e.g.\ Ref.\ \cite{Quevedo:1989},
\begin{subequations}
\begin{align}
	\label{Eq_FieldEqnWeyl}
	\Delta \psi &= 0 \, , \\
	\partial_\rho \gamma - \rho \, (\partial_\rho \psi + \partial_z \psi)(\partial_\rho \psi - \partial_z \psi) 
				&= 0 \, , \\
	\partial_z \gamma - 2 \rho \, \partial_\rho \psi \, \partial_z \psi 
				&= 0 \, .
\end{align}
\end{subequations}
The metric function $\gamma$ can be obtained by integration once the Laplace equation \eqref{Eq_FieldEqnWeyl} for $\psi$ has been solved. The general solution for all static, axisymmetric, and asymptotically flat spacetimes is given by \cite{Stephani:Book:2003}
\begin{subequations}
\begin{align}
\label{Eq_WeylMetricExpansion}
	\psi 	&= \sum_{l=0}^\infty c_l \dfrac{P_l(\cos \Theta)	}{R^{l+1}} \, , \\
	\gamma 	&= \sum_{l,i = 0}^\infty \dfrac{(i+1)(l+1)}{i+l+2} c_i c_l \notag \\
			&\times \dfrac{P_{l+1}(\cos \Theta) P_{i+1}(\cos \Theta) - P_l(\cos \Theta) P_i(\cos \Theta)}{R^{l+i+2}} \, ,
\end{align}
\end{subequations}
where $R^2 = \rho^2+z^2$ and $\cos \Theta = z/R$. The $P_l(\cos \Theta)$ are Legendre polynomials of degree $l$, and $c_l$ are constant expansion coefficients, sometimes called Weyl multipoles. 

The relativistic geoid is defined by the level sets of the time-independent redshift potential for observers that form an isometric congruence. Hence, their four-velocity field $u$ is proportional to a timelike Killing vector field $\xi$ as given by Eq.\ \eqref{Eq_CongruenceProportionality}. The relativistic potential $\phi$ is related to this Killing vector field by Eq.\ \eqref{Eq_gtt}. 

For the spacetime with line element \eqref{Eq_WeylMetric}, we have two linearly independent Killing vector fields, $\partial _t$ and $\partial _{\varphi}$. Note that any linear combination of these two Killing vector fields with \emph{constant} coefficients is again a Killing vector field. We consider I) the nonrotating congruence with worldlines that are integral curves of the timelike Killing vector field $\partial_t$ and II) a rotating congruence with worldlines that are integral curves of $\partial_t + \Omega \, \partial_\varphi$, with some $\Omega \in \mathbb{R}$. Note that the latter congruence is timelike only on a cylindrical domain about the symmetry axis; on the boundary of this domain, it becomes lightlike, and farther away from the axis, it is spacelike. The bigger the $\Omega$, the smaller the domain on which the congruence is timelike. Here, $\Omega$ has the dimension of an inverse time, i.e.\ the dimension of a frequency.

The first congruence, (I), is associated with observers of which the spatial Weyl coordinates $(\rho,\varphi,z)$ remain fixed; we can think of them as being attached to the surface of a ``nonrotating Earth''. The second congruence, (II), can be associated with observers attached to the surface of a ``rotating Earth'' where $\Omega$ is the angular velocity. As the metric is static, the gravitomagnetic field of the Earth is not taken into account. In the following, all quantities related to the first congruence, (I), will be denoted by the subscript $(\cdot)_{\text{stat}}$, while all quantities related to the second congruence, (II), will be denoted by the subscript $(\cdot)_{\text{rot}}$. We obtain, respectively, 
\begin{subequations}
\label{Eq_WeylMetricRedshiftStationary}
\begin{align}
	c^2 e^{2\phi_{\text{stat}}} &= -g(\partial_t,\partial_t) = c^2 e^{2\psi} \, , \\
	c^2 e^{2\phi_{\text{rot}}} &= - g(\partial_t + \Omega \, \partial_\varphi,\partial_t + \Omega \, \partial_\varphi) 
	\notag \\
	&= c^2 e^{2\psi} - \Omega^2 \rho^2 e^{-2\psi} \, .
\end{align}
\end{subequations}

The isochronometric surfaces for the respective congruence are defined by the level sets of $\phi$. Therefore we obtain
\begin{subequations}
\begin{align}
	e^{2\phi_{\text{stat}}} &= \text{constant} \Leftrightarrow e^{2\psi} =\text{constant} \, , \\
	e^{2\phi_{\text{rot}}} &= \text{constant} \Leftrightarrow e^{2\psi} - \dfrac{\Omega^2}{c^2} \rho^2 e^{-2\psi} = \text{constant}\, .
\end{align}
\end{subequations}
The relativistic geoid is one of these isochronometric surfaces, where the constant has to be chosen by a convention.  

Inserting the expansion \eqref{Eq_WeylMetricExpansion} gives the geoid in terms of the expansion coefficients $c_l$. However, this representation gives little insight into the geometry and the physical situation at hand: already for the simplest member of the Weyl class, the Schwarzschild spacetime, the coefficients must be chosen in a complicated way, such that the series \eqref{Eq_WeylMetricExpansion} converges to
\begin{align}
	\psi = \dfrac{1}{2} \log \left( \dfrac{r_+ + r_- - 2m}{r_+ + r_- + 2m} \right) \, , \quad r_\pm^2 := \rho^2 + (z\pm m)^2 \, .
\end{align}
The Schwarzschild metric in its usual form follows after the coordinate transformation
\begin{align}
	\label{Eq_CoordinateTrafo1}
	\dfrac{r}{m}-1 := \dfrac{r_+ + r_-}{2m} \, , \quad \cos \vartheta := \dfrac{r_+ - r_-}{2m}\, .
\end{align}
To obtain more physical insight, we introduce spheroidal coordinates $(x,y)$ by the coordinate transformation \cite{Quevedo:1989}
\begin{align}
	\label{Eq_CoordinateTrafo2}
	\rho^2 =: m^2(x^2-1)(1-y^2) \, , \quad z =: m x y \, ,
\end{align}
which is equivalent to
\begin{align}
\label{Eq_CoordinateTrafo3}
	x:=r/m-1 \, , \quad y := \cos \vartheta \, .
\end{align}
This yields the Weyl metric \eqref{Eq_WeylMetric} in spheroidal coordinates,
\begin{multline}
	\label{Eq_WeylMetricSpheroidal}
	g_{\mu \nu} dx^{\mu} dx^{\nu}  = -e^{2\psi} c^2 dt^2 + m^2 e^{-2\psi} (x^2-1)(1-y^2)d\varphi^2
	\\
	+ m^2 e^{-2\psi} e^{2\gamma} (x^2-y^2) 
	\left( \dfrac{dx^2}{x^2-1} + \dfrac{dy^2}{1-y^2} \right) \, .
\end{multline}
In these coordinates the relativistic potentials are, respectively,
\begin{subequations}
\label{Eq_WeylMetricSpheroidalRedshiftRotating}
\begin{align}
	e^{2\phi_{\text{stat}}} &= e^{2\psi} \, , \\
	e^{2\phi_{\text{rot}}} &= e^{2\psi} - \dfrac{\Omega^2}{c^2} \,  m^2 e^{-2\psi} (x^2-1)(1-y^2) \, .
\end{align}
\end{subequations}
The isochronometric surfaces and, thus, the geoid in these coordinates are, again, described by the respective level sets.

The vacuum field equation in the new coordinates can be found, e.g., in Refs.\ \cite{Quevedo:1989, Stephani:Book:2003}. In Ref.\ \cite{Quevedo:1989}, Quevedo has shown that the general asymptotically flat solution, with elementary flatness on the axis, in these coordinates is given by 
\begin{align}
	\label{Eq_WeylMetricExpansionSpheroidal}
	\psi = \sum_{l=0}^\infty (-1)^{l+1} q_l \, Q_l(x) \, P_l(y) \, ,	
\end{align}
where the $Q_l$ are the Legendre functions of the second kind as given in Ref.\ \cite{Bateman:1955}. The coefficients $q_l$ can be related to the $c_l$ in Eq.\ \eqref{Eq_WeylMetricExpansion}. Moreover, we will discuss in the next section how the $q_l$ are related to the relativistic multipole moments of the spacetime and, at the same time, to multipole moments of the Newtonian potential in the weak field limit. For the relativistic moments, we use those defined by Geroch and Hansen \cite{Geroch:1970b, Hansen:1974}.

In the representation \eqref{Eq_WeylMetricExpansionSpheroidal}, the Schwarzschild solution is obtained by simply choosing $q_0=1$ and $q_l = 0$ for all $l>0$; see Section \ref{Schwarzschild} below. For this choice of $q_0$, the parameter $m$ in \eqref{Eq_CoordinateTrafo1} is the usual mass parameter of the Schwarzschild solution, related to the Schwarzschild radius $r_s = 2m$.


\subsection{Newtonian limit}
 
Ehlers \cite{Ehlers:1981} gave a definition of the Newtonian limit that also yields a definition of the Newtonian multipole moments. For a Weyl spacetime, one has to assume that the potential $\psi$ depends on the parameter $\lambda = 1/c^2$. The Newtonian potential is then given by the limit
\begin{align}
	\label{Eq_NewtonianLimit:Ehlers}
	U (  \rho , z ) = \lim_{\lambda \to 0} \dfrac{1}{\lambda} \psi(\rho,z,\lambda) \, .	
\end{align}
Keeping the canonical coordinates $\rho$ and $z$ fixed during the limit procedure is motivated by the fact that, with respect to these cylindrical coordinates, $\psi$ satisfies the Laplace equation, which is supposed to hold also in the limit for the Newtonian potential $U$.  

It is then inevitable to assume that the coordinates $(x,y)$ depend on $\lambda$. This becomes clear if we consider the Schwarzschild case by choosing $q_0 = 1$ and $q_l = 0$ for all $l>0$. We see that the Newtonian limit leads to the potential
\begin{align}
	\label{Eq_}
  	U = -\dfrac{GM}{R} \, , \quad R^2 = \rho^2 + z^2 \, , 
\end{align}
if the parameter $m$ depends on $\lambda$ according to 
\begin{align}
\label{Eq_mM}
m = GM/c^2 = GM\lambda \, ,
\end{align}
where $G$ and $M$ are, of course, independent of $\lambda$. Inserting Eq.\ \eqref{Eq_mM} into Eq.\   \eqref{Eq_WeylMetricSpheroidal} clarifies how $x$ and $y$ depend on $\lambda$. 

Performing the limit \eqref{Eq_NewtonianLimit:Ehlers} of the expansion \eqref{Eq_WeylMetricExpansionSpheroidal} as was done in Ref.\ \cite{Quevedo:1989},\footnote{We perform the calculation here again, because in Ref.\ \cite{Quevedo:1989}, there are some minor errors in the limit procedure.} we have to calculate
\begin{align}
	\label{Eq_NewtonianLimit}
  	U = \lim_{\lambda \to 0} \dfrac{1}{\lambda} \sum_{l=0}^\infty (-1)^{l+1} q_l \, Q_l\left( \dfrac{r_+ + r_-}{2\lambda GM} \right) \, P_l\left( \dfrac{r_+ - r_-}{2\lambda GM} \right) \, .
\end{align}
For the coordinates $x$ and $y$, expressed in terms of $\rho$ and $z$, we calculate the limits
\begin{subequations}
	\begin{align}
		\label{Eq_NewtonianLimit:coordinateX}
  		\lim_{\lambda \to 0} x &= \lim_{\lambda \to 0} \dfrac{r_+ + r_-}{2\lambda GM} = \infty \, , \\
  		\label{Eq_NewtonianLimit:coordinateY}
  		\lim_{\lambda \to 0} y &= \lim_{\lambda \to 0} \dfrac{r_+ - r_-}{2\lambda GM} = \dfrac{z}{\sqrt{\rho^2+z^2}} \, .
	\end{align}
\end{subequations}
Using the fact that the Legendre polynomials are continuous, we obtain
\begin{align}
	\label{Eq_NewtonianLimit:LegendreP}
  	\lim_{\lambda \to 0} P_l \left( y \right) = P_l \left( \lim_{\lambda \to 0} y \right) = P_l \left( \dfrac{z}{\sqrt{\rho^2+z^2}} \right) \, .
\end{align}
As the limit $\lambda \to 0$ is equivalent to $x \to \infty$, we expand $Q_l(x)$ in powers of $1/x$ \cite{Bateman:1955, Quevedo:1989},
\begin{align}
	\label{Eq_NewtonianLimit:ExpansionLegendreQ}
  	Q_l(x) = Q_l\left( \dfrac{r_+ + r_-}{2\lambda GM} \right) = \sum_{k=0}^\infty b_{l+2k+1}^l \left( \dfrac{2\lambda GM}{r_+ + r_-} \right)^{l+2k+1} \, ,
\end{align}
where 
\begin{subequations}
	\begin{align}
		\label{Eq_}
  		b_{l+2k+1}^l &= \dfrac{(l+2k-1)(l+2k)}{2k(2l+2k+1)} b_{l+2k-1}^l \, , \\ 
  		b_{l+1}^l &= \dfrac{l!}{(2l+1)!!} \, .
	\end{align}
\end{subequations}

The limit of each summand of Eq.\ \eqref{Eq_NewtonianLimit} exists and is finite. Absolute convergence allows us to interchange the sum and the limit \cite{Quevedo:1986}. We insert the series expansion for $Q_l(x)$ and calculate the remaining limit
\begin{align}
	\label{Eq_Uexp1}
  	U 
  	&= \sum_{l=0}^\infty (-1)^{l+1} P_l\left( \dfrac{z}{\sqrt{\rho^2+z^2}} \right) \lim_{\lambda \to 0} \dfrac{1}{\lambda} q_l \, Q_l\left( \dfrac{r_+ + r_-}{2\lambda GM} \right) \notag \\
  	&= \sum_{l=0}^\infty (-1)^{l+1} P_l \left( \dfrac{z}{\sqrt{\rho^2+z^2}} \right) \notag \\
  	&\times \lim_{\lambda \to 0} \dfrac{1}{\lambda} q_l \sum_{k=0}^\infty b_{l+2k+1}^l \left( \dfrac{2\lambda GM}{r_+ + r_-} \right)^{l+2k+1} \, .
\end{align}
This limit exists and is nonzero if the dimensionless coefficients $q_l$ are of the form \cite{Quevedo:1989}
\begin{align}
	\label{Eq_NewtonianLimit:qParameter}
  	q_l = (G/c^2)^{-l} \bar{q}_l 
\end{align}
with new coefficients $\bar{q}{}_l$ that are independent of $\lambda$ and have dimension $[\bar{q}_l] = (\mathrm{m/kg})^l$.
Then, only the $k=0$ term in (\ref{Eq_Uexp1}) gives a nonzero limit. We finally obtain the Newtonian potential
\begin{align}
	\label{Eq_NewtonianLimit:final}
  	U =& \sum_{l=0}^\infty (-1)^{l+1} b_{l+1}^l P_l \left( \dfrac{z}{\sqrt{\rho^2+z^2}} \right) 
 	\notag \\
  	&\times 
    \lim_{\lambda \to 0} q_l \lambda^l \left( \dfrac{2GM}{r_+ + r_-} \right)^{l+1}
	\notag \\
  	&= G \sum_{l=0}^\infty (-1)^{l+1} b_{l+1}^l \bar{q}_l M^{l+1} P_l \left( \dfrac{z}{\sqrt{\rho^2+z^2}} \right) 
   \notag \\
  	&\times 
    \lim_{\lambda \to 0} \left( \dfrac{2}{r_+ + r_-} \right)^{l+1} \notag \\
  	&= -G \, \sum_{l=0}^\infty (-1)^{l} \dfrac{l!}{(2l+1)!!} \bar{q}_l \, M^{l+1} \, \dfrac{P_l(\cos \Theta)}{R^{l+1}}
\end{align}
where 
\begin{align}
	\label{Eq_}
  	\cos \Theta = \dfrac{z}{\sqrt{\rho^2+z^2}} \, , \quad R^2 = \rho^2 + z^2 \, . 
\end{align}


\subsection{Multipole moments}

If we compare Eq.\ \eqref{Eq_NewtonianLimit:final} with  Eq.\ \eqref{Eq_NewtonianPotenialAxialSymDecomposition} for the Newtonian multipole moments $N_l$ in the axisymmetric case, we see that 
\begin{align}
	\label{NewtonianMultipoles}
	N_l = (-1)^{l} \dfrac{l!}{(2l+1)!!} \bar{q}_l \, M^{l+1} \, .
\end{align}
Choosing $q_0 = \bar{q}_0 = 1$, we identify $M$ as the total mass of the source (in kg) that gives the monopole moment $N_0 = M$. A dipole moment can always be made to vanish by transforming the origin of the coordinate system into the center of mass. The quadrupole moment is given by $N_2 = -2/15 \, \bar{q}_2 M^3$. The $l$th-order multipole moment has the dimension $[N_l] = \mathrm{kg} \, \mathrm{m}^l$ such that for each moment $N_l$ we get $[N_l/N_0] = {\mathrm{m}}^l$.

From this identification, we deduce that the parameters $\bar{q}_l$, which are independent of $\lambda$, determine the Newtonian moments of the gravitating source of which the exterior we describe by the metric \eqref{Eq_WeylMetricSpheroidal}. On the other hand, the parameters $\bar{q}_l$ also determine the relativistic Geroch-Hansen moments $R_l$ uniquely. The latter, which depend of course on $\lambda = c^{-2}$, can be written in the form
\begin{align}
	R_l = N_l + C_l	\, ,
\end{align}
as a sum of the Newtonian moments and relativistic corrections $C_l$, where the $C_l$ can be calculated exactly, i.e.\ with no approximation involved. Following Quevedo \cite{Quevedo:1989}, we obtain
\begin{subequations}
\begin{align}
	\label{Eq_MultipolesRelativisticCorrections}
	C_0 &= C_1 = C_2 = 0 \, , \\
	C_3 &= - \dfrac{2}{5} m^2 N_1 \, , \\
	C_4 &= - \dfrac{2}{7} m^2 N_2 - \dfrac{6}{7} m \dfrac{G}{c^2}  N_1^2 \, .
\end{align}
\end{subequations}
In general, the correction terms $C_l$ are of the form $C_l = C_l(N_{l-2}, N_{l-3}, \dots, N_0)$. The octupole correction $C_3$ can be made to vanish by transforming away the Newtonian dipole. Then, a difference between the relativistic and the Newtonian multipole moments occurs for the first time at the 16-pole moment $R_4$, which is a surprising result that was first derived in Ref.\ \cite{Quevedo:1989}.


\subsection{Examples}

In this section, we apply our definition of the relativistic geoid to particular axisymmetric and static vacuum solutions to Einstein's field equation. We choose three examples, all of which are asymptotically flat: the Schwarzschild metric, the Erez-Rosen metric, and the q-metric (Zipoy-Vorhees metric). 


\subsubsection{Monopole: Schwarzschild metric}\label{Schwarzschild}

Choosing $q_0 =1, \, q_l = 0$ for all $l > 0$ in the expansion \eqref{Eq_WeylMetricExpansionSpheroidal}, we obtain a spacetime which possesses only a monopole moment $R_0 = M$, and the metric functions become
\begin{align}
	\psi = \dfrac{1}{2} \log \left( \dfrac{x-1}{x+1} \right)	\, , \quad \gamma &= \dfrac{1}{2} \log \left( \dfrac{x^2-1}{x^2-y^2} \right) \, .
\end{align}
The relativistic potential $\phi$ in this spacetime is given by Eqs.\ \eqref{Eq_WeylMetricRedshiftStationary} and \eqref{Eq_WeylMetricSpheroidalRedshiftRotating} for the two different congruences, respectively. We obtain
\begin{subequations}
\begin{align}
	e^{2\phi_{\text{stat}}} &= \left( \dfrac{x-1}{x+1} \right) \, , \\
	e^{2\phi_{\text{rot}}} & = \left( \dfrac{x-1}{x+1} \right) - \dfrac{\Omega^2}{c^2} \, m^2 (x+1)^2(1-y^2) \, .
\end{align}
\end{subequations}
The metric \eqref{Eq_WeylMetricSpheroidal} then yields the well-known Schwarzschild metric after the coordinate transformation ${x=r/m-1}$ and  $y = \cos \vartheta$:
\begin{multline}
	\label{Eq_Schwarzschild:metricUsual}
	g = - \left( 1- \dfrac{2m}{r} \right) c^2 dt^2 + \left( 1- \dfrac{2m}{r} \right)^{-1} dr^2 \\
	+ r^2 d\vartheta^2 + r^2 \sin^2 \vartheta d\varphi^2 \, .
\end{multline}
Hence, the relativistic potential for static and rotating observers becomes, respectively,
\begin{subequations}
\begin{align}
	e^{2\phi_{\text{stat}}} &= \left( 1- \dfrac{2m}{r} \right) \, , \\
	e^{2\phi_{\text{rot}}} &= \left( 1- \dfrac{2m}{r} \right) - \dfrac{\Omega^2}{c^2} \, r^2 \sin^2 \vartheta \, . \label{Eq_Schwarzschild:phiRot}
\end{align}
\end{subequations}
Their equipotential surfaces determine the isochronometric surfaces
\begin{subequations}
\begin{align}
	e^{2\phi_{\text{stat}}} &= \text{constant} \Leftrightarrow r = \text{constant} \, , \\
	e^{2\phi_{\text{rot}}} &= \text{constant} \notag \\
	&\Leftrightarrow \left( 1- \dfrac{2m}{r} \right) - \dfrac{\Omega^2}{c^2} r^2 \sin^2 \vartheta 
 = \text{constant} \, , \label{Eq_GeoidSchwarzschild}
\end{align}
\end{subequations}
one of which is the relativistic geoid in this spacetime. Figure \ref{Fig_Geoid-a} shows the level sets of the relativistic potential for both cases in a coordinate contour plot. 

We now compare the relativistic geoid defined by Eq.\ \eqref{Eq_GeoidSchwarzschild} with its Newtonian analog. For the Newtonian potential $U=-GM/R$ of a spherically symmetric mass distribution, the geoid is defined by an equipotential surface, see Eq.\ \eqref{Eq_NewtonianGeoid},
\begin{align}
	\label{Eq_GeoidNewtonianMonopole}
	W = -\dfrac{GM}{R} - \dfrac{1}{2} \Omega ^2 R^2 \sin^2 \vartheta = W_0 = \text{constant} \, .
\end{align}
Using the relation $m = GM/c^2$, we get from \eqref{Eq_GeoidSchwarzschild} the condition for the relativistic geoid,
\begin{align}
	1 + \dfrac{2}{c^2} \left( - \dfrac{GM}{r} - \dfrac{1}{2} \Omega ^2 r^2 \sin^2 \vartheta \right) = \text{constant}\, .
\end{align}
Hence, the term in brackets must be constant. This is, formally, the same result as for the nonrelativistic geoid \eqref{Eq_GeoidNewtonianMonopole}. Of course, the Newtonian geoid is defined in a flat geometry, while the spatial part of the Schwarzschild metric is not flat. Therefore, the intrinsic geometry of a surface in the Schwarzschild geometry is in general different from that of a surface with the same coordinate representation in flat space. However, as the spheres $r = r_0$ in the Schwarzschild geometry have area $4 \pi r_0^2$, the intrinsic geometry of the Schwarzschild geoid for the nonrotating observers is the same as that of the corresponding Newtonian geoid. 

In Figs.\ \ref{Fig_embedding2} and \ref{Fig_embedding2_3D} in the bottom row on the right, we show an isometric embedding into Euclidean space $\mathbb{R}^3$ of the isochronometric surfaces as seen by the rotating observers. This isometric embedding reveals the intrinsic geometry of these surfaces; close to the source the surfaces are ``squashed spheres,'' whereas farther away, they deform into cylinders due to the increasing influence of the rotation term that is proportional to $r^2$; see Eq.\ \eqref{Eq_Schwarzschild:phiRot}. For details on the embedding procedure, we refer to Appendix \ref{sec_Embedding}.


\subsubsection{Quadrupole I: Erez-Rosen metric}

Choosing $q_0 = 1, \, q_1 = 0, \, q_2 \neq 0$, and $q_l = 0$ for all $l > 2$, we obtain a metric that possesses a monopole moment $R_0 = M$ and, additionally, an independent quadrupole moment 
\begin{align}
	R_2 = \dfrac{2}{15} \bar{q}_2 \, M^{3} \, .
\end{align}
The metric functions $\psi$ and $\gamma$ in Eq.\ \eqref{Eq_WeylMetricSpheroidal} become
\begin{multline}
	2 \psi = \log \left( \dfrac{x-1}{x+1} \right) + q_2 (3y^2-1)\left( \dfrac{(3x^2-1)}{4} \right. \\
	\times \left. \log \left( \dfrac{x-1}{x+1} \right) + \dfrac{3}{2} x \right) \, ,
\end{multline}
and 
\begin{multline}
	\gamma 	= \dfrac{1}{2} (1+q_2)^2 \log \left( \dfrac{x^2-1}{x^2-y^2} \right) \\
			- \dfrac{3}{2} q_2 (1-y^2) \left( x \log \left(\dfrac{x-1}{x+1}\right) + 2 \right) + \dfrac{9}{16} q_2^2 (1-y^2) \\
			\times \left[ x^2 + 4y^2 -9x^2y^2 - \dfrac{4}{3} + x \left( x^2 +7y^2 -9x^2y^2 - \dfrac{5}{3} \right) \right. \\
			\times \log \left( \dfrac{x-1}{x+1} \right) + \dfrac{1}{4} (x^2-1)(x^2+y^2-9x^2y^2-1) \\
			\left. \times \log \left(\dfrac{x-1}{x+1}\right)^2 \right] \, .
\end{multline}

This metric is the vacuum solution found by Erez and Rosen \cite{Erez:1959}\footnote{As pointed out in Ref.\ \cite{Young1969}, the original work by Erez and Rosen contains some mistakes concerning numerical factors within the expression for the metric functions. A corrected version can be found, for example, in Ref.\ \cite{Young1969}.}. If the quadrupole moment vanishes, $q_2 \to 0$, we reobtain the Schwarzschild metric. 

\begin{figure*}
	\centering
	\subfigure[ \quad Schwarzschild spacetime \label{Fig_Geoid-a}]{\includegraphics[width=0.3\textwidth]{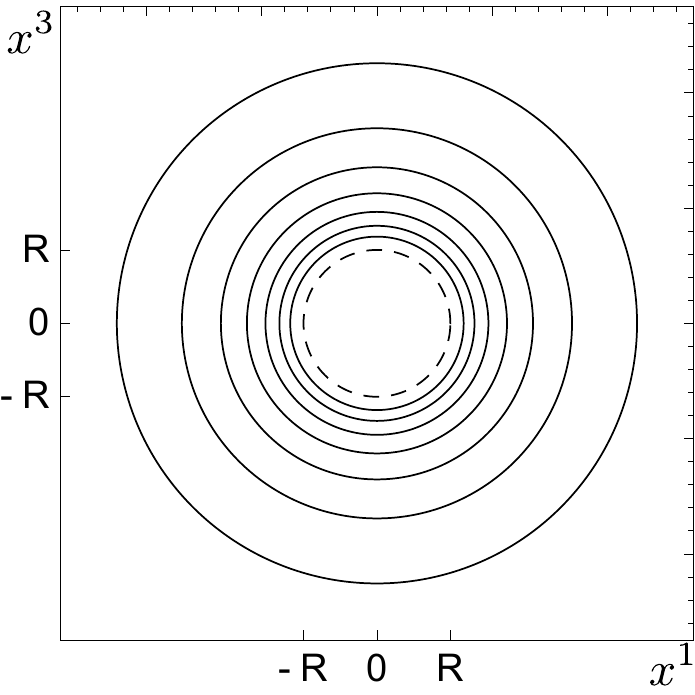} \qquad \includegraphics[width=0.3\textwidth]{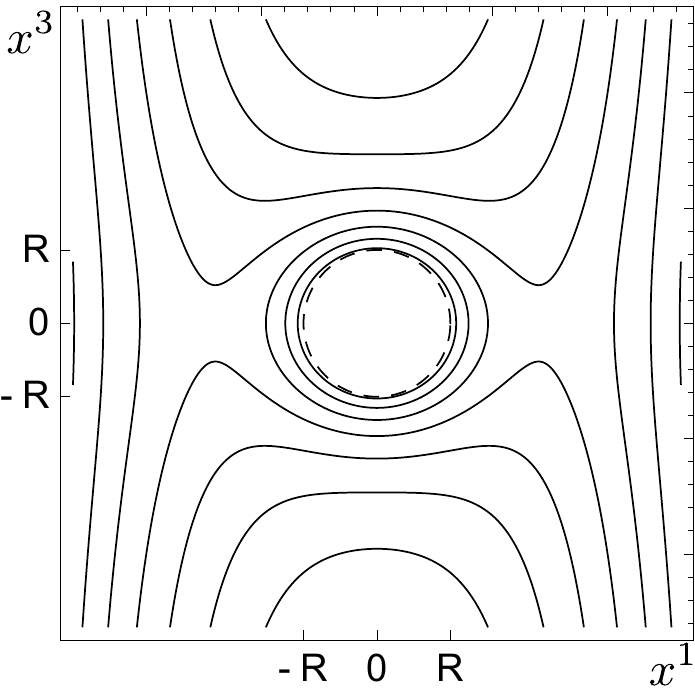}} \\
	\subfigure[ \quad Erez-Rosen spacetime \label{Fig_Geoid-b}]{\includegraphics[width=0.3\textwidth]{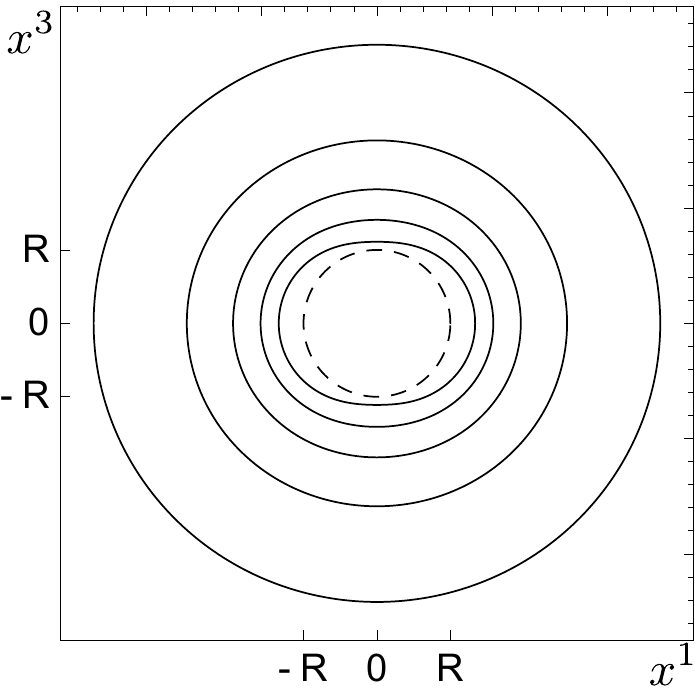} \quad
	\includegraphics[width=0.3\textwidth]{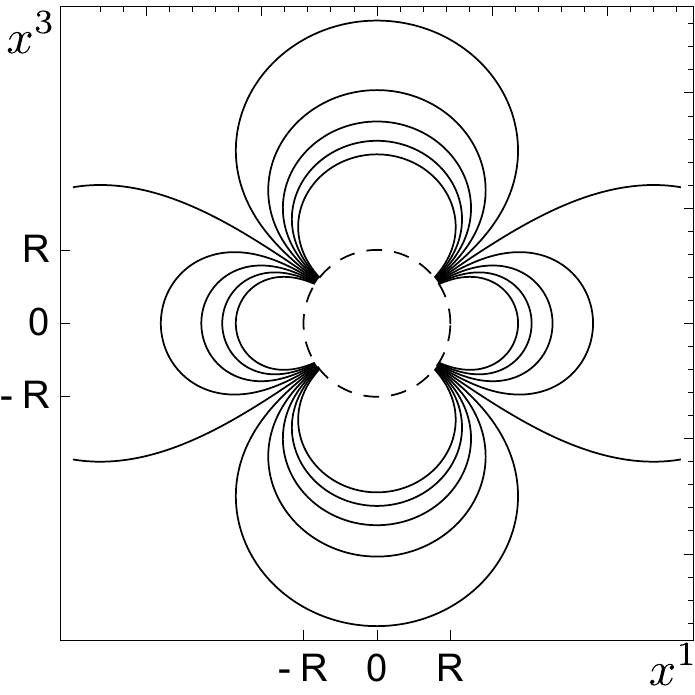} \quad 
	\includegraphics[width=0.3\textwidth]{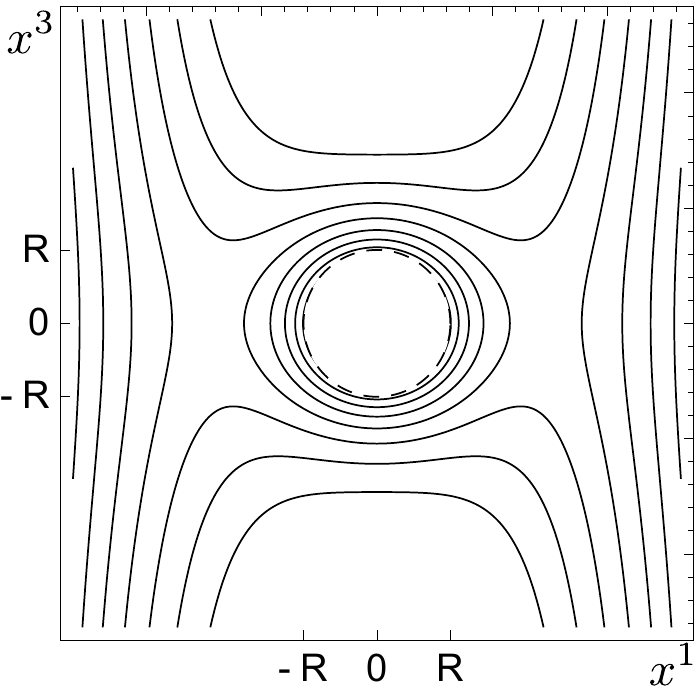}} \\
	\subfigure[ \quad Kerr spacetime \label{Fig_Geoid-c}]{\includegraphics[width=0.3\textwidth]{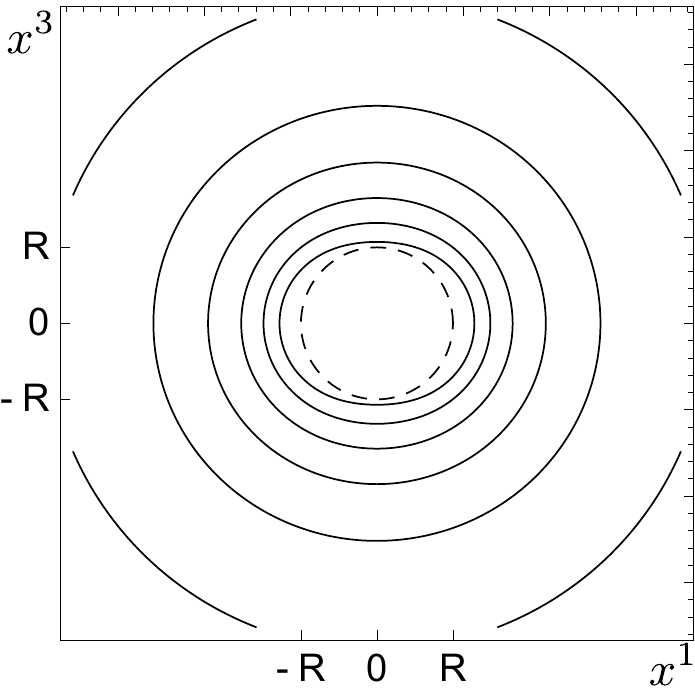} \qquad
	\includegraphics[width=0.3\textwidth]{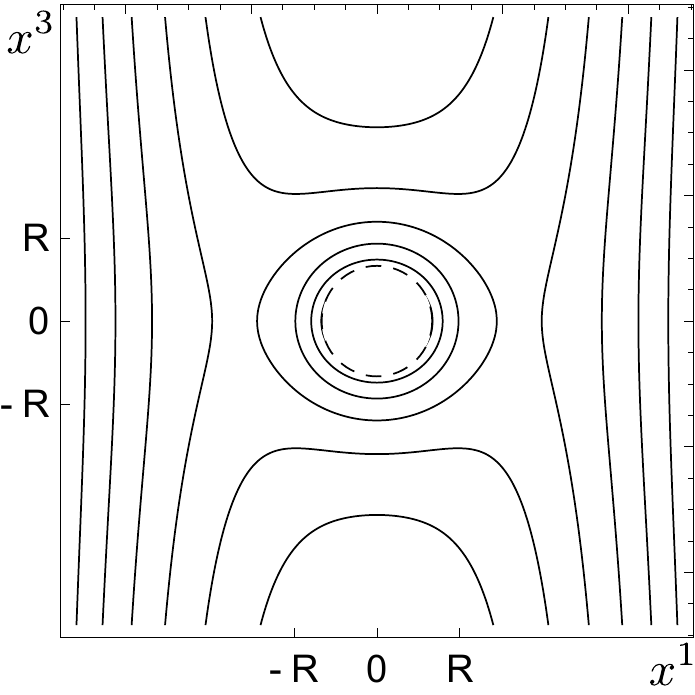}}
	\caption{\label{Fig_Geoid}
The level sets of the relativistic potential in a plane $\varphi= \mathrm{constant}$. (a): Level sets in the Schwarzschild spacetime for the static congruence (left) and the rotating congruence (right). (b): Redshift potential in the Erez-Rosen spacetime and a negative quadrupole parameter (oblate case) for the static congruence (left) and the rotating congruence (right). The pure quadrupolar contribution as difference to the monopole contribution is shown in the middle. 
(c): Level sets in Kerr spacetime for the stationary congruence (left) and the rotating congruence (right). For all plots we introduced pseudo-Cartesian coordinates $(x^1,x^3)$ by the usual relations to spherical coordinates $(r,\vartheta)$. In either case the dashed line is a circle in these coordinates, corresponding to $r=\text{constant}$ surfaces in the respective spacetime.}
\end{figure*}

The relativistic potential for static and rotating observers is, respectively,
\begin{subequations}
\label{Eq_RedshiftPotentialErezRosen}
\begin{align}
	e^{2\phi_{\text{stat}}} &= e^{2\psi} = \left( \dfrac{x-1}{x+1} \right)  \exp \left \lbrace q_2 (3y^2-1) \left( \dfrac{(3x^2-1)}{4} \right. \right.  \notag \\
	&\times \left. \left. \log \left( \dfrac{x-1}{x+1} \right) + \dfrac{3}{2} x \right) \right \rbrace \, , \\
	e^{2\phi_{\text{rot}}} &= e^{2\phi_{\text{stat}}} - \dfrac{\Omega^2}{c^2} m^2 (x^2-1)(1-y^2) e^{- 2\phi_{\text{stat}}}\, .
\end{align}
\end{subequations}
The isochronometric surfaces are shown in Fig.\ \ref{Fig_Geoid-b}. We also show the effect of the quadrupole term alone by subtracting the monopole contribution, i.e.\ subtracting the Schwarzschild term. 

Using the coordinate transformation \eqref{Eq_CoordinateTrafo3}, we can switch to the coordinates $(r,\vartheta)$ and obtain
\begin{subequations}
\label{Eq_RedshiftPotentialErezRosen2}
\begin{align}
	e^{2\phi_{\text{stat}}} &= \left( 1-\dfrac{2m}{r} \right)  \exp \left \lbrace q_2 (3\cos^2\vartheta-1) \right. \notag \\
	&\left. \times \left[ \left( \dfrac{3}{4} \left(\dfrac{r}{m}-1 \right)^2- \dfrac{1}{4} \right) \log \left( 1-\dfrac{2m}{r} \right) \right. \right. \notag \\
	&+ \left. \left. \dfrac{3}{2} \left(\dfrac{r}{m}-1\right) \right] \right \rbrace \, , \\
	e^{2\phi_{\text{rot}}} &= e^{2\phi_{\text{stat}}} - \dfrac{\Omega^2}{c^2} \, r^2\sin^2\vartheta \, e^{- 2\phi_{\text{stat}}}\, .
\end{align}
\end{subequations}
Thereupon, the geoid can also be determined in terms of the coordinates $(r, \vartheta)$.

We expand $\exp(2\phi_{\text{stat}})$ up to cubic order in $m/r$ because this is where quadrupole corrections appear. We obtain
\begin{align}
	e^{2\phi_{\text{stat}}} &= 1 - \dfrac{2m}{r} - \dfrac{2}{15} q_2 m^3 \dfrac{3\cos^2\vartheta -1}{r^3} + \mathcal{O}(m^4/r^4) \notag \\
	&= 1 - \dfrac{2}{c^2} \left( \dfrac{GM}{r} + G M m^2 \dfrac{2}{15} q_2 \dfrac{3\cos^2\vartheta - 1}{2r^3} \right) \notag \\
	&+ \mathcal{O}(m^4/r^4) \notag \\
	&= 1 - \dfrac{2}{c^2} \left( \dfrac{GM}{r} + G N_2 \dfrac{3\cos^2\vartheta - 1}{2r^3} \right) \notag \\
	&+ \mathcal{O}(m^4/r^4) \, .
\end{align}
For
\begin{align}
  	N_2= \dfrac{2}{15} M m^2 q_2 = \dfrac{2}{15} \bar{q}_2 M^3 \, ,
\end{align}
the term in brackets is the Newtonian potential of a quadrupolar gravitational source; see Eq.\ \eqref{Eq_NewtonianPotenialAxialSymDecomposition} for comparison. 

This result shows that, indeed, the Newtonian limit of the Erez-Rosen spacetime yields the Newtonian gravitational potential of a source that possesses only a monopole and a quadrupole moment. Hence, the relativistic geoid for the Erez-Rosen spacetime in terms of the level sets of Eq.\ \eqref{Eq_RedshiftPotentialErezRosen2} reproduces the Newtonian expression in lowest order. Higher orders are, however, different. Moreover, one has to keep in mind that in the Erez-Rosen spacetime the coordinates do not have the same geometric meaning as in the Newtonian theory. The metric on a surface $t=\text{constant}$ and $r=\text{constant}$ is not the usual metric on the 2-sphere $S^2$, and $r$ is not an area coordinate as it was in the Schwarzschild spacetime. We can visualize the intrinsic geometry of isochronometric surfaces by isometrically embedding them into the Euclidean space $\mathbb{R}^3$. These surfaces are defined by an equation of the form
\begin{align}
	\label{Eq_}
  	e^{ 2 \phi (r,\vartheta) } = f _0 = \text{constant}  \, .
\end{align}
The value $f_0 > 0$ labels these surfaces. For $f_0 \to 0$, the surface of infinite redshift for observers on integral curves of $\partial_t$ is approached. For static spacetimes, this surface is a horizon. The relevant equations for constructing the embeddings are given in Appendix \ref{sec_Embedding}. For the Schwarzschild spacetime, the embedding yields standard spheres in $\mathbb{R}^3$ for the congruence on integral curves of $\partial_t$, and for the congruence on integral curves of $\partial_t + \Omega \partial_\varphi$, the embedding yields deformed spheres close to the horizon and deformed cylinders further away, cf.\ Figs.\ \ref{Fig_embedding2} and \ref{Fig_embedding2_3D} on the right in the bottom row.

For the Erez-Rosen spacetime, we have to consider two different signs of the quadrupole parameter. Hence, the embedded surfaces are either prolate or oblate; see the middle rows of Figs.\ \ref{Fig_embedding1} -- \ref{Fig_embedding2_3D}. We see that the isochronometric surfaces in the Erez-Rosen spacetime for negative quadrupole parameter develop ``bulges'' around the poles close to the horizon. Farther away, the embedded surfaces become oblate or prolate squashed spheres. With non-zero rotation, the embedded surfaces deform into cylinders farther away from the source, analogously to the rotating Schwarzschild case.

\subsubsection{Quadrupole II: q-metric}
Another example of a two-parameter family of metrics that is actually the simplest generalization of the Schwarzschild metric is the $q$-metric \cite{Quevedo:2011, Quevedo:2013, Toktarbay:2014, Quevedo:2016, Quevedo:2016b}. The $q$-metric, as constructed by Quevedo, is obtained by a Zipoy-Voorhees transformation of the Schwarzschild solution. Zipoy \cite{Zipoy:1966} and Voorhees \cite{Voorhees:1970} considered such solutions of the vacuum field equation in their papers. A similar transformation was also used before in the work of Bach (and Weyl) \cite{Bach:1922}. For a discussion of the Zipoy-Voorhees ($q$-)metric, we refer the reader to, e.g.\ the book by Griffiths and Podolsk{\'y} \cite{Griffiths:2009}.

The $q$-metric possesses independent monopole and quadrupole moments, and all higher multipole moments are determined by these two. The metric functions read
\begin{align}
	e^{2\psi} = \left( \dfrac{x-1}{x+1} \right)^{1+q} \, , \quad e^{2\gamma} = \left( \dfrac{x^2-1}{x^2-y^2} \right)^{(1+q)^2} \, .
\end{align}
The relativistic monopole and quadrupole moments of this spacetime are given by $R_0 = (1+q)M$ and $R_2 = -Mm^2 q(1+q)(2+q)/3$ \cite{Quevedo:2016}. The limit $q \to 0$ yields the Schwarzschild metric. The relativistic potential for static and rotating observers is, respectively,
\begin{subequations}
\begin{align}
	e^{2\phi_{\text{stat}}} &= \left( \dfrac{x-1}{x+1} \right)^{1+q} \, , \\	
	e^{2\phi_{\text{rot}}} &= \left( \dfrac{x-1}{x+1} \right)^{1+q} - \dfrac{\Omega^2}{c^2} \, m^2 \left( \dfrac{x-1}{x+1} \right)^{-(1+q)} \notag \\
	&\times (x^2-1)(1-y^2) \, .
\end{align}
\end{subequations}
With the coordinate transformation \eqref{Eq_CoordinateTrafo3}, the equations that define the isochronometric surfaces read
\begin{subequations}
\begin{align}
		e^{2\phi_{\text{stat}}} &= \left( 1-\dfrac{2m}{r} \right)^{1+q} \, , \\	
	e^{2\phi_{\text{rot}}} &= \left( 1-\dfrac{2m}{r} \right)^{1+q}  \notag \\
	& - \dfrac{\Omega^2}{c^2} \, \left( 1-\dfrac{2m}{r} \right)^{-q} r^2\sin^2\vartheta \, .
\end{align}
\end{subequations}
Even though the level sets of the redshift potential $\phi_{\text{stat}}$ coincide with the surfaces $x=\text{constant}$ and thus with the surfaces $r=\text{constant}$, this does not mean that the geoid is spherically symmetric. The metric on the surfaces $t=\text{constant}$ and $r=\text{constant}$ is not the usual metric on the $S^2$, and $r$ is not an area coordinate as it was in the Schwarzschild spacetime. To put this into geometrical terms, one can use the relativistic flattening \cite{Philipp:2017b} that measures the deviation from spherical symmetry
\begin{align}
	f := 1 - \dfrac{C_\vartheta}{C_\varphi} \, ,
\end{align}
where $C_\vartheta$ and $C_\varphi$ are the circumferences, measured with the metric, of circles at $r=r_0$ in the $\vartheta$-direction (polar circles) and $\varphi$-direction (azimuthal circles), respectively. The circumference $C_\varphi$ is measured in the equatorial plane $\vartheta=\pi/2$, whereas for $C_\vartheta$, the azimuthal angle $\varphi$ is arbitrary due to the symmetry. For the Schwarzschild spacetime, this flattening is zero, whereas for the $q$-metric, we obtain
\begin{multline}
	f = 1 - (x^2-1)^{\frac{q}{2}(2+q)} \\
	\times x^{-q(2+q)} {_2F_1}\left( \dfrac{1}{2}, \dfrac{1}{2} q(2+q), 1, 1/x^2 \right) \, .
\end{multline}
Here, ${_2F_1}$ is one of the hypergeometric functions. In the limits $r\to \infty$ and $q \to 0$, the flattening becomes zero. For a positive $q$, the flattening is positive, and the surfaces $x=\text{constant}$ are oblate, because circles in the $\varphi$-direction are larger. For a negative value of $q$, these surfaces are prolate.

\begin{figure*}[hbt]
  \includegraphics[width=0.4\textwidth]{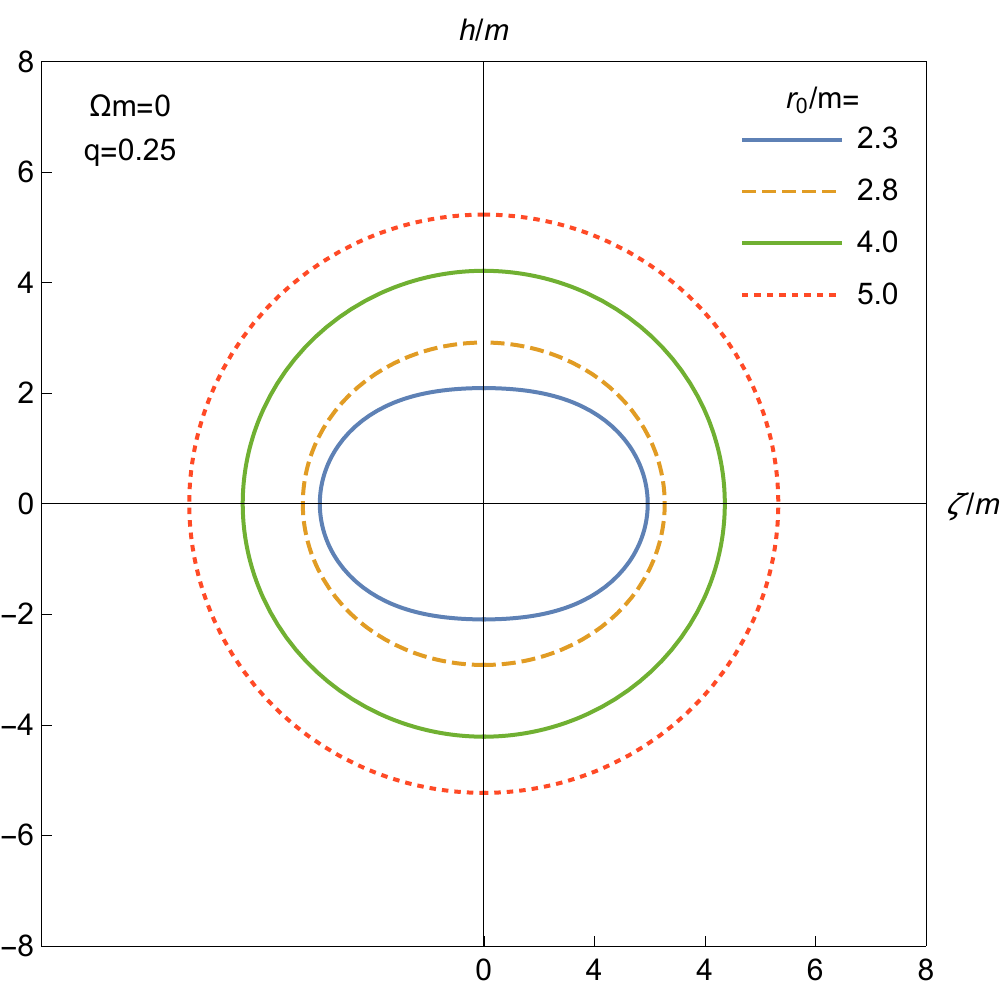} \hfill
  \includegraphics[width=0.4\textwidth]{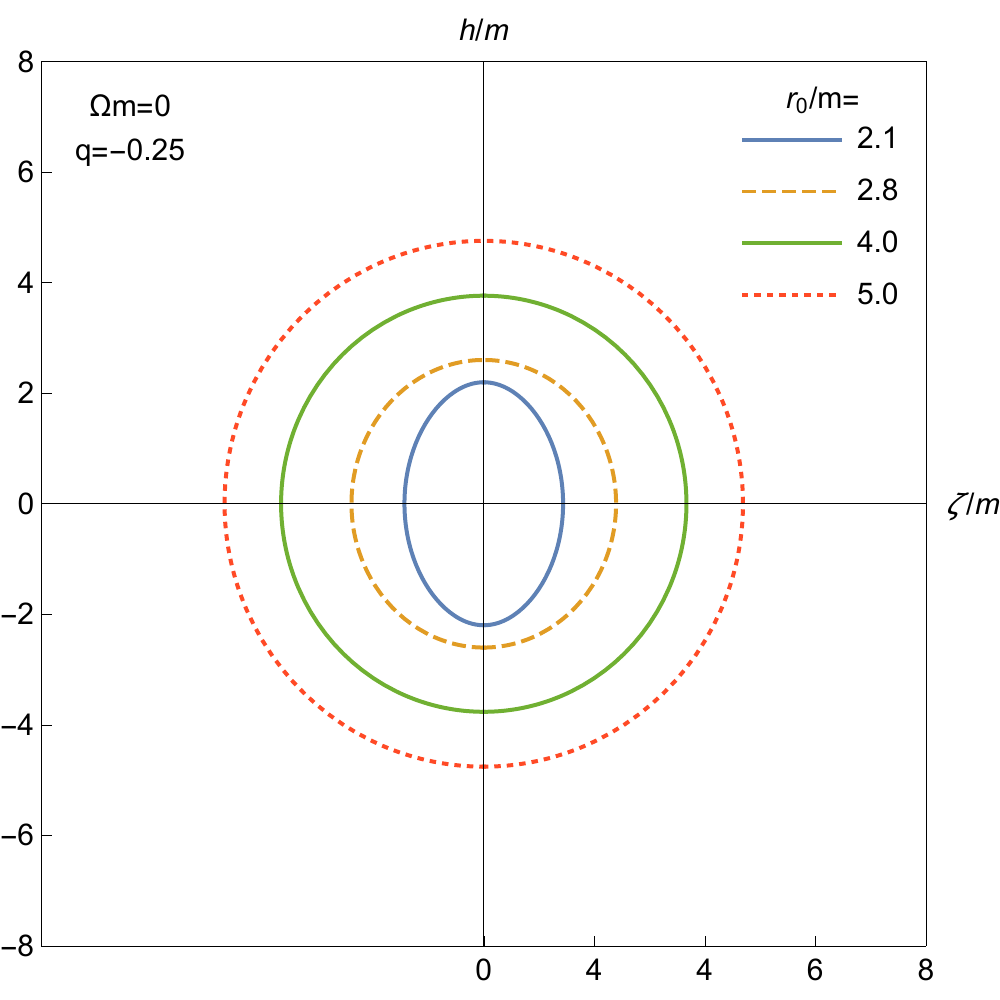}
  \includegraphics[width=0.4\textwidth]{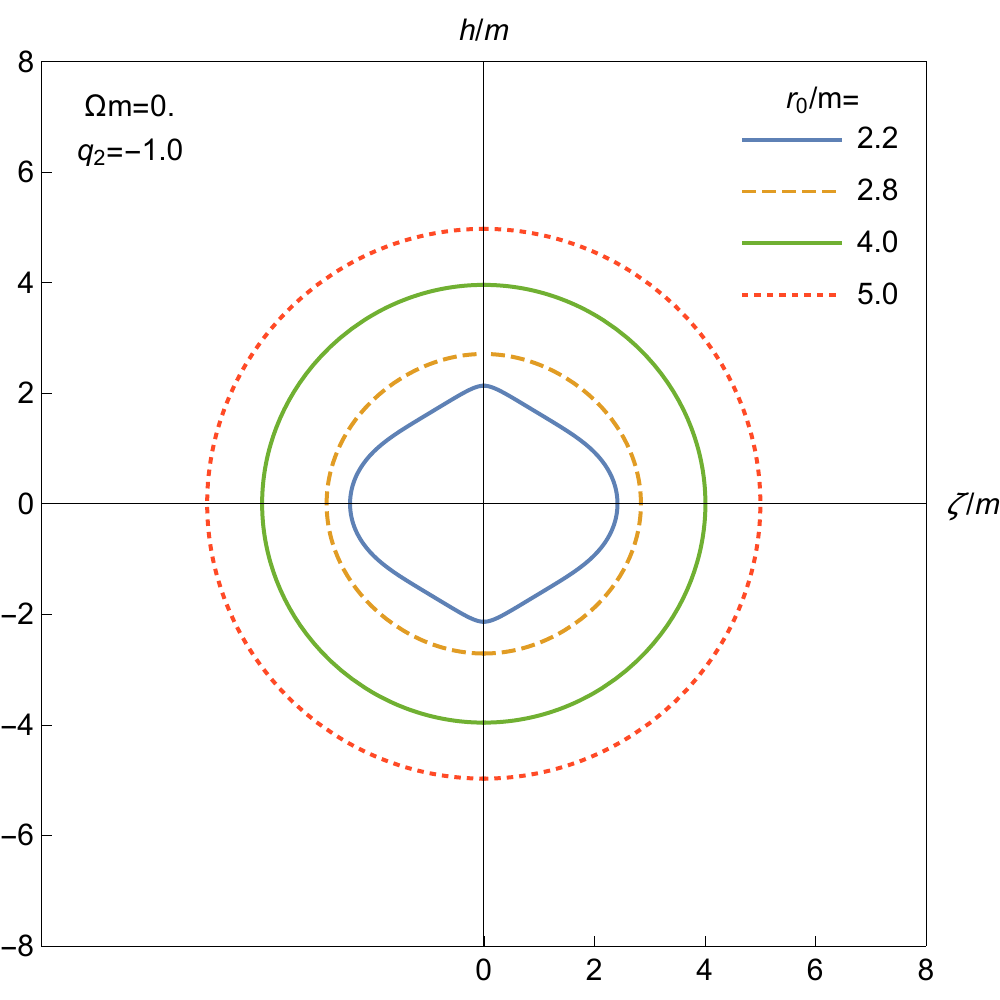} \hfill
  \includegraphics[width=0.4\textwidth]{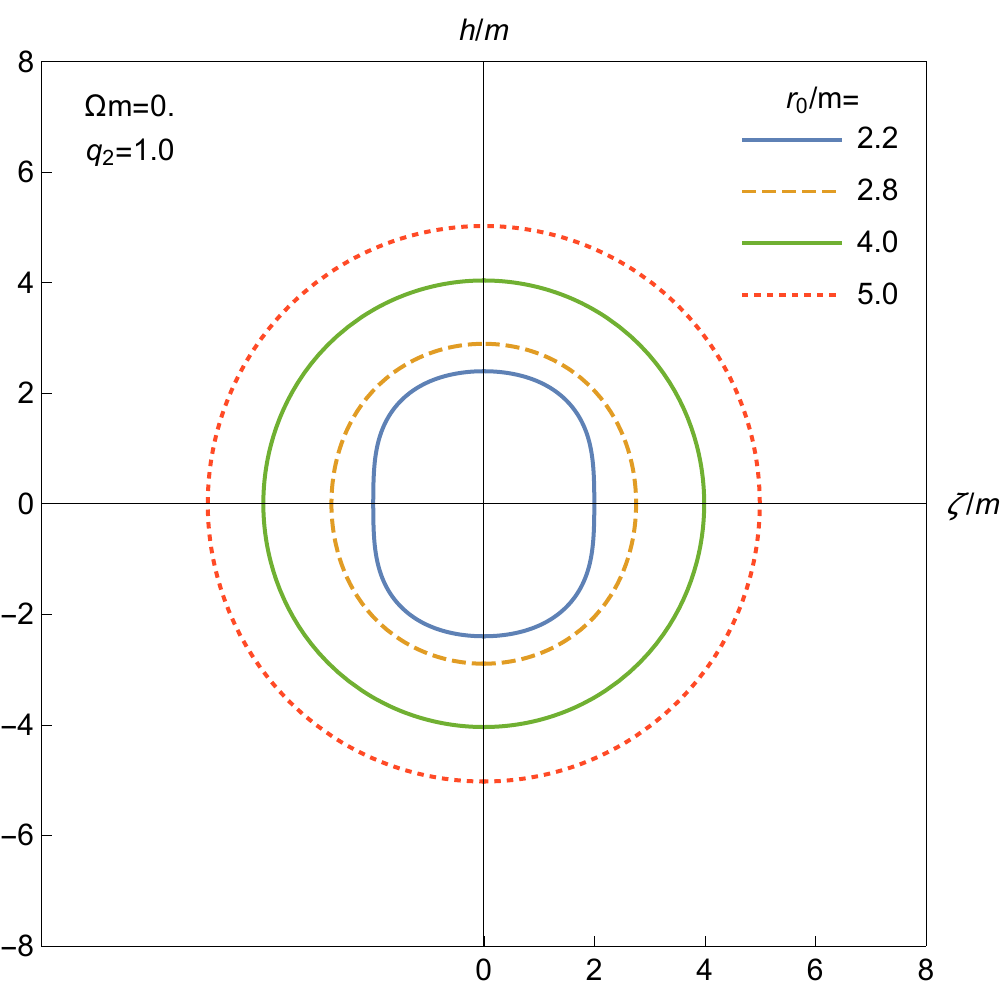}
  \includegraphics[width=0.4\textwidth]{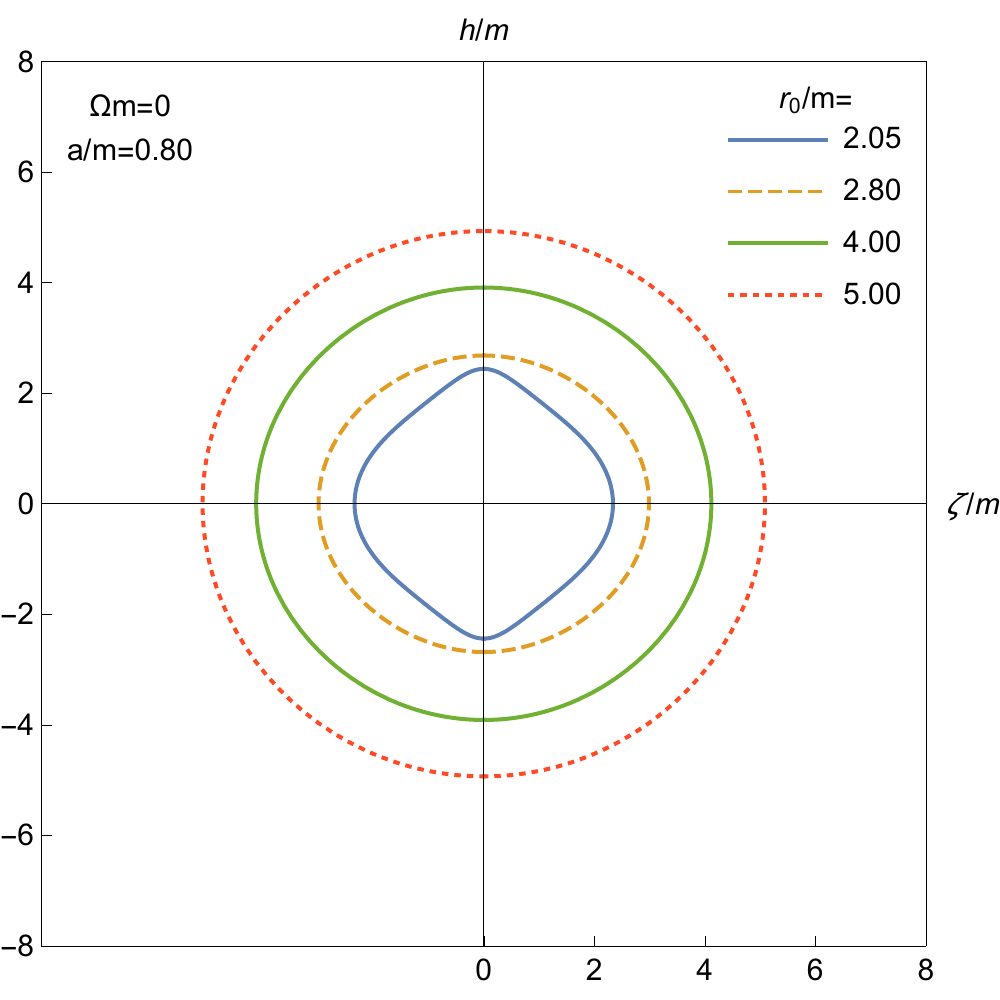} \hfill
  \includegraphics[width=0.4\textwidth]{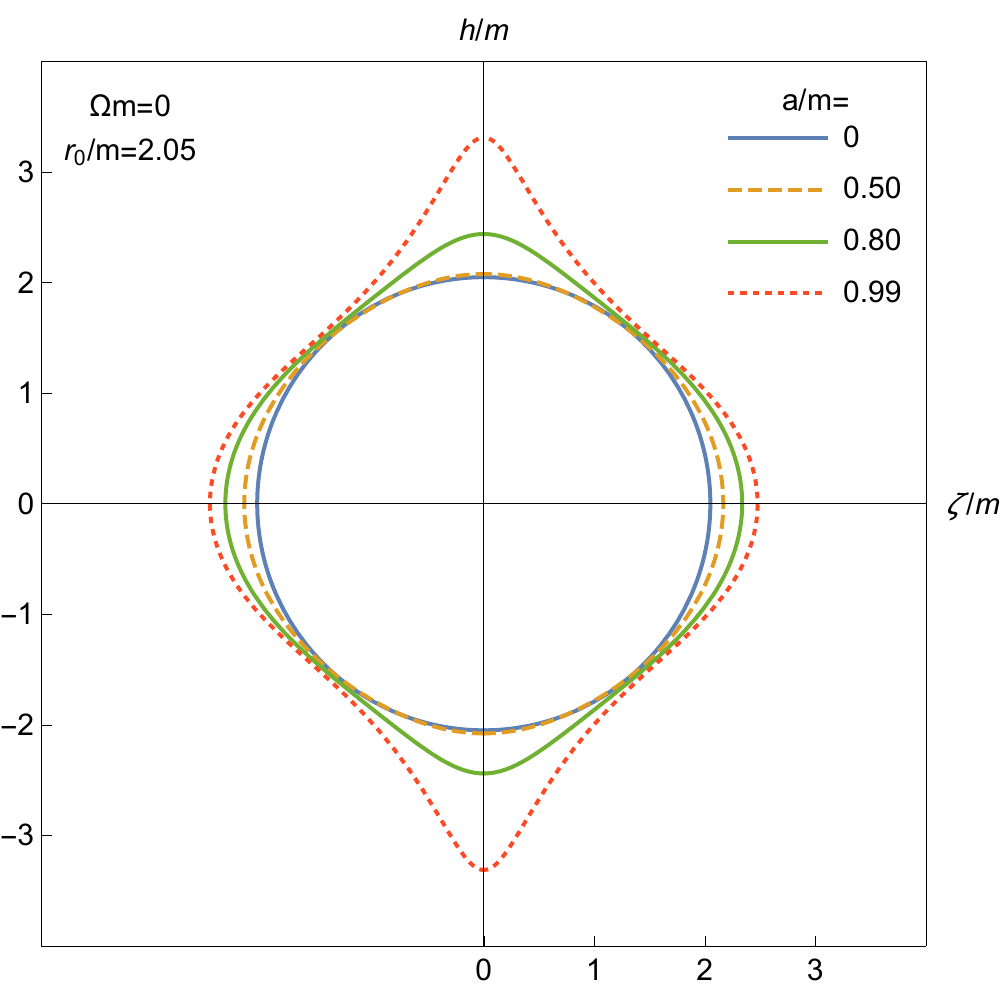}
  \caption{Isometric embedding of isochronometric surfaces $\exp \big( 2 \phi_{\text{stat}} \big)=f_0$ into the Euclidean space $\mathbb{R}^3$. The relativistic geoid is by definition one of these surfaces. The value of $r_0$ in the plots is the intersection of the level surface $f_0$ with the radial lines in the equatorial plane. Upper row: q-metric results for oblate (left) and prolate (right) quadrupole configuration. Middle row: Erez-Rosen metric results for oblate (left) and prolate (right) quadrupole configuration. Lower row: Kerr metric results for fixed $a=0.8m$ but different level surfaces (left) and the same level surface close to the ergoregion but different values $a=(0,\, 0.5m,\, 0.8m,\, m)$. The smaller the value of $r_0 > 2m$, the closer the level surface is to the surface of infinite redshift for observers on integral curves of $\partial_t$. All necessary parameters are depicted in the respective plots.}
  \label{Fig_embedding1}
\end{figure*}

\begin{figure*}[hbt]
  \includegraphics[width=0.4\textwidth]{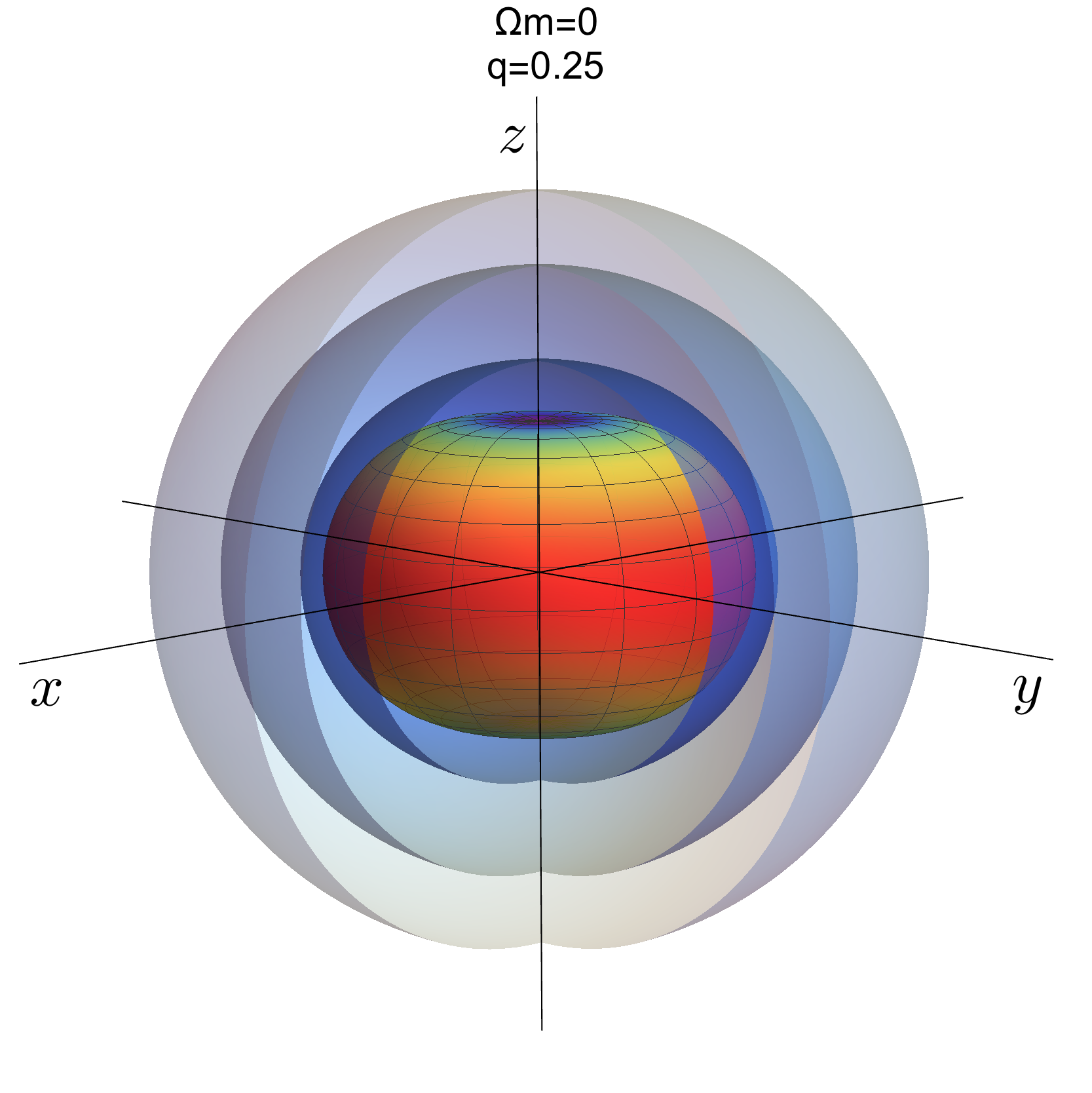} \hfill
  \includegraphics[width=0.4\textwidth]{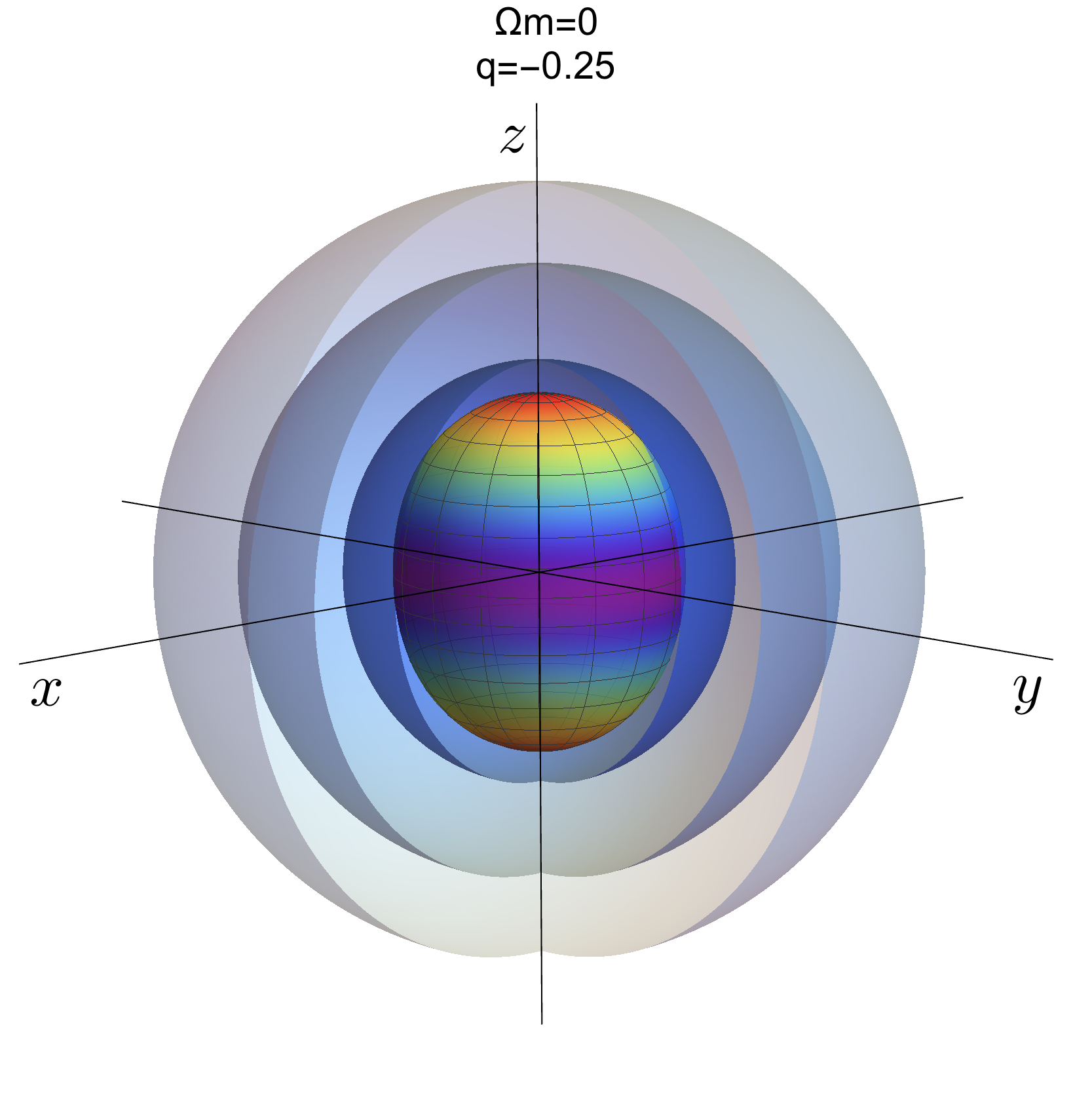}
  \includegraphics[width=0.4\textwidth]{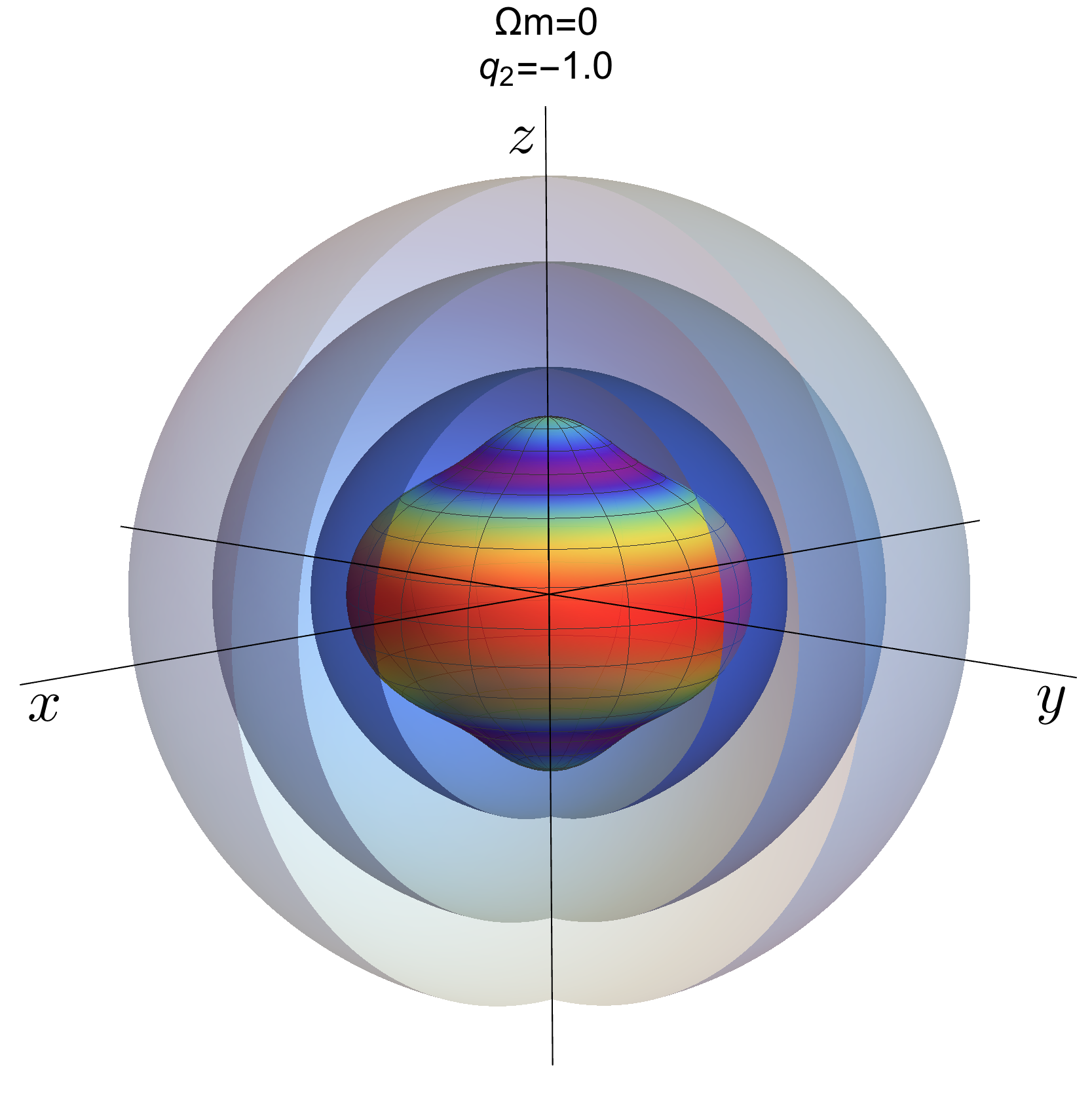} \hfill
  \includegraphics[width=0.4\textwidth]{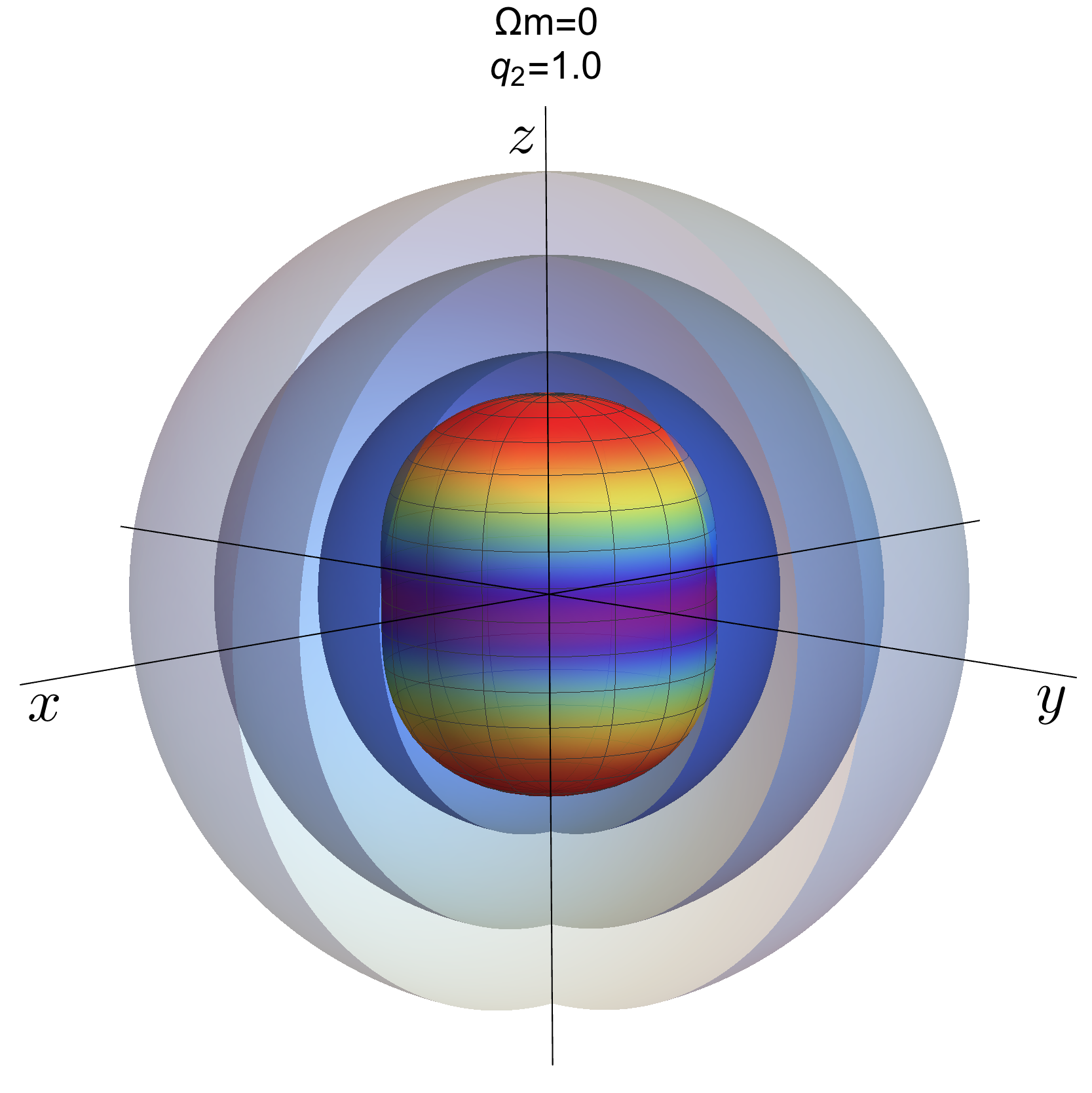}
  \includegraphics[width=0.4\textwidth]{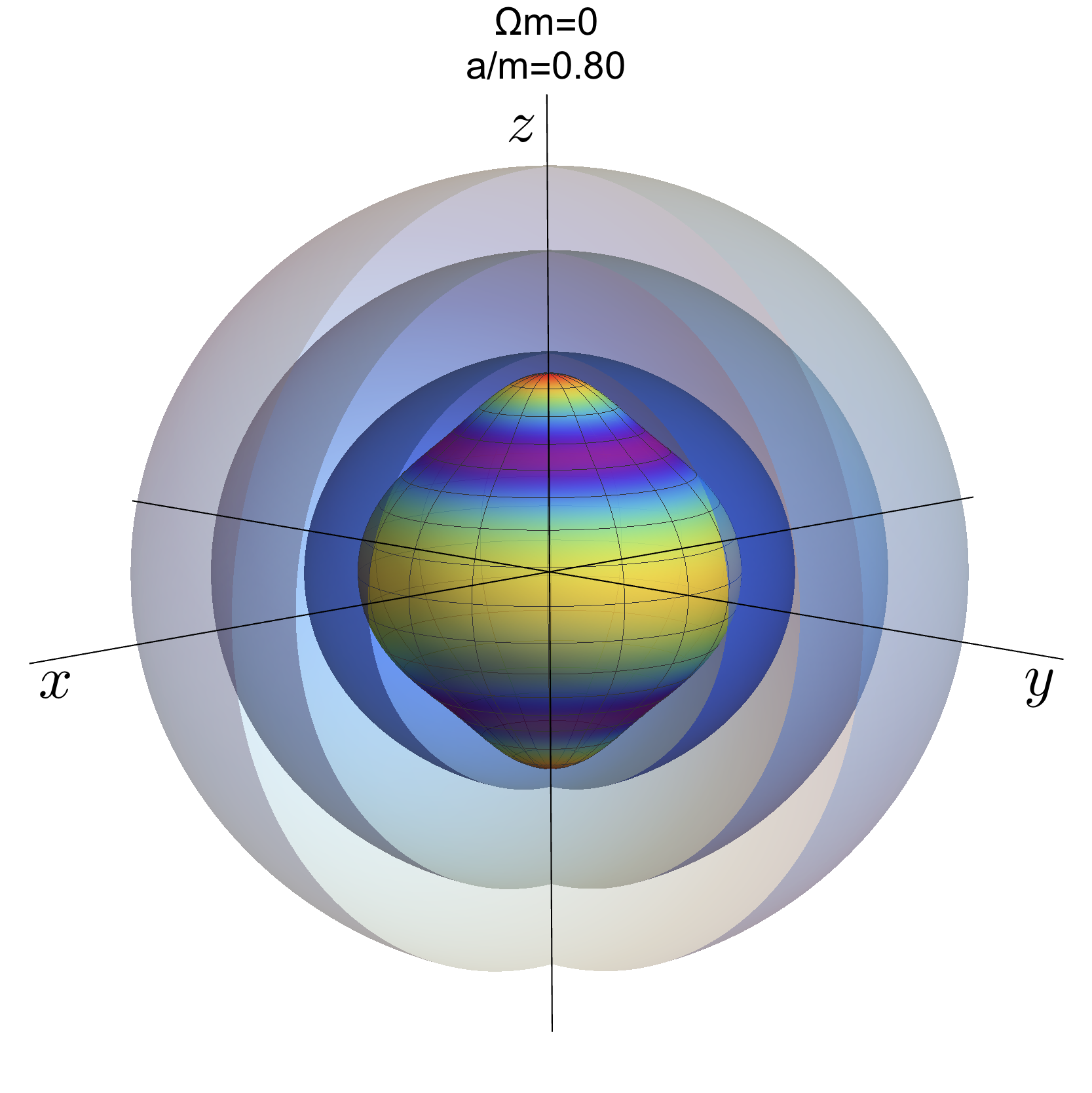} \hfill
  \includegraphics[width=0.4\textwidth]{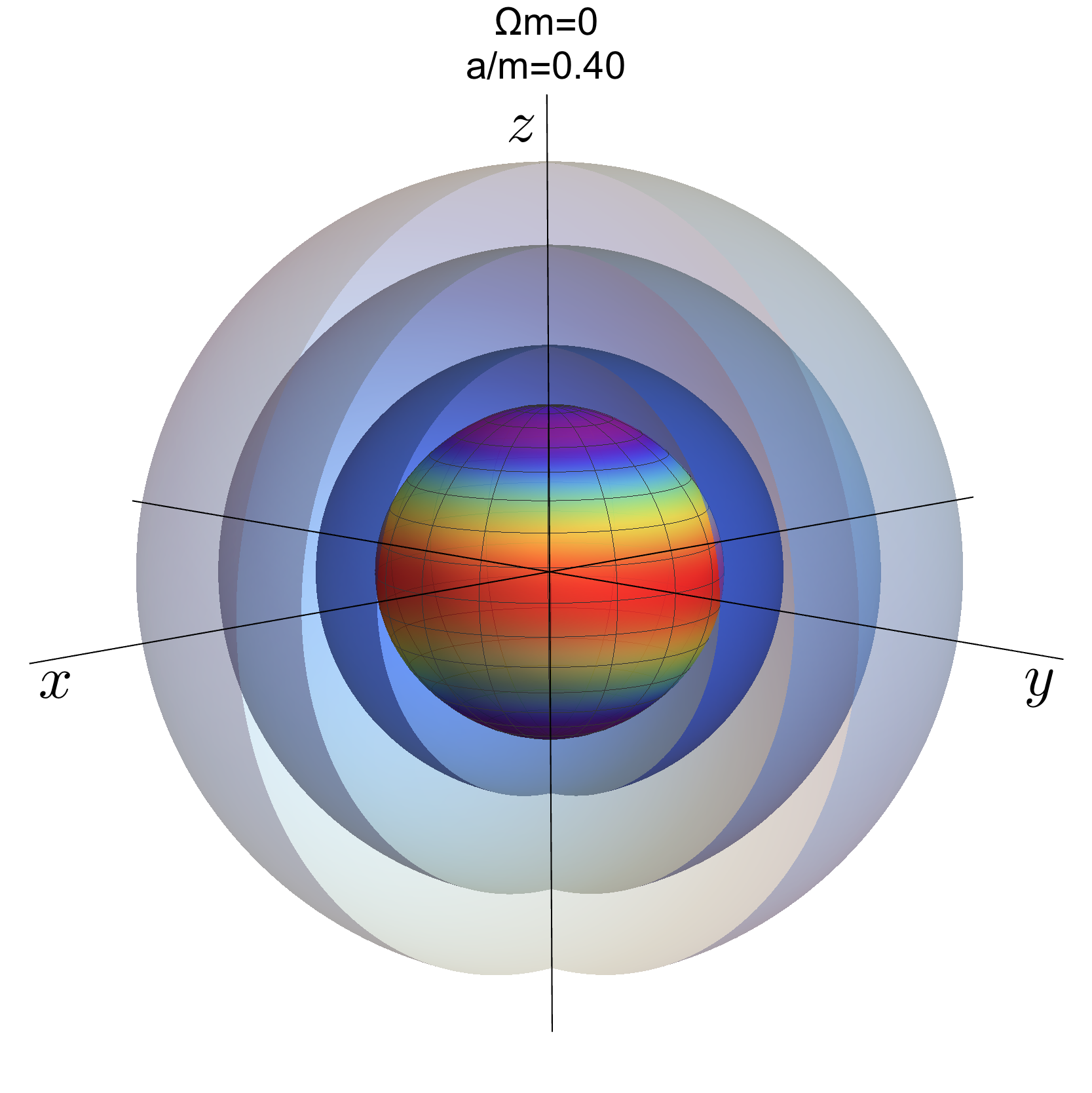}
  \caption{Isometric embedding of isochronometric surfaces $\exp \big( 2 \phi_{\text{stat}} \big)= f_0$ into the Euclidean space $\mathbb{R}^3$. We show the level surfaces in 3-dimensional plots. The level surfaces and their order correspond to those shown in Fig.\ \ref{Fig_embedding1}. In the bottom row on the right we additionally show the result for the Kerr spacetime and $a=0.4m$. For each plot, the innermost level surface is color coded to depict the actual shape such that red corresponds to the farthest distance and purple corresponds to the closest distance to the origin of $\mathbb{R}^3$.}
  \label{Fig_embedding1_3D}
\end{figure*}

\begin{figure*}[hbt]
  \includegraphics[width=0.4\textwidth]{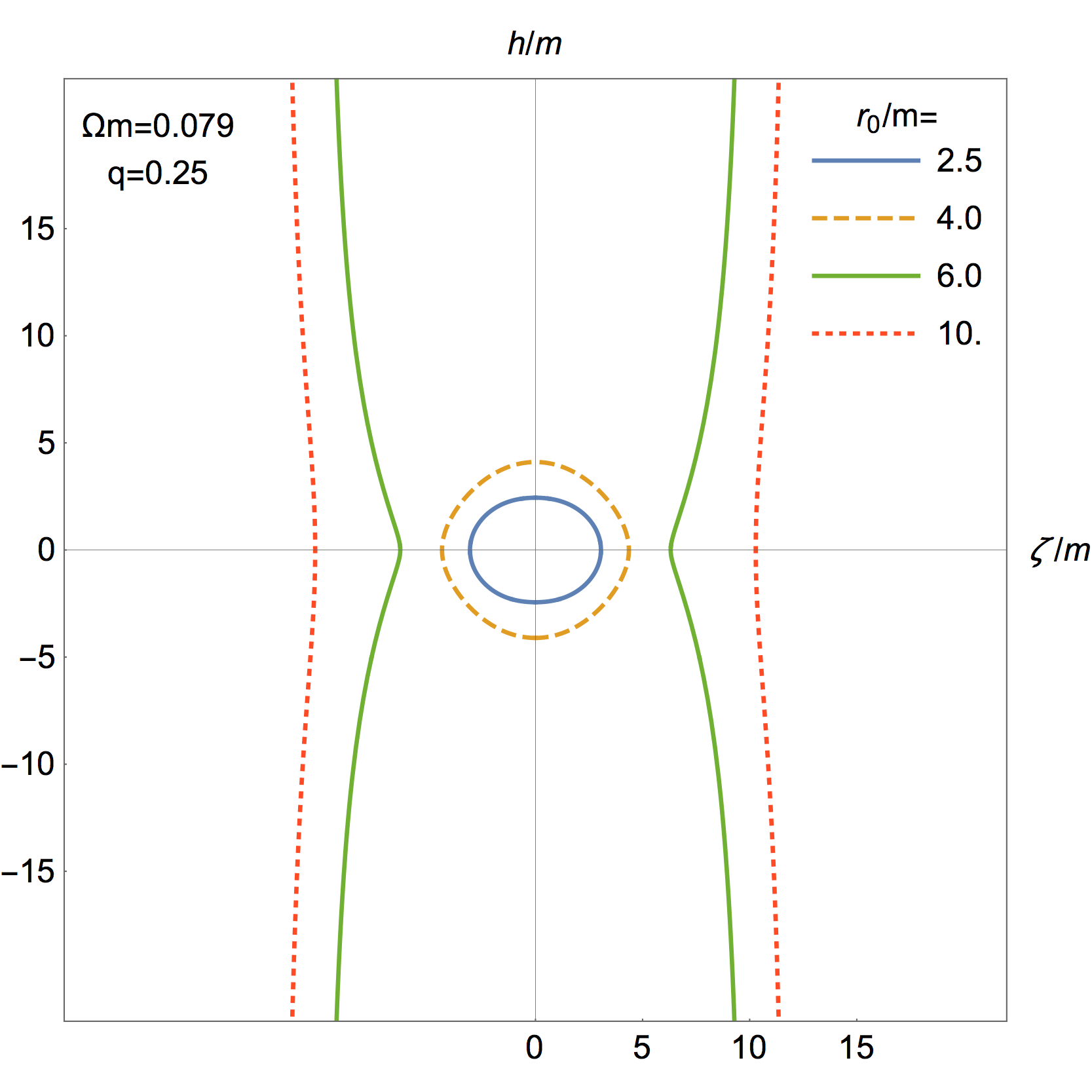} \hfill
  \includegraphics[width=0.4\textwidth]{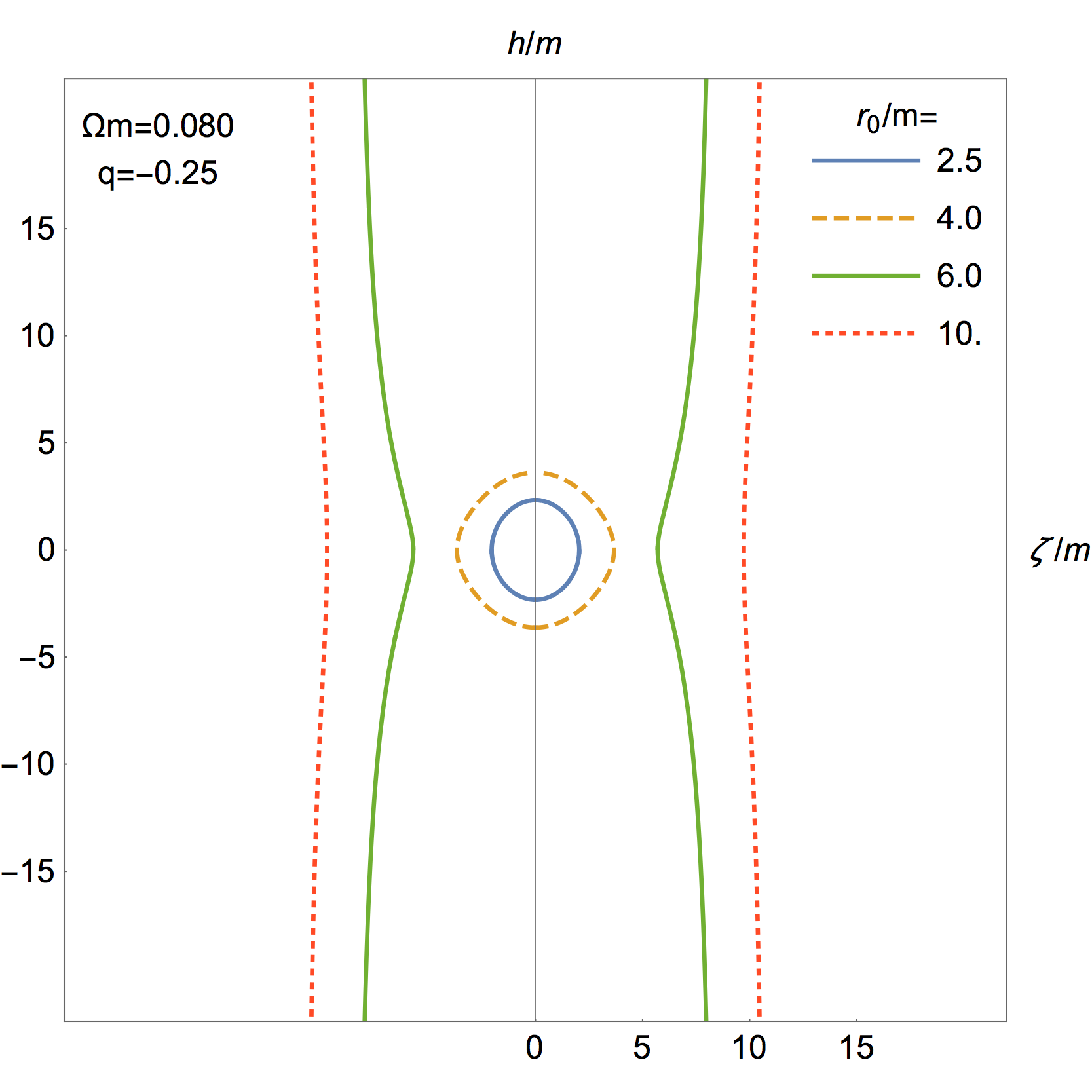}
  \includegraphics[width=0.4\textwidth]{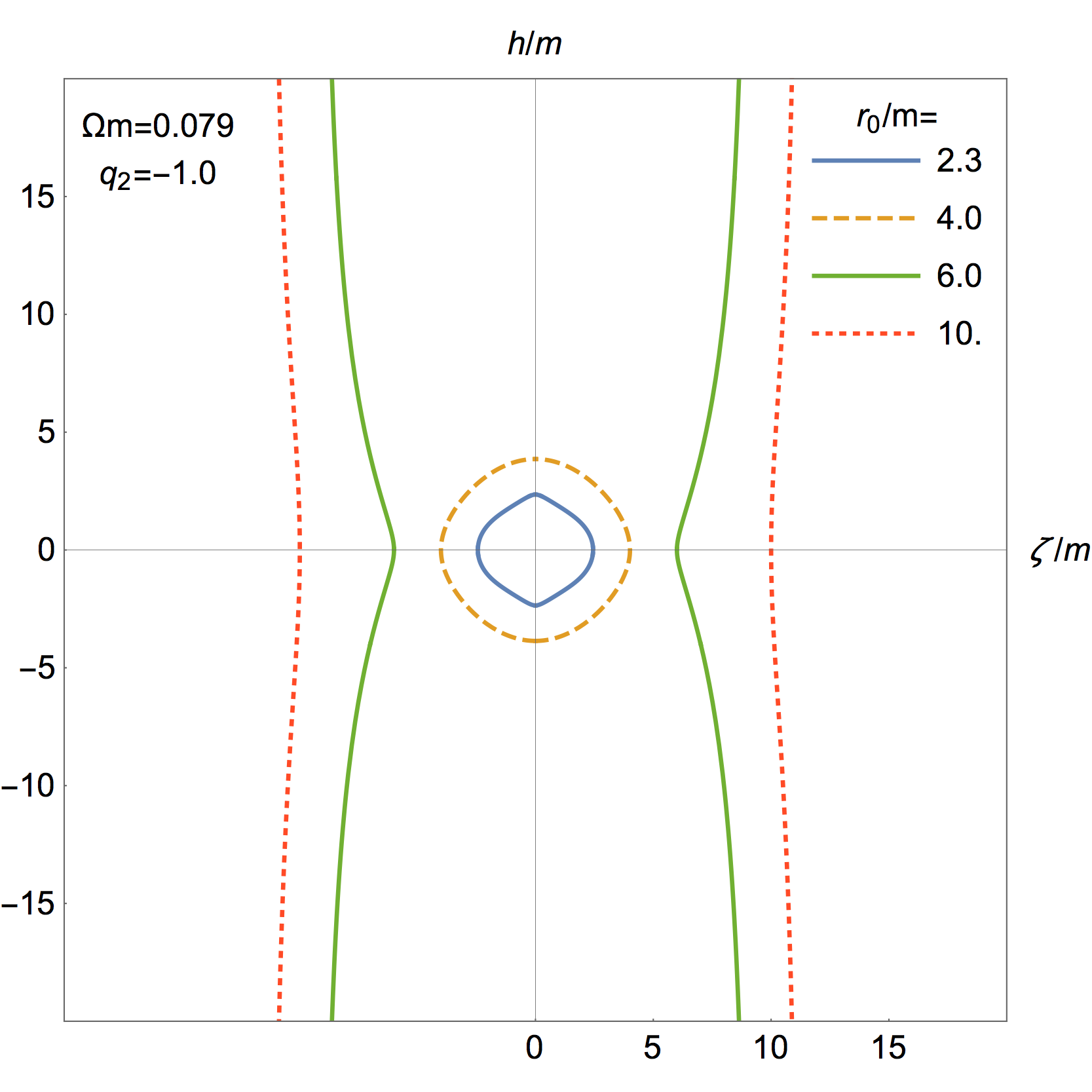} \hfill
  \includegraphics[width=0.4\textwidth]{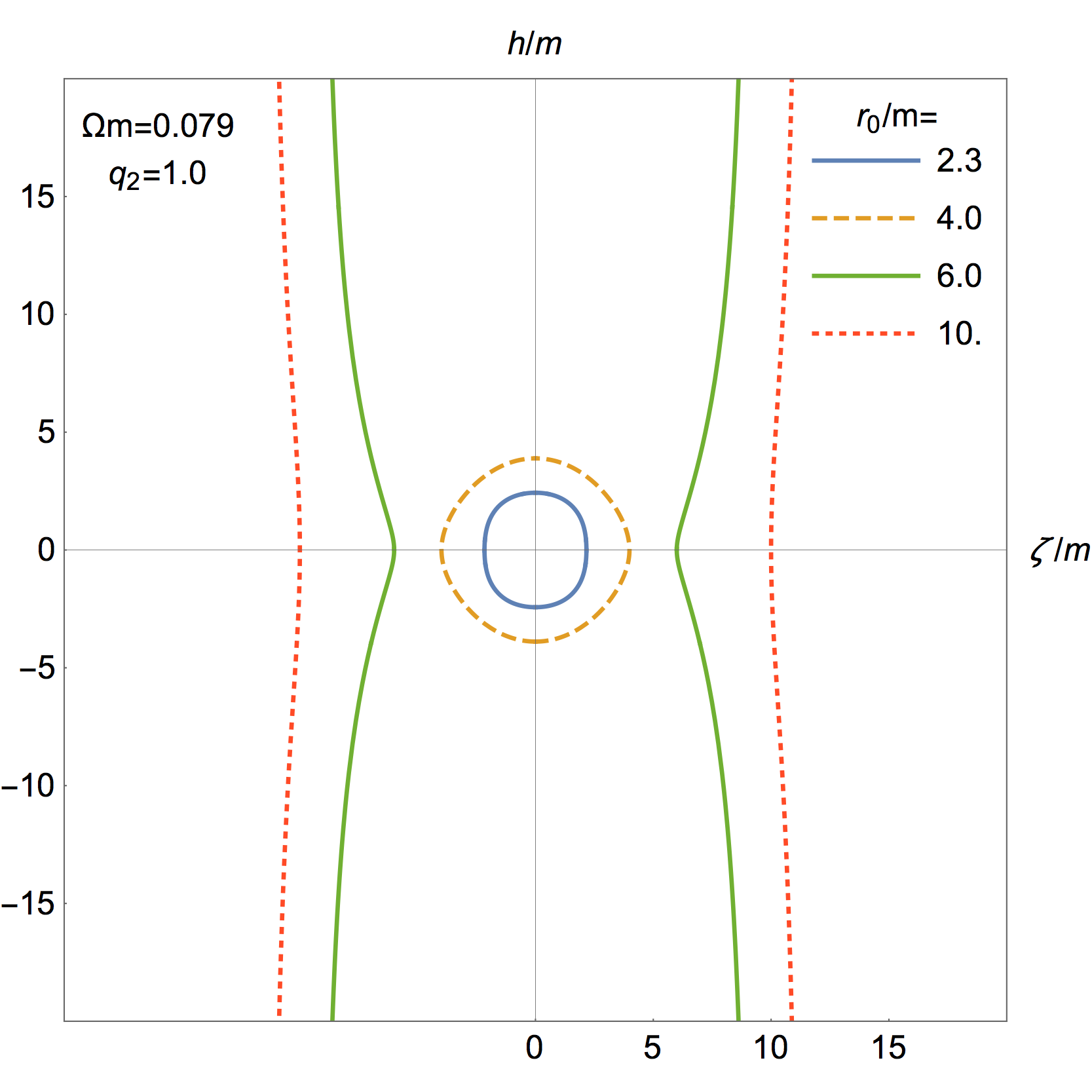}
  \includegraphics[width=0.4\textwidth]{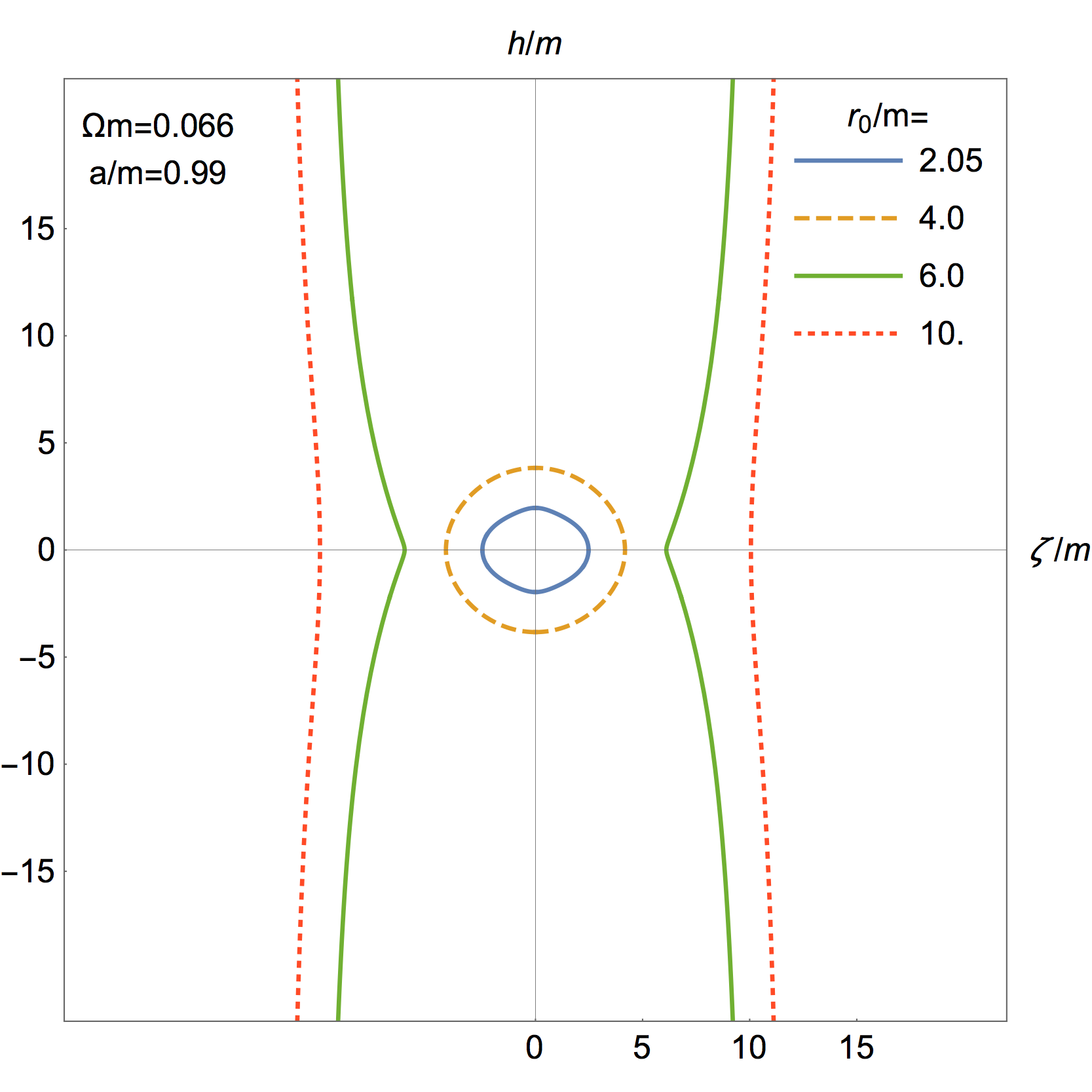} \hfill
  \includegraphics[width=0.4\textwidth]{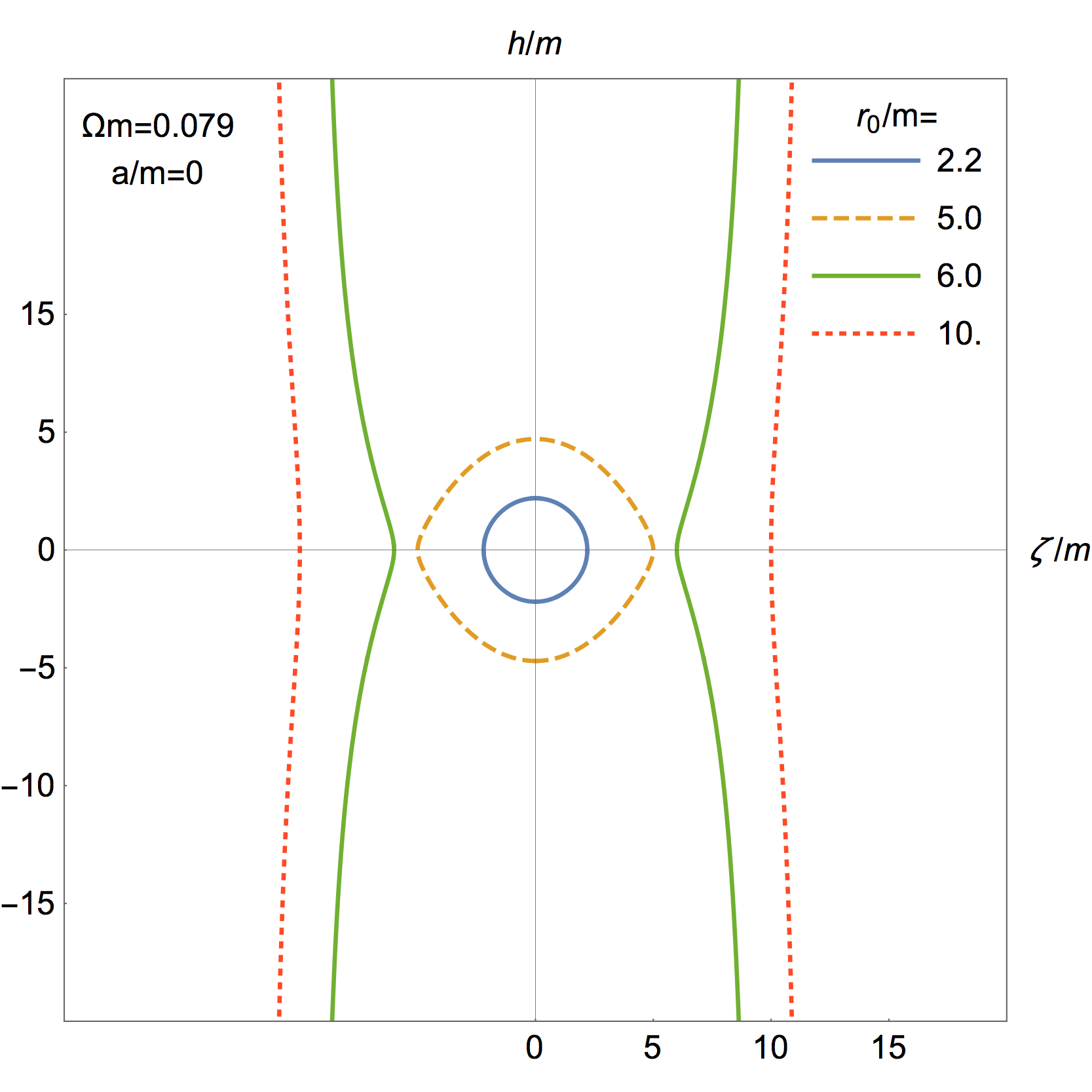}
  \caption{Isometric embedding of isochronometric surfaces $\exp \big( 2 \phi_{\text{rot}} \big)=f_0$ into the Euclidean space $\mathbb{R}^3$. The relativistic geoid as seen by observers on the rotating congruence is by definition one of these surfaces. The value of $r_0$ in the plots is the intersection of the level surface $f_0$ with the radial lines in the equatorial plane. Upper row: q-metric results for oblate (left) and prolate (right) quadrupole configuration. Middle row: Erez-Rosen metric results for oblate (left) and prolate (right) quadrupole configuration. Lower row: Kerr metric results for fixed $a=0.99m$ but different level surfaces (left) and the Schwarzschild result for $a=0$ (right). All necessary parameters are depicted in the respective plots.}
  \label{Fig_embedding2}
\end{figure*}

\begin{figure*}[hbt]
  \includegraphics[width=0.4\textwidth]{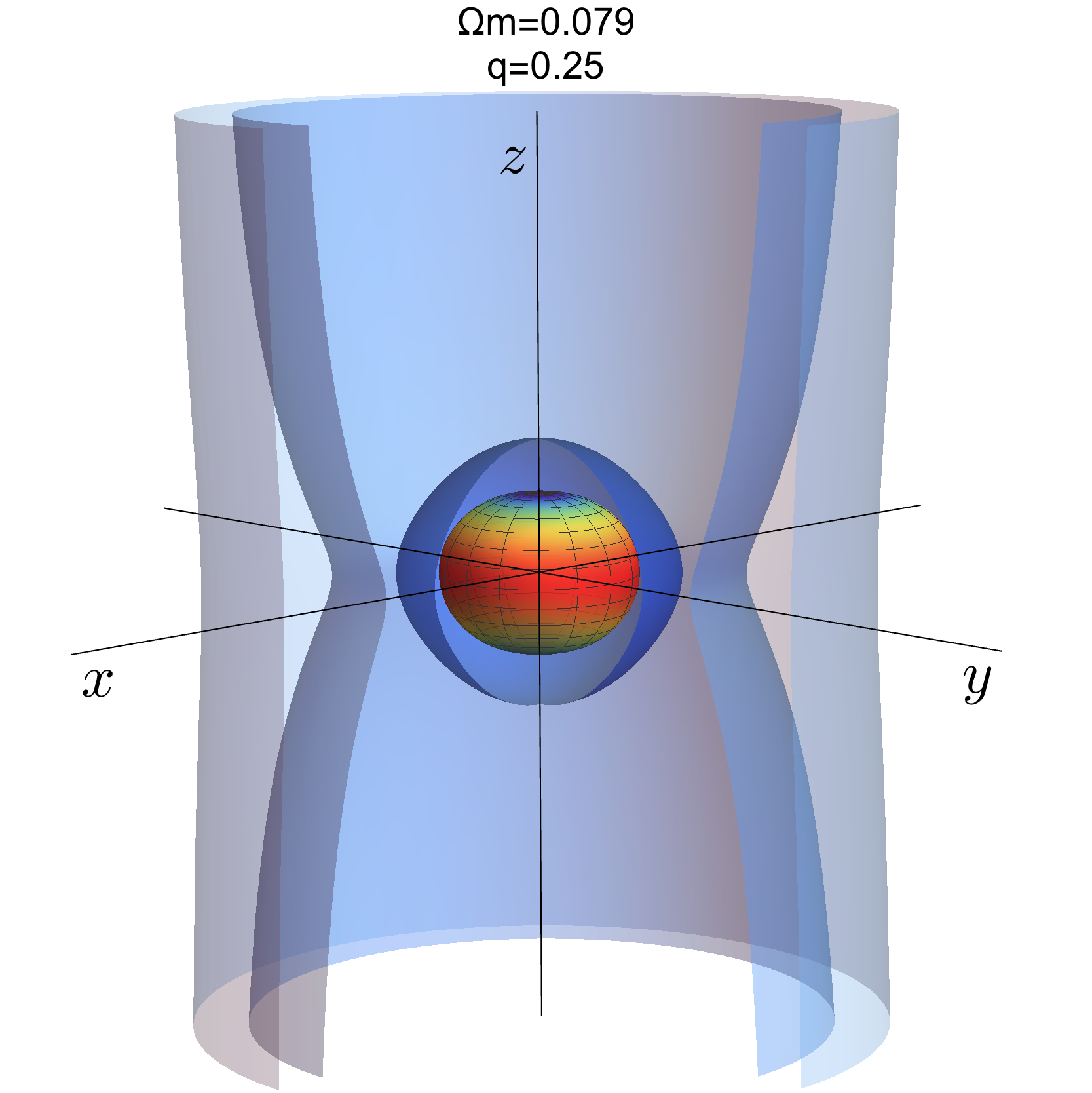} \hfill
  \includegraphics[width=0.4\textwidth]{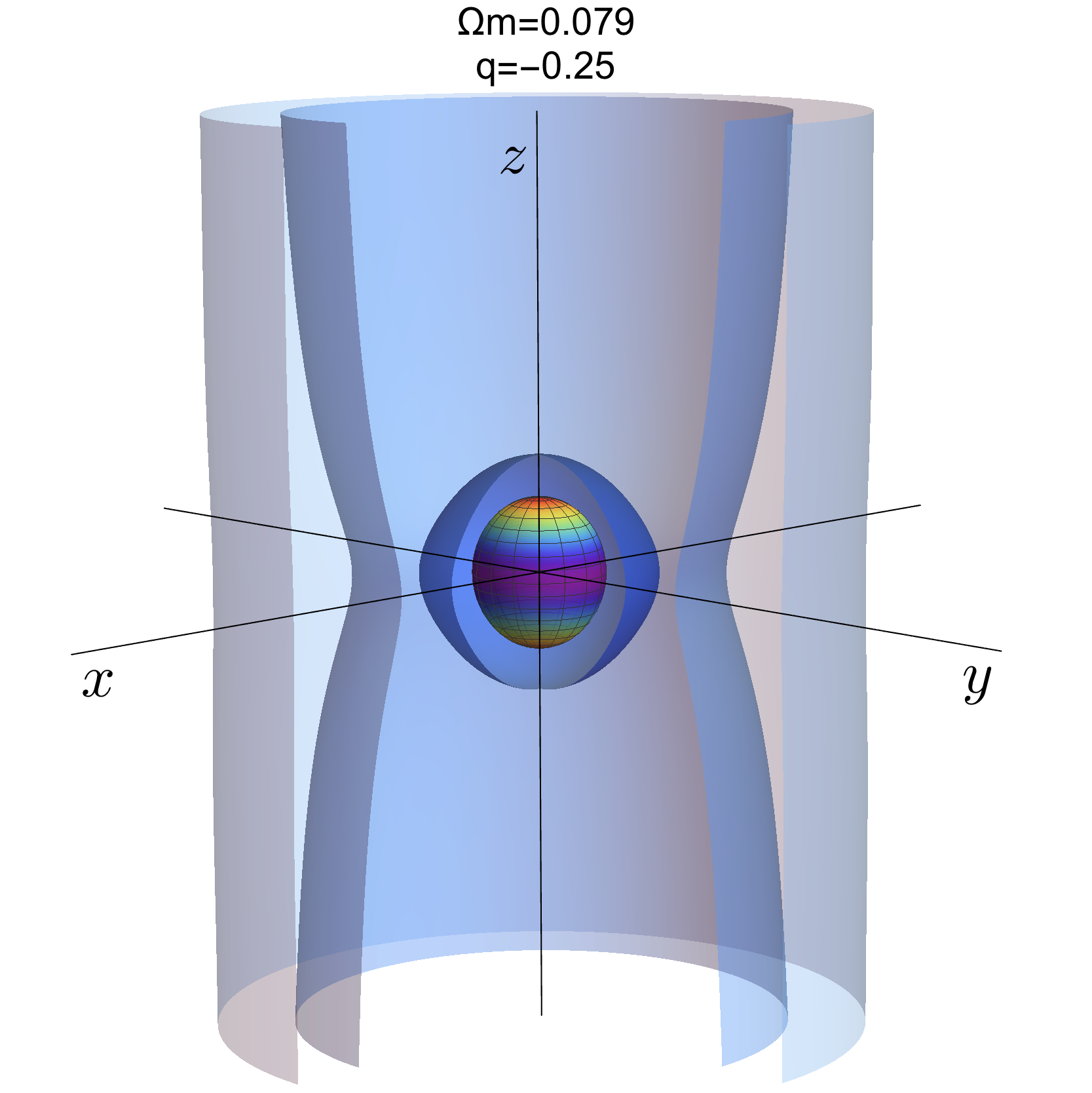}
  \includegraphics[width=0.4\textwidth]{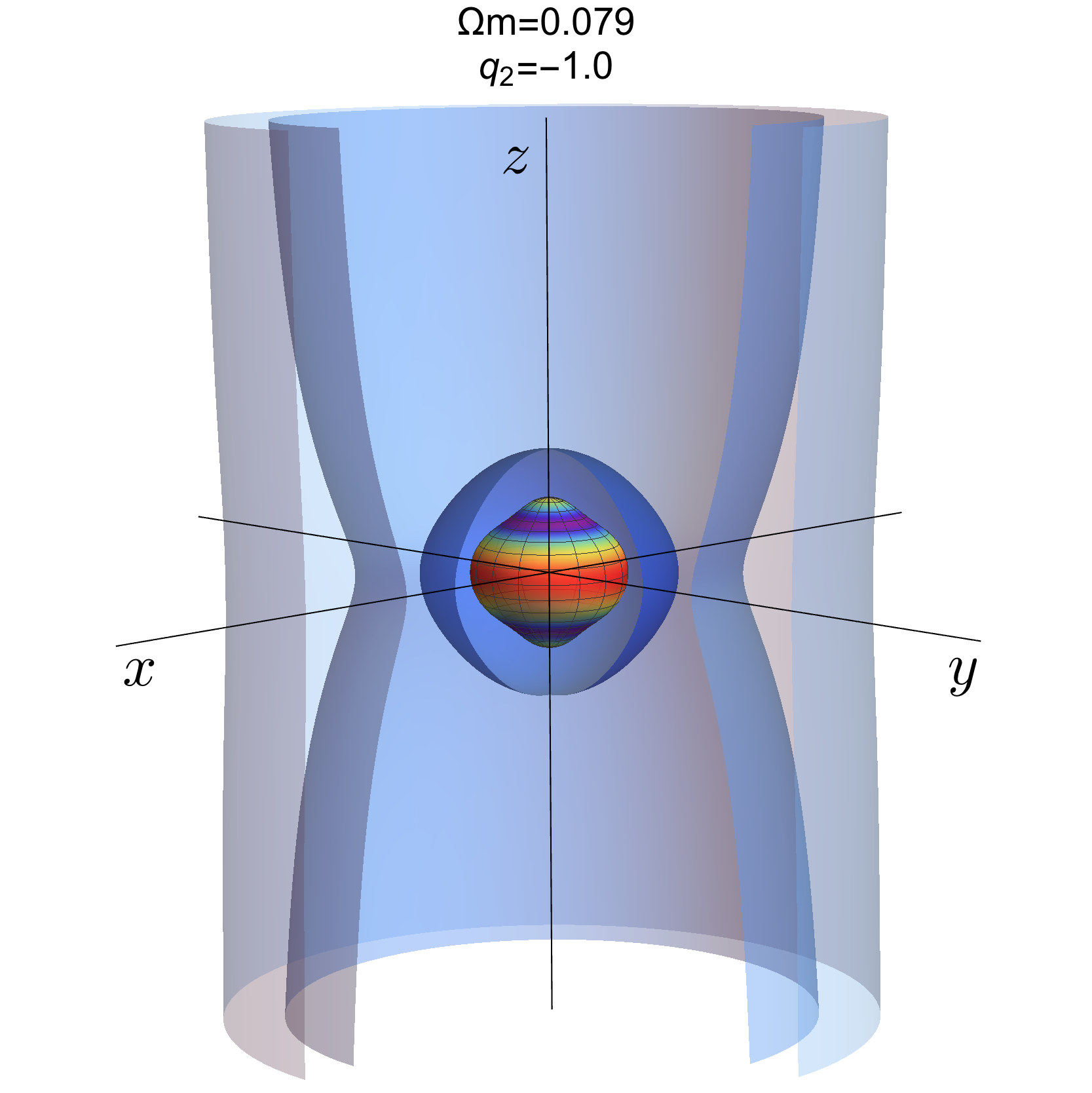} \hfill
  \includegraphics[width=0.4\textwidth]{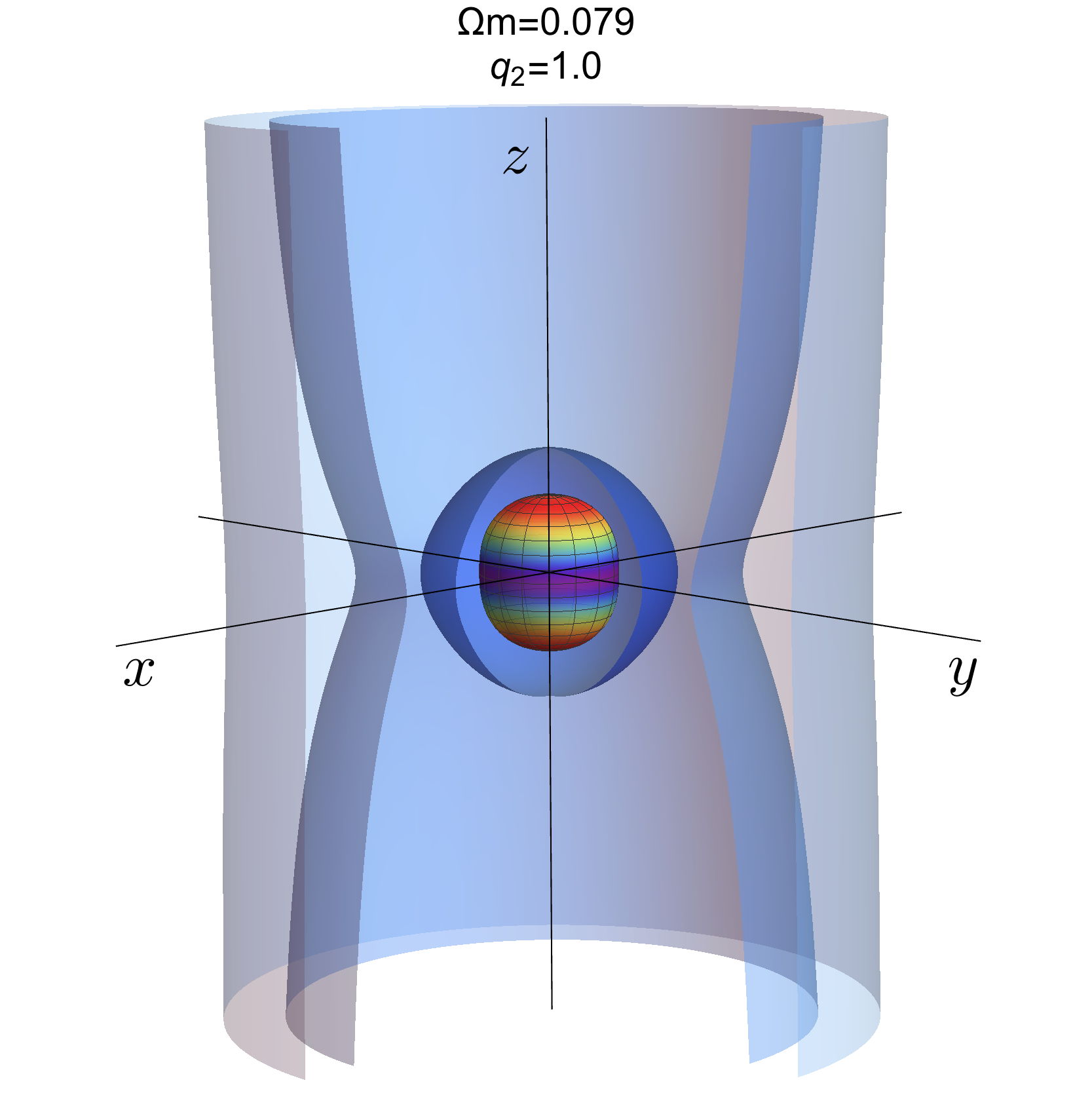}
  \includegraphics[width=0.4\textwidth]{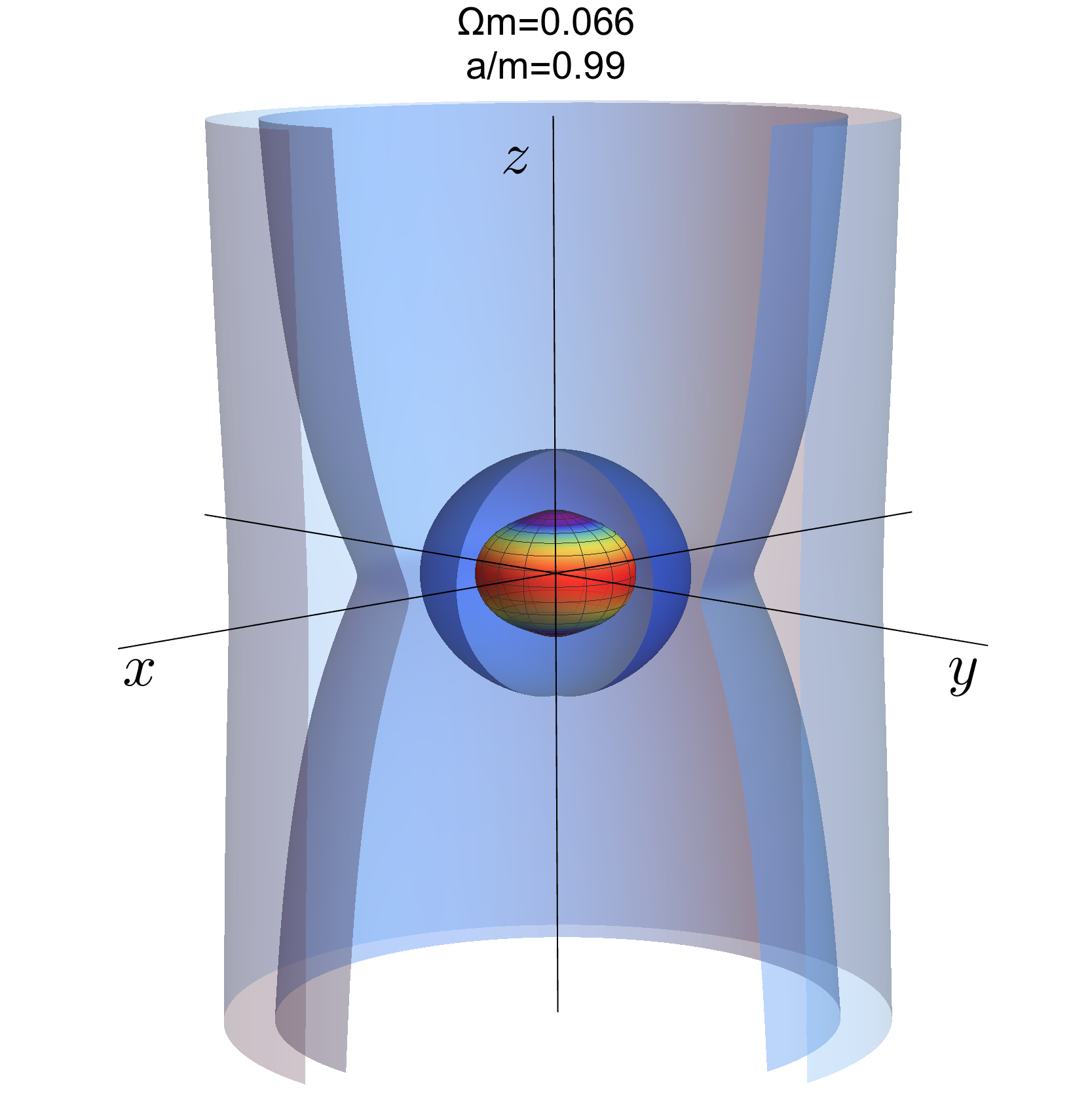} \hfill
  \includegraphics[width=0.4\textwidth]{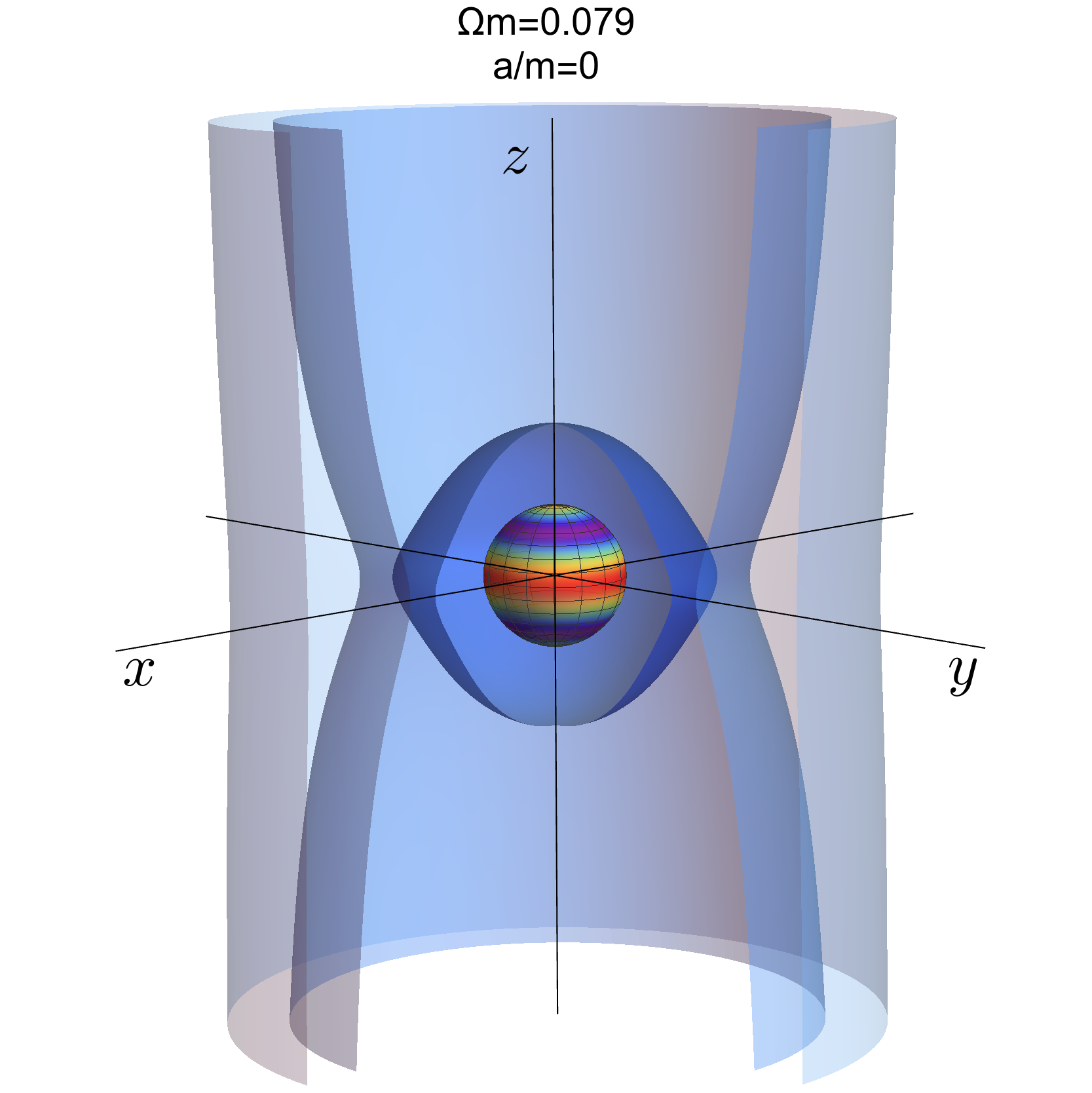}
  \caption{Isometric embedding of isochronometric surfaces $\exp \big( 2 \phi_{\text{rot}} \big)= f_0$ into the Euclidean space $\mathbb{R}^3$. We show the level surfaces in 3-dimensional plots. The level surfaces and their order correspond to those shown in Fig.\ \ref{Fig_embedding2}. For each plot, the innermost level surface is color coded to depict the actual shape such that red corresponds to the farthest distance and purple corresponds to the closest distance to the origin of $\mathbb{R}^3$.}
  \label{Fig_embedding2_3D}
\end{figure*}

As for the Erez-Rosen metric, we may also visualize the isochronometric surfaces of the $q$-metric by isometrically embedding them into the Euclidean space $\mathbb{R}^3$. The result is shown in the top rows of Figs.\ \ref{Fig_embedding1} -- \ref{Fig_embedding2_3D}. Again, we refer to Appendix \ref{sec_Embedding} for details about the construction of the embeddings. As for the Erez-Rosen metric, we have two different signs of the quadrupole parameter. Hence, the embedded surfaces are either oblate or prolate as can be seen in the plots. However, in contrast to the Erez-Rosen metric, the isochronometric surfaces do not develop bulges near the poles in the oblate case; see Fig.\ \ref{Fig_embedding1_3D} in the top row on the left. For the rotating case, the embedding yields cylinders farther away from the source, and the results are qualitatively similar to those obtained for the Schwarzschild and Erez-Rosen cases.

\section{Axisymmetric stationary spacetimes}

\subsection{Axisymmetric stationary solutions to Einstein's vacuum field equation}
\label{Sec_stationarySpacetimes}

All axisymmetric and stationary solutions to Einstein's vacuum field equation can be transformed into the Weyl-Lewis-Papapetrou form. Here, we use spheroidal coordinates since they have proven to be useful in the last section. The metric in these coordinates reads 
\begin{multline}
	\label{Eq_WLPMetric}
	g = -e^{2\psi} (c \, dt+\omega d\varphi)^2 + e^{-2\psi} \sigma^2 
	\left[ e^{2\gamma} (x^2-y^2) \begin{matrix} \, \\ \, \end{matrix} \right. \\
	\times \left. \left( \dfrac{dx^2}{x^2-1} + \dfrac{dy^2}{1-y^2} \right) + (x^2-1)(1-y^2) d\varphi^2 \right] 
\end{multline}
where $\psi$, $\gamma$, and $\omega$ are functions of $x$ and $y$ while $\sigma$ is a constant.
Defining the complex Ernst potential
\begin{align}
E := e^{2\psi} + i \, \Sigma \, , \quad \epsilon := \dfrac{1-E}{1+E} \, ,
\end{align}
where $\Sigma$ is given by
\begin{subequations}
\begin{align}
	\sigma (x^2-1) \partial_x \Sigma &= -e^{4\psi} \partial_y \omega \, , \\
	\sigma (1-y^2) \partial_y \Sigma &= e^{4\psi} \partial_x \omega \, ,
\end{align}
\end{subequations}
reduces the vacuum field equation to a complex equation for the Ernst potential, which can be found, for example, in Ref.\ \cite{Quevedo:2016}. For static spacetimes, the Ernst potential becomes real, and the formalism of Sec.\ \ref{Sec_staticSpacetimes} may be used for constructing solutions. 
We again construct the relativistic potentials
\begin{subequations}
\begin{align}
	e^{2\phi_{\text{stat}}} = & e^{2\psi} \, ,\\
	e^{2\phi_{\text{rot}}} = & e^{2\psi} + 2 \, \dfrac{\Omega}{c} \, \omega e^{2\psi} - 
	\dfrac{\Omega^2}{c^2}  \left[ e^{-2\psi} \sigma^2 (x^2-1)(1-y^2) \right. \notag \\
	&\left. - \omega^2 e^{2\psi} \right] \, ,
\end{align}
\end{subequations}
for the Killing vector fields $\partial _t$ and $\partial _t + \Omega \, \partial _{\varphi}$. The relativistic potential $\phi_{\text{rot}}$ is now defined by the metric function $\psi$ and the twist potential $\omega$, leading to gravitomagnetic contributions.

A simple solution to the Ernst equation for $\omega = 0$ is $\xi = 1/x$. This yields the Schwarzschild solution in spheroidal coordinates, which we considered in the last section. 


\subsection{Example: Kerr spacetime}
The best known and most important stationary and axisymmetric solution to Einstein's vacuum field equation is the Kerr metric. In this case, the Ernst potential depends on the mass parameter $m$ and the spin parameter $a$,
\begin{align}
	\epsilon^{-1} = \dfrac{\sigma}{m} x + i \dfrac{a}{m} y \, , \quad \sigma = \sqrt{m^2-a^2} \, ,
\end{align}
and the metric functions in the Weyl-Lewis-Papapetrou representation become
\begin{subequations}
\begin{align}
	e^{2\psi} &= \dfrac{\sigma^2 x^2 + a^2 y^2 - m^2}{(\sigma x  + m)^2 + a^2 y^2} \, , \\
	\omega &= \dfrac{2a m\, (\sigma x  + m)(1-y^2)}{\sigma^2 x^2  + a^2 y^2 - m^2} \, , \\
	\gamma &= \dfrac{1}{2} \log \left( \dfrac{\sigma^2 x^2 + a^2 y^2 - m^2}{\sigma^2(x^2-y^2)} \right) \, .
\end{align}
\end{subequations}
After the coordinate transformation 
\begin{align}
	\sigma x = r-m\, , \quad y=\cos \vartheta \, ,
\end{align}
we obtain the Kerr metric in its well-known form given in Boyer-Lindquist coordinates $(t,r,\vartheta,\varphi)$,
\begin{align}
	g 	&= - \left( 1-\dfrac{2mr}{\rho^2} \right) c^2 dt^2 + \dfrac{\rho^2}{\Delta} dr^2 + \rho^2 d\vartheta^2 \notag \\
		&+ \sin^2 \vartheta \left( r^2 + a^2 + \dfrac{2m r a^2\sin^2\vartheta}{\rho^2} \right) d\varphi^2 \notag \\
		&- \dfrac{4mra\sin^2\vartheta}{\rho^2} \, c \, dt d\varphi \, , 
\end{align}
where
\begin{align}
	\rho^2 = r^2 + a^2 \cos^2 \vartheta \, , \quad \Delta = r^2 + a^2 -2mr \, .
\end{align}
The relativistic potential for the congruence of Killing observers on integral curves of $\partial_t$  is now given by
\begin{align}
	\label{Eq_KerrRedshiftStationary}
	e^{2\phi_{\text{stat}}} = 1-\dfrac{2mr}{\rho^2} = 1-\dfrac{2mr}{r^2+a^2\cos^2\vartheta} \, .
\end{align}
For Killing observers on a rotating congruence, i.e.\ on integral curves of $\partial _t + \Omega \partial _{\varphi}$ with $\Omega \neq 0$, the relativistic potential $\phi$ satisfies
\begin{multline}
	\label{Eq_KerrRedshiftRotating}
	e^{2\phi_{\text{rot}}} = 1-\dfrac{2mr}{r^2+a^2\cos^2\vartheta} + 
	4 \, \dfrac{\Omega}{c} \,  \dfrac{amr\sin^2\vartheta}{\big( r^2+a^2\cos^2\vartheta \big)} \\
	- \, \dfrac{\Omega^2}{c^2} \,  
	\sin^2 \vartheta \left( r^2+a^2+\dfrac{2m r a^2\sin^2\vartheta}{r^2+a^2\cos^2\vartheta} \right) \, .
\end{multline}
In either case, for any two observers within such a congruence at positions $(r,\vartheta)$ and $(\tilde{r},\tilde{\vartheta})$, respectively, the redshift is 
\begin{align}
\label{Eq_zKerr2}
	1+z = \dfrac{\nu}{\tilde{\nu}} = \dfrac{ e^{\phi(\tilde{r},\tilde{\vartheta})} }{ e^{\phi(r,\vartheta)} } \, .
\end{align}
Figure \ref{Fig_Geoid-c} shows a contour plot of the functions $\exp\big(2\phi_{\text{stat}}\big)$ and  $\exp\big(2\phi_{\text{rot}}\big)$ in pseudo-Cartesian coordinates. To infer more about the intrinsic geometry of the isochronometric surfaces Figs.\ \ref{Fig_embedding1} -- \ref{Fig_embedding2_3D} show their isometric embeddings into Euclidean 3-space. The embedding of the surface $\exp\big(2\phi_{\text{stat}}\big) = f_0$ exists for all $0 < f_0 < 1$ and all values of $a/m$. In the limit $f_0 \to 0$, the isochronometric surfaces approach the ergosurface, i.e.\ the boundary of the ergoregion. An isometric embedding of the ergosurface was first discussed by Sharp \cite{Sharp:1981}. It is known that the ergosurface starts to develop bulges around the poles if $a^2$ approaches its extremal value  $m^2$; for a picture, see Pelavas \cite{Pelavas:2001}. Our plots show a similar behavior of the isochronometric surfaces near the ergosurface.

As an aside, we mention that our formalism may also be used for calculating the gravitomagnetic redshift on the surface of the Earth if the spacetime geometry outside of the Earth is approximated by the Kerr metric. For satellite orbits, the gravitomagnetic redshift (or gravitomagnetic clock effect) has been studied before; see Ref. \cite{Hackmann:2014} for the case of arbitrary orbits. For clocks on the surface of the Earth, we may use the redshift potential \eqref{Eq_KerrRedshiftRotating}. If one clock rotates on the equator, $(r, \vartheta = \pi/2)$, and the other one is situated at the north pole, $(\tilde{r}, \tilde{\vartheta} = 0)$, the redshift becomes
\begin{multline}
	1+z = \dfrac{\nu}{\tilde{\nu}} \\
	= \dfrac{\sqrt{1-\dfrac{2m\tilde{r}}{\tilde{r}^2+a^2}}}{\sqrt{1-\dfrac{2m}{r} + 4 \, \dfrac{\Omega}{c} \, \dfrac{am}{r}
	- \, \dfrac{\Omega^2}{c^2} \,  \left( r^2+a^2+\dfrac{2ma^2}{r} \right)}} \, .
\end{multline}
Subtracting the gravitoelectric part, i.e.\ the same expression for $a=0$, the remainder gives the gravitomagnetic redshift between these two clocks. Inserting the values for all parameters leads to a gravitomagnetic redshift of\footnote{For the calculation we used the following values for the Earth: $m = 0.0044\,$m, $a=743\,m=3.3\,$m, $\Omega=2\pi/86400\,$s, equatorial radius $r=6378.137\,$km$\, $ and polar radius $\tilde{r} = 6356.752\,$km.}
\begin{align}
	z_{\text{grav.magn.}} \sim 10^{-21} \, ,	
\end{align}
which is about 3 orders of magnitude away from contemporary precision but might be measured in the foreseeable future with further improved clocks.


\section{Post-Newtonian approximation of the geoid}

In this section, we consider the PN approximation of the relativistic geoid, and we demonstrate that, indeed, the familiar expression is reproduced at the 1PN level.
    
According to the most recent resolution of the International Astronomical Union (IAU), see, e.g.\ Refs.\ \cite{Soffel:2003, Kaplan:2006}, the PN approximation of the metric of the Earth in geocentric coordinates $(cT,X^i)$ and under the assumption of stationarity reads
\begin{subequations}
\begin{align}
	g_{00} &= - \left( 1 - \dfrac{2 U}{c^2} + \dfrac{2 U^2}{c^4} \right) + \mathcal{O}(c^6) \, , \\
	g_{0i} &= - \, \dfrac{4 U^i}{c^3} + \mathcal{O}(c^5) \, , \\
	g_{ij} &= \delta_{ij} \left( 1 + \dfrac{2U}{c^2} \right) + \mathcal{O}(c^4) \, ,
\end{align}
\end{subequations}
where the potentials $U,U^i$ fulfill the equations
\begin{subequations}
\begin{align}
	\Delta U(X) &= - 4\pi G \rho(X) \, ,\\
	\Delta U^i(X) &= -4\pi G \rho^i(X) \, .
\end{align}
\end{subequations}
The quantities $\rho, \rho^i$ are related to the energy-momentum tensor of the Earth by $\rho = (T^{00} + T^{ii})/c^2$ and $\rho^i = T^{0i}/c$, evaluated in the Geocentric Celestial Reference System (GCRS). For the scalar and vector potentials, one obtains
\begin{subequations}
\begin{align}
	U(X) &= G \int d^3 X' \, \dfrac{\rho(\mathbf{X}')}{|\mathbf{X}-\mathbf{X}'|} \, , \\
	U^i(X) &= G \int d^3 X' \, \dfrac{\rho^i(\mathbf{X}')}{|\mathbf{X}-\mathbf{X}'|} \, .
\end{align}
\end{subequations}
Changing to corotating geocentric coordinates $(c\bar{T},\bar{X}^i)$, the metric becomes \cite{Soffel:1988}
\begin{subequations}
\label{Eq_PNRotating}
\begin{align}
	g_{00} &= - \left( 1 - \dfrac{2 U}{c^2} + \dfrac{2 U^2}{c^4} \right) + \Omega^2(\bar{X}^2+\bar{Y}^2)/c^2 \, , \\
	g_{0i} &= \mathbf{L} - \mathbf{\bar{X}}\times \mathbf{\Omega}/c \, , \\
	g_{ij} &= \delta_{ij} \left( 1 + \dfrac{2U}{c^2} \right) \, ,
\end{align}
\end{subequations}
where 
\begin{align}
	\mathbf{L} = -2 G \dfrac{\mathbf{J}\times \mathbf{\bar{X}}}{c^3 R^3} \, ,
\end{align}
and $\mathbf{\Omega}, \mathbf{J}$ are the angular velocity and angular momentum of the Earth. We use the usual three-vector notation only as a shorthand notation. The vector field $\partial_{\bar{T}}$ is a Killing vector field of the spacetime \eqref{Eq_PNRotating}. Observers on the Earth's surface move on its integral curves since for them $d\bar{X}^i = 0$. These observers form an isometric congruence. The corresponding relativistic potential $ \phi_{PN}$ is given by
\begin{align}
	e^{2\phi_{PN}} =  -g_{00}=1 - \dfrac{2U}{c^2} + \dfrac{2 U^2}{c^4} - \Omega^2(\bar{X}^2+\bar{Y}^2)/c^2 \, .
\end{align}
The defining condition for the relativistic geoid as a level set of the relativistic potential $\phi_{PN}$ yields
\begin{align}
	U + \dfrac{1}{2}\Omega^2(\bar{X}^2+\bar{Y}^2) - \dfrac{U^2}{c^2} = \text{constant}\, ,
\end{align}
which is exactly the expression given by Soffel \emph{et al.}\ in Ref.\ \cite{Soffel:1988}; see their Eq.\ (4). The first two terms reproduce the classical definition of the Newtonian geoid, whereas the last term adds a relativistic correction at the 1PN level.


\section{Conclusion}
In this work, we have generalized the Newtonian and post-Newtonian definitions of the geoid to a fully general relativistic setting. As this definition is not restricted to weak gravitational fields, it makes sense not only for the Earth and other planets but also for compact objects such as neutron stars. Just as the former definitions of the geoid, our definition is based on the assumption that the Earth rotates rigidly with constant angular velocity about a fixed axis. Under this assumption, the Earth is associated with an isometric congruence of worldlines, i.e.\ with a family of Killing observers. We have defined the geoid in terms of isochronometric surfaces that are the level sets of the redshift potential for this isometric observer congruence. As the isochronometric surfaces may be realized with networks of standard clocks that are connected by fiber links, this is an operational definition of the geoid.

While we consider the definition of the geoid in terms of clocks as primary, we have also emphasized that the redshift potential associated with an isometric congruence is, at the same time, an acceleration potential. This observation generalizes the equality of the u- and a-geoid, which was known to hold in a PN setting, into the full formalism of general relativity.

In practical geodesy, our stationary gravitational field is the time average of the real gravitational field of the Earth. The real gravitational field of the Earth contains time-dependent parts which have to be treated through, e.g., an appropriate reduction. Here, we focus on the correct and fully relativistic definition of the geoid without time dependence.

We have illustrated our definition of the geoid by calculating the isochronometric surfaces of axisymmetric and static spacetimes, with the Schwarzschild metric, the Erez-Rosen metric, and the $q$-metric as particular examples. We have then considered the case of axisymmetric and stationary spacetimes, with the Kerr metric as a particular example. As the shape of the isochronometric surfaces in a chosen coordinate system has no invariant meaning, we have isometrically embedded these surfaces into Euclidean 3-space to show their intrinsic geometry. As an aside, we have mentioned that the redshift potential for rotating observers in the Kerr metric may be used for estimating the gravitomagnetic redshift for clocks on the surface of the Earth. 

Finally, we have derived the redshift potential and the relativistic geoid in a 1PN spacetime and recovered the previously known result.

An important task for the future is to express the geoid of a rotating and non-axisymmetric body in terms of multipole moments. This is conceptually challenging because in this case the spacetime is not stationary near infinity; the Killing vector field associated with the rotating body becomes spacelike outside of a cylindrical region about the rotation axis. For this reason, the time-independent asymptotically defined Geroch-Hansen multipole moments do not exist. In future work, we are planning to tackle the question of how local measurements in the neighborhood of a gravitating body are to be related to appropriately defined multipole moments in a relativistic formalism without approximations.      

We emphasize again that our formalism is valid for stationary non-axisymmetric objects as well, as long as the backreaction from gravitational radiation and the resulting slowdown of the rotation can be ignored. In this sense, our geoid can be constructed for any irregularly shaped rotating body.


\begin{acknowledgments}
This work was supported by the Deutsche Forschungsgemeinschaft (DFG) through the Collaborative Research Center (SFB) 1128 ``geo-Q'' and the Research Training Group 1620 ``Models of Gravity.'' We also acknowledge support by the German Space Agency DLR with funds provided by the Federal Ministry of Economics and Technology (BMWi) under Grant No.\ DLR 50WM1547.

The authors would like to thank Pac{\^ o}me Delva, Heiner Denker, Domenico Giulini, Norman G\"urlebeck, Sergei Kopeikin, J\"urgen M\"uller, and Michael Soffel for helpful discussions and for reading the manuscript. The first author acknowledges insightful discussions with Vojt{\v e}ch Witzany and Michael Fennen.
\end{acknowledgments}

\appendix


\section{Isometric embedding of isochronometric surfaces} \label{sec_Embedding}
As the coordinate representation of the geoid has no invariant geometric meaning, it is recommendable to isometrically embed the isochronometric surfaces into Euclidean 3-space. If such an embedding is possible, it represents the intrinsic geometry of the geoid. 

In all examples that we considered in this paper, the geoid was defined by the level sets of a function
\begin{align}
	\label{Eq_appendix:Embedding1}
  	f(x,y) = f_0 = \text{constant} \, ,
\end{align}
where $x$ and $y$ are spheroidal coordinates. As an alternative, we may use the coordinates $(r,\vartheta)$, which are related to $(x,y)$ by the coordinate transformation $x=r/m-1, \, y=\cos \vartheta$; see Eq.\ \eqref{Eq_CoordinateTrafo3}. 

On the two-dimensional surface defined by \eqref{Eq_appendix:Embedding1}, we must have 
\begin{align}
	\label{Eq_appendix:Embedding2}
  	0 = df =  \partial_x f(x,y) dx + \partial_y f(x,y) dy \, ;
\end{align}
hence,
\begin{align}
	\label{Eq_appendix:Embedding3}
  	dx^2 = \left( \dfrac{\partial_y f(x,y)}{\partial_x f(x,y)} \right)^2 dy^2 \, .
\end{align}
As a consequence, the two-dimensional Riemannian metric on the surface $f=f_0$ is
\begin{multline}
	\label{Eq_appendix:Embedding4}
	g^{(2)} = \left[ g_{xx}(x,y) \left( \dfrac{\partial_y f(x,y)}{\partial_x f(x,y)} \right)^2 + g_{yy} \right] dy^2 \\
	+ g_{\varphi \varphi}(x,y) d\varphi^2 \, .
\end{multline} 

We want to isometrically embed this surface into Euclidean 3-space with cylindrical coordinates $(\zeta,\varphi,h)$,
\begin{align}
	\label{Eq_appendix:embedding:EuclideanMetric}
  	g_E^{(3)} = dh^2 + d\zeta^2 + \zeta^2 d\varphi^2 \, .
\end{align}
The embedding functions $h(y)$ and $\zeta(y)$ are to be determined from the equation
\begin{multline}
	\label{Eq_appendix:Embedding5}
  	\left[ g_{xx}(x,y) \left( \dfrac{\partial_y f(x,y)}{\partial_x f(x,y)} \right)^2 + g_{yy} \right] dy^2 + g_{\varphi \varphi}(x,y) d\varphi^2 \\
  	= \left( h'(y)^2 + \zeta'(y)^2 \right) \, dy^2 + \zeta(y)^2 d\varphi^2 \, .
\end{multline}
If Eq.\ \eqref{Eq_appendix:Embedding1} can be explicitly solved for $x=x(y)$, we may insert this expression into (\ref{Eq_appendix:Embedding5}). Comparing coefficients results in
\begin{subequations}
	\begin{align}
		\label{Eq_zeta}
  		\zeta(y) 	&= \left. \sqrt{g_{\varphi \varphi}(x,y)} \right|_{x=x(y)} \, ,\\
         \label{Eq_h}
  	h(y) 		&= \pm \int_{0}^{y} dy \, \left( g_{xx}(x,y) \left( \dfrac{\partial_y f(x,y)}{\partial_x f(x,y)} \right)^2 + g_{yy}(x,y) \right. \notag \\
  	&-\left. \dfrac{g'_{\varphi \varphi}(x,y)^2}{4 g_{\varphi \varphi}(x,y)} \right)^{1/2}_{x=x(y)} \, .
	\end{align}
\end{subequations}
In Eq.\ \eqref{Eq_h}, the expression $g'_{\varphi \varphi}$, by abuse of notation, is understood to mean that first $x(y)$ is to be inserted and then the derivative with respect to $y$ is to be taken. The integral in Eq.\ \eqref{Eq_h} has to be calculated either analytically, if this is possible, or numerically. 
 
Equations \eqref{Eq_zeta} and \eqref{Eq_h} give us the cylindrical radius coordinate $\zeta$ and the cylindrical height coordinate $h$ in Euclidean 3-space as functions of the parameter $y$ of which the allowed range is given by $y \in [-1,1]$, corresponding to $\vartheta \in [0,\pi]$. In this way, we get a meridional section of the embedded surface in parametrized form; by letting this figure rotate about the axis $\zeta = 0$, we get the entire embedded surface. The embedding is possible near all $y$ values for which 
\begin{align}
	\label{Eq_}
  	g_{xx}(x,y) \left( \dfrac{\partial_y f(x,y)}{\partial_x f(x,y)} \right)^2 + g_{yy}(x,y)  > \dfrac{g'_{\varphi \varphi}(x,y)^2}{4 g_{\varphi \varphi}(x,y)} \, .
\end{align}
If this condition is violated, the surface cannot be isometrically embedded into Euclidean 3-space, which means that its intrinsic geometry is hard to visualize. 

This direct construction of the embedded surface in parametrized form is possible if Eq.\ \eqref{Eq_appendix:Embedding1} can be explicitly solved for $x=x(y)$. If this cannot be done, we have at least an expression for the derivative of this function, as Eq.\ \eqref{Eq_appendix:Embedding2} implies that
\begin{align}
	\label{Eq_}
  	x'(y) = \dfrac{dx}{dy} = - \dfrac{\partial_y f(x,y)}{\partial_x f(x,y)} \, .
\end{align}
Using Eq.\ (\ref{Eq_h}), we obtain a coupled system of ordinary differential equations,
\begin{subequations}
	\begin{align}
		\label{Eq_}
  		x'(y) &= \left. - \dfrac{\partial_y f(x,y)}{\partial_x f(x,y)} \right|_{x=x(y)} \, ,\\
  		h'(y) &= \left( g_{xx}(x,y) \left( \dfrac{\partial_y f(x,y)}{\partial_x f(x,y)} \right)^2 + g_{yy}(x,y) \right. \notag \\
  	&-\left. \dfrac{g'_{\varphi \varphi}(x,y)^2}{4 g_{\varphi \varphi}(x,y)} \right)^{1/2}_{x=x(y)} \, ,
\end{align}
\end{subequations}
for the functions $x(y)$ and $h(y)$, which is to be solved numerically with initial conditions $x(0) = x_0$, $h(0) =0$. Of course, this is possible only if an embedding exists. If $x(y)$ and $h(y)$ have been determined, the function $\zeta(y)$ is given by Eq.\ \eqref{Eq_zeta}.


\section{Conventions and Symbols \label{app_conventions}}

In the following, we summarize our conventions and collect some frequently used formulas. A directory of symbols used throughout the text can be found in Table \ref{tab_symbols}. For an arbitrary $k$-tensor $T_{\mu_1 \dots \mu_k}$, the symmetrization and antisymmetrization are defined by
\begin{eqnarray}
T_{(\mu_1\dots \mu_k)} &:=& {\frac 1{k!}}\sum_{I=1}^{k!}T_{\pi_I\!\{\mu_1\dots \mu_k\}},\label{S}\\
T_{[\mu_1\dots \mu_k]} &:=& {\frac 1{k!}}\sum_{I=1}^{k!}(-1)^{|\pi_I|}T_{\pi_I\!\{\mu_1\dots \mu_k\}},\label{A}
\end{eqnarray}
where the sum is taken over all possible permutations (symbolically denoted by $\pi_I\!\{\mu_1\dots \mu_k\}$) of its $k$ indices. 

The signature of the spacetime metric is assumed to be $(-,+,+,+)$. Greek indices $\mu, \nu ,\lambda , \dots$ are spacetime indices and take values $0 \dots 3$. Latin indices $i,j,k$ are spatial indices and take values $1\dots 3$. 

\begin{widetext}
\begin{table*}[]
\caption{\label{tab_symbols}Directory of symbols.}
\begin{ruledtabular}
\begin{tabular}{llllll}
Symbol & Unit & Explanation & Symbol & Unit & Explanation \\
\hline
&\\
\hline
$g_{\mu\nu}$ 				& 1 & Metric & 
$M$ 						& kg & Mass of the central object\\

$\sqrt{-g}$ 					& 1 & Determinant of the metric & 
$m$ 						& m & Mass of the central object \\

$\delta^\mu_\nu$ 			& 1 & Kronecker symbol & 
$\rho$						& kg\,m$^{-3}$ & Mass density \\

$\gamma, \tilde{\gamma}$ 	& 1 & Observer worldlines & 
$G$ 						& m$^3\,$kg$^{-1}\,$s$^{-2}$ & Newton's gravitational constant\\

$u^\mu$ 						& m\,s$^{-1}$ & Observer four-velocity & 
$c$ 						& m\,s$^{-1}$ & Speed of light\\

$a^\mu$ 						& m\,s$^{-2}$ & Observer four-acceleration & 
$N_l$ 						& kg$\,$m$^l$ & Newtonian multipole moments\\

$\phi$ 						& 1 & (Redshift, acceleration) potential & 
$R_l$ 						& kg$\,$m$^l$ & Geroch-Hansen multipole moments\\

$\xi^\mu$ 					& m\,s$^{-1}$ & Killing vector field & 
$C_l$ 						& kg$\,$m$^l$ & Relativistic multipole moment corrections\\

$\psi, \gamma$ 				& 1 & Weyl's metric functions & 
$\tau, \tilde{\tau}$ 		& s & Proper times\\

$\omega_{\mu\nu}, \omega^\mu$ & s$^{-1}$ & Rotation (tensor, vector) & 
$\nu_1,\, \nu_2$ 			& s$^{-1}$ & Measured frequencies\\

$\sigma_{\mu\nu}$ 			& s$^{-1}$ & Shear tensor & 
$\Omega$ 					& s$^{-1}$ & Angular velocity\\

$\theta$ 					& s$^{-1}$ & Congruence expansion & 
$P_{lm}, P_l$ 				& 1 & (Associated) Legendre polynomials\\

$E, \epsilon$ 				& 1 & Ernst potentials &
$Q_l$						& 1 & Legendre functions of 2nd kind\\

$P^\mu_\nu$ 					& 1 & Projection operator &
$c_l$ 						& m$^{l+1}$ & Series expansion coefficients\\

$\partial_\mu\,$, $D_\mu$ 	& m$^{-1}$ & (Partial, covariant) derivative &
$(q_l, \, \bar{q}_l)$ 		& (1,$\,$m$^l$\,kg$^{-l}$) & Series expansion coefficients\\

$\frac{D}{ds} = $``$\dot{\phantom{a}}$'' & s$^{-1}$ & Total covariant derivative &
$C_{\vartheta}, C_\varphi$ 	& m & (Polar, azimuthal) circumferences\\

$(r, \vartheta, \varphi)$ 	& (m,1,1) & Spherical coordinates & 
$f$ 						& 1 & Flattening parameter\\

$(x, y, \varphi)$ 			& 1 & Spheroidal coordinates & 
$U$ 						& m$^2$\,s$^{-2}$ & Newtonian gravitational potential\\

$(\rho, z, \varphi)$ 		& (m,m,1) & Canonical Weyl coordinates & 
$V$ 						& m$^2$\,s$^{-2}$ & Centrifugal potential\\

$(X,Y,Z)$ 					& m & PN geocentric coordinates &
$W$ 						& m$^2$\,s$^{-2}$ & Total potential\\

$(\bar{X},\bar{Y},\bar{Z})$ & m & PN geocentric corotating coordinates & $f(x,y)$ & 1 & Geoid embedding functions\\

$(\zeta,h,\varphi)$ 		& (m,m,1) & Cylindrical coordinates in $\mathbb{R}^3$ & $n$ & 1 & Index of refraction\\
\end{tabular} 
\end{ruledtabular}
\end{table*}
\end{widetext}

\clearpage
\bibliography{geoid.bib}

\begin{thebibliography}{51}%
\makeatletter
\providecommand \@ifxundefined [1]{%
 \@ifx{#1\undefined}
}%
\providecommand \@ifnum [1]{%
 \ifnum #1\expandafter \@firstoftwo
 \else \expandafter \@secondoftwo
 \fi
}%
\providecommand \@ifx [1]{%
 \ifx #1\expandafter \@firstoftwo
 \else \expandafter \@secondoftwo
 \fi
}%
\providecommand \natexlab [1]{#1}%
\providecommand \enquote  [1]{``#1''}%
\providecommand \bibnamefont  [1]{#1}%
\providecommand \bibfnamefont [1]{#1}%
\providecommand \citenamefont [1]{#1}%
\providecommand \href@noop [0]{\@secondoftwo}%
\providecommand \href [0]{\begingroup \@sanitize@url \@href}%
\providecommand \@href[1]{\@@startlink{#1}\@@href}%
\providecommand \@@href[1]{\endgroup#1\@@endlink}%
\providecommand \@sanitize@url [0]{\catcode `\\12\catcode `\$12\catcode
  `\&12\catcode `\#12\catcode `\^12\catcode `\_12\catcode `\%12\relax}%
\providecommand \@@startlink[1]{}%
\providecommand \@@endlink[0]{}%
\providecommand \url  [0]{\begingroup\@sanitize@url \@url }%
\providecommand \@url [1]{\endgroup\@href {#1}{\urlprefix }}%
\providecommand \urlprefix  [0]{URL }%
\providecommand \Eprint [0]{\href }%
\providecommand \doibase [0]{http://dx.doi.org/}%
\providecommand \selectlanguage [0]{\@gobble}%
\providecommand \bibinfo  [0]{\@secondoftwo}%
\providecommand \bibfield  [0]{\@secondoftwo}%
\providecommand \translation [1]{[#1]}%
\providecommand \BibitemOpen [0]{}%
\providecommand \bibitemStop [0]{}%
\providecommand \bibitemNoStop [0]{.\EOS\space}%
\providecommand \EOS [0]{\spacefactor3000\relax}%
\providecommand \BibitemShut  [1]{\csname bibitem#1\endcsname}%
\let\auto@bib@innerbib\@empty
\bibitem [{\citenamefont {Torge}\ and\ \citenamefont
  {M{\"u}ller}(2012)}]{TorgeMueller:2012}%
  \BibitemOpen
  \bibfield  {author} {\bibinfo {author} {\bibfnamefont {W.}~\bibnamefont
  {Torge}}\ and\ \bibinfo {author} {\bibfnamefont {J.}~\bibnamefont
  {M{\"u}ller}},\ }\href@noop {} {\emph {\bibinfo {title} {Geodesy}}}\
  (\bibinfo  {publisher} {De Gruyter},\ \bibinfo {address} {Berlin},\ \bibinfo
  {year} {2012})\BibitemShut {NoStop}%
\bibitem [{\citenamefont {Loomis}\ \emph {et~al.}(2012)\citenamefont {Loomis},
  \citenamefont {Nerem},\ and\ \citenamefont {Luthcke}}]{Loomis:2012}%
  \BibitemOpen
  \bibfield  {author} {\bibinfo {author} {\bibfnamefont {B.~D.}\ \bibnamefont
  {Loomis}}, \bibinfo {author} {\bibfnamefont {R.~S.}\ \bibnamefont {Nerem}}, \
  and\ \bibinfo {author} {\bibfnamefont {S.~B.}\ \bibnamefont {Luthcke}},\
  }\href {\doibase 10.1007/s00190-011-0521-8} {\bibfield  {journal} {\bibinfo
  {journal} {J. Geod.}\ }\textbf {\bibinfo {volume} {86}},\ \bibinfo {pages}
  {319} (\bibinfo {year} {2012})}\BibitemShut {NoStop}%
\bibitem [{\citenamefont {Flechtner}\ \emph {et~al.}(2016)\citenamefont
  {Flechtner}, \citenamefont {Neumayer}, \citenamefont {Dahle}, \citenamefont
  {Dobslaw}, \citenamefont {Fagiolini}, \citenamefont {Raimondo},\ and\
  \citenamefont {Guentner}}]{Flechtner:2016}%
  \BibitemOpen
  \bibfield  {author} {\bibinfo {author} {\bibfnamefont {F.}~\bibnamefont
  {Flechtner}}, \bibinfo {author} {\bibfnamefont {K.-H.}\ \bibnamefont
  {Neumayer}}, \bibinfo {author} {\bibfnamefont {C.}~\bibnamefont {Dahle}},
  \bibinfo {author} {\bibfnamefont {H.}~\bibnamefont {Dobslaw}}, \bibinfo
  {author} {\bibfnamefont {E.}~\bibnamefont {Fagiolini}}, \bibinfo {author}
  {\bibfnamefont {J.-C.}\ \bibnamefont {Raimondo}}, \ and\ \bibinfo {author}
  {\bibfnamefont {A.}~\bibnamefont {Guentner}},\ }\href {\doibase
  10.1007/s10712-015-9338-y} {\bibfield  {journal} {\bibinfo  {journal} {Surv.
  Geophys.}\ }\textbf {\bibinfo {volume} {37}},\ \bibinfo {pages} {453}
  (\bibinfo {year} {2016})}\BibitemShut {NoStop}%
\bibitem [{\citenamefont {Bloom}\ \emph {et~al.}(2014)\citenamefont {Bloom},
  \citenamefont {Nicholson}, \citenamefont {Williams}, \citenamefont
  {Campbell}, \citenamefont {Bishof}, \citenamefont {Zhang}, \citenamefont
  {Zhang}, \citenamefont {Bromley},\ and\ \citenamefont {Ye}}]{Bloom:2014}%
  \BibitemOpen
  \bibfield  {author} {\bibinfo {author} {\bibfnamefont {B.~J.}\ \bibnamefont
  {Bloom}}, \bibinfo {author} {\bibfnamefont {T.~L.}\ \bibnamefont
  {Nicholson}}, \bibinfo {author} {\bibfnamefont {J.~R.}\ \bibnamefont
  {Williams}}, \bibinfo {author} {\bibfnamefont {S.~L.}\ \bibnamefont
  {Campbell}}, \bibinfo {author} {\bibfnamefont {M.}~\bibnamefont {Bishof}},
  \bibinfo {author} {\bibfnamefont {X.}~\bibnamefont {Zhang}}, \bibinfo
  {author} {\bibfnamefont {W.}~\bibnamefont {Zhang}}, \bibinfo {author}
  {\bibfnamefont {S.~L.}\ \bibnamefont {Bromley}}, \ and\ \bibinfo {author}
  {\bibfnamefont {J.}~\bibnamefont {Ye}},\ }\href {\doibase
  10.1038/nature12941} {\bibfield  {journal} {\bibinfo  {journal} {Nature}\
  }\textbf {\bibinfo {volume} {506}},\ \bibinfo {pages} {71} (\bibinfo {year}
  {2014})}\BibitemShut {NoStop}%
\bibitem [{\citenamefont {Soffel}\ \emph {et~al.}(1988)\citenamefont {Soffel},
  \citenamefont {Herold}, \citenamefont {Ruder},\ and\ \citenamefont
  {Schneider}}]{Soffel:1988}%
  \BibitemOpen
  \bibfield  {author} {\bibinfo {author} {\bibfnamefont {M.~H.}\ \bibnamefont
  {Soffel}}, \bibinfo {author} {\bibfnamefont {H.}~\bibnamefont {Herold}},
  \bibinfo {author} {\bibfnamefont {H.}~\bibnamefont {Ruder}}, \ and\ \bibinfo
  {author} {\bibfnamefont {M.}~\bibnamefont {Schneider}},\ }\href@noop {}
  {\bibfield  {journal} {\bibinfo  {journal} {Manuscripta Geodaetica}\ }\textbf
  {\bibinfo {volume} {13}},\ \bibinfo {pages} {143} (\bibinfo {year}
  {1988})}\BibitemShut {NoStop}%
\bibitem [{\citenamefont {Kopeikin}\ \emph {et~al.}(2015)\citenamefont
  {Kopeikin}, \citenamefont {Mazurova},\ and\ \citenamefont
  {Karpik}}]{Kopeikin:2015}%
  \BibitemOpen
  \bibfield  {author} {\bibinfo {author} {\bibfnamefont {S.~M.}\ \bibnamefont
  {Kopeikin}}, \bibinfo {author} {\bibfnamefont {E.~M.}\ \bibnamefont
  {Mazurova}}, \ and\ \bibinfo {author} {\bibfnamefont {A.~P.}\ \bibnamefont
  {Karpik}},\ }\href {https://doi.org/10.1016/j.physleta.2015.02.046}
  {\bibfield  {journal} {\bibinfo  {journal} {Phys. Lett. A}\ }\textbf
  {\bibinfo {volume} {379}},\ \bibinfo {pages} {1555} (\bibinfo {year}
  {2015})}\BibitemShut {NoStop}%
\bibitem [{\citenamefont {Bjerhammar}(1985)}]{Bjerhammar:1985}%
  \BibitemOpen
  \bibfield  {author} {\bibinfo {author} {\bibfnamefont {A.}~\bibnamefont
  {Bjerhammar}},\ }\href@noop {} {\bibfield  {journal} {\bibinfo  {journal}
  {Bull. G{\'e}od{\'e}sique}\ }\textbf {\bibinfo {volume} {59}},\ \bibinfo
  {pages} {207} (\bibinfo {year} {1985})}\BibitemShut {NoStop}%
\bibitem [{\citenamefont {Bjerhammar}(1986)}]{Bjerhammar:1986}%
  \BibitemOpen
  \bibfield  {author} {\bibinfo {author} {\bibfnamefont {A.}~\bibnamefont
  {Bjerhammar}},\ }\href@noop {} {\emph {\bibinfo {title} {{Relativistic
  Geodesy}}}},\ \bibinfo {type} {Tech. Rep.}\ \bibinfo {number} {NOS 118 NGS
  36}\ (\bibinfo {year} {1986})\BibitemShut {NoStop}%
\bibitem [{\citenamefont {Kopeikin}\ \emph {et~al.}(2016)\citenamefont
  {Kopeikin}, \citenamefont {Han},\ and\ \citenamefont
  {Mazurova}}]{Kopeikin:2016}%
  \BibitemOpen
  \bibfield  {author} {\bibinfo {author} {\bibfnamefont {S.}~\bibnamefont
  {Kopeikin}}, \bibinfo {author} {\bibfnamefont {W.}~\bibnamefont {Han}}, \
  and\ \bibinfo {author} {\bibfnamefont {E.}~\bibnamefont {Mazurova}},\ }\href
  {\doibase 10.1103/PhysRevD.93.044069} {\bibfield  {journal} {\bibinfo
  {journal} {Phys. Rev. D}\ }\textbf {\bibinfo {volume} {93}},\ \bibinfo
  {pages} {044069} (\bibinfo {year} {2016})}\BibitemShut {NoStop}%
\bibitem [{\citenamefont {Oltean}\ \emph {et~al.}(2016)\citenamefont {Oltean},
  \citenamefont {Epp}, \citenamefont {McGrath},\ and\ \citenamefont
  {Mann}}]{Oltean:2015}%
  \BibitemOpen
  \bibfield  {author} {\bibinfo {author} {\bibfnamefont {M.}~\bibnamefont
  {Oltean}}, \bibinfo {author} {\bibfnamefont {R.~J.}\ \bibnamefont {Epp}},
  \bibinfo {author} {\bibfnamefont {P.~L.}\ \bibnamefont {McGrath}}, \ and\
  \bibinfo {author} {\bibfnamefont {R.~B.}\ \bibnamefont {Mann}},\ }\href
  {http://stacks.iop.org/0264-9381/33/i=10/a=105001} {\bibfield  {journal}
  {\bibinfo  {journal} {Classical Quantum Grav.}\ }\textbf {\bibinfo {volume}
  {33}} (\bibinfo {year} {2016})}\BibitemShut {NoStop}%
\bibitem [{\citenamefont {M{\"u}ller}\ \emph {et~al.}(2008)\citenamefont
  {M{\"u}ller}, \citenamefont {Soffel},\ and\ \citenamefont
  {Klioner}}]{Mueller:2008}%
  \BibitemOpen
  \bibfield  {author} {\bibinfo {author} {\bibfnamefont {J.}~\bibnamefont
  {M{\"u}ller}}, \bibinfo {author} {\bibfnamefont {M.~H.}\ \bibnamefont
  {Soffel}}, \ and\ \bibinfo {author} {\bibfnamefont {S.~A.}\ \bibnamefont
  {Klioner}},\ }\href {\doibase 10.1007/s00190-007-0168-7} {\bibfield
  {journal} {\bibinfo  {journal} {J. Geod.}\ }\textbf {\bibinfo {volume}
  {82}},\ \bibinfo {pages} {133} (\bibinfo {year} {2008})}\BibitemShut
  {NoStop}%
\bibitem [{\citenamefont {Kopeikin}\ \emph {et~al.}(2011)\citenamefont
  {Kopeikin}, \citenamefont {Efroimsky},\ and\ \citenamefont
  {Kaplan}}]{Kopeikin:Book:2011}%
  \BibitemOpen
  \bibfield  {author} {\bibinfo {author} {\bibfnamefont {S.~M.}\ \bibnamefont
  {Kopeikin}}, \bibinfo {author} {\bibfnamefont {M.}~\bibnamefont {Efroimsky}},
  \ and\ \bibinfo {author} {\bibfnamefont {G.}~\bibnamefont {Kaplan}},\ }\href
  {\doibase 10.1002/9783527634569.fmatter} {\emph {\bibinfo {title}
  {{Relativistic Celestial Mechanics of the Solar System}}}}\ (\bibinfo
  {publisher} {Wiley-VCH, Weinheim, Germany},\ \bibinfo {year}
  {2011})\BibitemShut {NoStop}%
\bibitem [{\citenamefont {Delva}\ and\ \citenamefont {Ger{\v
  s}l}(2017)}]{Delva:2017}%
  \BibitemOpen
  \bibfield  {author} {\bibinfo {author} {\bibfnamefont {P.}~\bibnamefont
  {Delva}}\ and\ \bibinfo {author} {\bibfnamefont {J.}~\bibnamefont {Ger{\v
  s}l}},\ }\href {\doibase 10.3390/universe3010024} {\bibfield  {journal}
  {\bibinfo  {journal} {Universe}\ }\textbf {\bibinfo {volume} {3}},\ \bibinfo
  {pages} {24} (\bibinfo {year} {2017})}\BibitemShut {NoStop}%
\bibitem [{NGS()}]{NGS}%
  \BibitemOpen
  \href@noop {} {}\bibinfo {howpublished}
  {\url{https://www.ngs.noaa.gov/GEOID/geoid_def.html}}\BibitemShut {NoStop}%
\bibitem [{\citenamefont {Barthelmes}(2013)}]{Barthelmes:2013}%
  \BibitemOpen
  \bibfield  {author} {\bibinfo {author} {\bibfnamefont {F.}~\bibnamefont
  {Barthelmes}},\ }\href {\doibase 10.2312/GFZ.b103-09026} {\emph {\bibinfo
  {title} {{Definition of functionals of the geopotential and their calculation
  from spherical harmonic models : theory and formulas used by the calculation
  service of the International Centre for Global Earth Models (ICGEM)}}}},\
  \bibinfo {type} {Tech. Rep.}\ \bibinfo {number} {STR09/02}\ (\bibinfo {year}
  {2013})\BibitemShut {NoStop}%
\bibitem [{\citenamefont {Perlick}(1987)}]{Perlick:1987}%
  \BibitemOpen
  \bibfield  {author} {\bibinfo {author} {\bibfnamefont {V.}~\bibnamefont
  {Perlick}},\ }\href {\doibase 10.1007/BF00759142} {\bibfield  {journal}
  {\bibinfo  {journal} {Gen. Relativ. Gravit.}\ }\textbf {\bibinfo {volume}
  {19}},\ \bibinfo {pages} {1059} (\bibinfo {year} {1987})}\BibitemShut
  {NoStop}%
\bibitem [{\citenamefont {Kermack}\ \emph {et~al.}(1934)\citenamefont
  {Kermack}, \citenamefont {McCrea},\ and\ \citenamefont
  {Whittaker}}]{Kermack:1934}%
  \BibitemOpen
  \bibfield  {author} {\bibinfo {author} {\bibfnamefont {W.~O.}\ \bibnamefont
  {Kermack}}, \bibinfo {author} {\bibfnamefont {W.~H.}\ \bibnamefont {McCrea}},
  \ and\ \bibinfo {author} {\bibfnamefont {E.~T.}\ \bibnamefont {Whittaker}},\
  }\href {\doibase 10.1017/S0370164600015479} {\bibfield  {journal} {\bibinfo
  {journal} {Proc. R. Soc. Edinburgh}\ }\textbf {\bibinfo {volume} {53}},\
  \bibinfo {pages} {31} (\bibinfo {year} {1934})}\BibitemShut {NoStop}%
\bibitem [{\citenamefont {Brill}(1972)}]{Brill:1972}%
  \BibitemOpen
  \bibfield  {author} {\bibinfo {author} {\bibfnamefont {D.~R.}\ \bibnamefont
  {Brill}},\ }in\ \href@noop {} {\emph {\bibinfo {booktitle} {Methods of Local
  and Global Differential Geometry in General Relativity}}},\ \bibinfo {editor}
  {edited by\ \bibinfo {editor} {\bibfnamefont {D.}~\bibnamefont {Farnsworth}},
  \bibinfo {editor} {\bibfnamefont {J.}~\bibnamefont {Fink}}, \bibinfo {editor}
  {\bibfnamefont {J.}~\bibnamefont {Porter}}, \ and\ \bibinfo {editor}
  {\bibfnamefont {A.}~\bibnamefont {Thompson}}},\ \bibinfo {organization} {Yale
  University}\ (\bibinfo  {publisher} {Springer},\ \bibinfo {address}
  {Berlin},\ \bibinfo {year} {1972})\ pp.\ \bibinfo {pages}
  {45--47}\BibitemShut {NoStop}%
\bibitem [{\citenamefont {Straumann}(1984)}]{Straumann:1984}%
  \BibitemOpen
  \bibfield  {author} {\bibinfo {author} {\bibfnamefont {N.}~\bibnamefont
  {Straumann}},\ }\href@noop {} {\emph {\bibinfo {title} {{General relativity
  and relativistic astrophysics}}}}\ (\bibinfo  {publisher} {Springer},\
  \bibinfo {address} {Berlin},\ \bibinfo {year} {1984})\BibitemShut {NoStop}%
\bibitem [{\citenamefont {Hasse}\ and\ \citenamefont
  {Perlick}(1988)}]{Hasse:1988}%
  \BibitemOpen
  \bibfield  {author} {\bibinfo {author} {\bibfnamefont {W.}~\bibnamefont
  {Hasse}}\ and\ \bibinfo {author} {\bibfnamefont {V.}~\bibnamefont
  {Perlick}},\ }\href {\doibase 10.1063/1.527863} {\bibfield  {journal}
  {\bibinfo  {journal} {J. Math. Phys.}\ }\textbf {\bibinfo {volume} {29}},\
  \bibinfo {pages} {2064} (\bibinfo {year} {1988})}\BibitemShut {NoStop}%
\bibitem [{\citenamefont {Perlick}(1990)}]{Perlick:1990}%
  \BibitemOpen
  \bibfield  {author} {\bibinfo {author} {\bibfnamefont {V.}~\bibnamefont
  {Perlick}},\ }\href {\doibase 10.1063/1.528645} {\bibfield  {journal}
  {\bibinfo  {journal} {J. Math. Phys.}\ }\textbf {\bibinfo {volume} {31}},\
  \bibinfo {pages} {1962} (\bibinfo {year} {1990})}\BibitemShut {NoStop}%
\bibitem [{\citenamefont {Perlick}(2000)}]{Perlick:2000}%
  \BibitemOpen
  \bibfield  {author} {\bibinfo {author} {\bibfnamefont {V.}~\bibnamefont
  {Perlick}},\ }\href {https://books.google.de/books?id=Un3my55rV6YC} {\emph
  {\bibinfo {title} {{Ray Optics, Fermat's Principle, and Applications to
  General Relativity}}}},\ \bibinfo {series} {Lecture Notes in Physics
  Monographs}, Vol.~\bibinfo {volume} {61}\ (\bibinfo  {publisher} {Springer},\
  \bibinfo {year} {2000})\BibitemShut {NoStop}%
\bibitem [{\citenamefont {Ehlers}(1961)}]{Ehlers:1961}%
  \BibitemOpen
  \bibfield  {author} {\bibinfo {author} {\bibfnamefont {J.}~\bibnamefont
  {Ehlers}},\ }\href {http://hdl.handle.net/11858/00-001M-0000-0013-5F19-F}
  {\bibfield  {journal} {\bibinfo  {journal} {{Abhandlungen der
  Mathmatisch-Naturwissenschaftlichen Klasse}}\ }\textbf {\bibinfo {volume}
  {11}} (\bibinfo {year} {1961})}\BibitemShut {NoStop}%
\bibitem [{\citenamefont {Ehlers}(1993)}]{Ehlers:1993}%
  \BibitemOpen
  \bibfield  {author} {\bibinfo {author} {\bibfnamefont {J.}~\bibnamefont
  {Ehlers}},\ }\href {\doibase 10.1007/BF00759031} {\bibfield  {journal}
  {\bibinfo  {journal} {Gen. Relativ. Gravit.}\ }\textbf {\bibinfo {volume}
  {25}},\ \bibinfo {pages} {1225} (\bibinfo {year} {1993})}\BibitemShut
  {NoStop}%
\bibitem [{\citenamefont {Salzman}\ and\ \citenamefont
  {Taub}(1954)}]{Salzmann:1954}%
  \BibitemOpen
  \bibfield  {author} {\bibinfo {author} {\bibfnamefont {G.}~\bibnamefont
  {Salzman}}\ and\ \bibinfo {author} {\bibfnamefont {A.~H.}\ \bibnamefont
  {Taub}},\ }\href {\doibase 10.1103/PhysRev.95.1659} {\bibfield  {journal}
  {\bibinfo  {journal} {Phys. Rev.}\ }\textbf {\bibinfo {volume} {95}},\
  \bibinfo {pages} {1659} (\bibinfo {year} {1954})}\BibitemShut {NoStop}%
\bibitem [{\citenamefont {Soffel}\ and\ \citenamefont
  {Frutos}(2016)}]{Soffel:2016}%
  \BibitemOpen
  \bibfield  {author} {\bibinfo {author} {\bibfnamefont {M.}~\bibnamefont
  {Soffel}}\ and\ \bibinfo {author} {\bibfnamefont {F.}~\bibnamefont
  {Frutos}},\ }\href {\doibase 10.1007/s00190-016-0927-4} {\bibfield  {journal}
  {\bibinfo  {journal} {J. Geod.}\ }\textbf {\bibinfo {volume} {90}},\ \bibinfo
  {pages} {1345} (\bibinfo {year} {2016})}\BibitemShut {NoStop}%
\bibitem [{\citenamefont {Weyl}(1917)}]{Weyl:1917}%
  \BibitemOpen
  \bibfield  {author} {\bibinfo {author} {\bibfnamefont {H.}~\bibnamefont
  {Weyl}},\ }\href {\doibase 10.1002/andp.19173591804} {\bibfield  {journal}
  {\bibinfo  {journal} {Ann. Phys.}\ }\textbf {\bibinfo {volume} {359}},\
  \bibinfo {pages} {117} (\bibinfo {year} {1917})}\BibitemShut {NoStop}%
\bibitem [{\citenamefont {Erez}\ and\ \citenamefont {Rosen}(1959)}]{Erez:1959}%
  \BibitemOpen
  \bibfield  {author} {\bibinfo {author} {\bibfnamefont {G.}~\bibnamefont
  {Erez}}\ and\ \bibinfo {author} {\bibfnamefont {N.}~\bibnamefont {Rosen}},\
  }\href@noop {} {\bibfield  {journal} {\bibinfo  {journal} {Bull. Res. Coun.
  Isr.}\ }\textbf {\bibinfo {volume} {8F}},\ \bibinfo {pages} {47} (\bibinfo
  {year} {1959})}\BibitemShut {NoStop}%
\bibitem [{\citenamefont {Quevedo}\ \emph {et~al.}(2012)\citenamefont
  {Quevedo}, \citenamefont {Toktarbay},\ and\ \citenamefont
  {Yerlan}}]{Quevedo:2013}%
  \BibitemOpen
  \bibfield  {author} {\bibinfo {author} {\bibfnamefont {H.}~\bibnamefont
  {Quevedo}}, \bibinfo {author} {\bibfnamefont {S.}~\bibnamefont {Toktarbay}},
  \ and\ \bibinfo {author} {\bibfnamefont {A.}~\bibnamefont {Yerlan}},\
  }\href@noop {} {\bibfield  {journal} {\bibinfo  {journal} {Int. J. Math.
  Phys.}\ }\textbf {\bibinfo {volume} {3}} (\bibinfo {year}
  {2012})}\BibitemShut {NoStop}%
\bibitem [{\citenamefont {{Zipoy}}(1966)}]{Zipoy:1966}%
  \BibitemOpen
  \bibfield  {author} {\bibinfo {author} {\bibfnamefont {D.~M.}\ \bibnamefont
  {{Zipoy}}},\ }\href {\doibase 10.1063/1.1705005} {\bibfield  {journal}
  {\bibinfo  {journal} {J. Math. Phys.}\ }\textbf {\bibinfo {volume} {7}},\
  \bibinfo {pages} {1137} (\bibinfo {year} {1966})}\BibitemShut {NoStop}%
\bibitem [{\citenamefont {Voorhees}(1970)}]{Voorhees:1970}%
  \BibitemOpen
  \bibfield  {author} {\bibinfo {author} {\bibfnamefont {B.~H.}\ \bibnamefont
  {Voorhees}},\ }\href {\doibase 10.1103/PhysRevD.2.2119} {\bibfield  {journal}
  {\bibinfo  {journal} {Phys. Rev. D}\ }\textbf {\bibinfo {volume} {2}},\
  \bibinfo {pages} {2119} (\bibinfo {year} {1970})}\BibitemShut {NoStop}%
\bibitem [{\citenamefont {Quevedo}(1989)}]{Quevedo:1989}%
  \BibitemOpen
  \bibfield  {author} {\bibinfo {author} {\bibfnamefont {H.}~\bibnamefont
  {Quevedo}},\ }\href {\doibase 10.1103/PhysRevD.39.2904} {\bibfield  {journal}
  {\bibinfo  {journal} {Phys. Rev. D}\ }\textbf {\bibinfo {volume} {39}},\
  \bibinfo {pages} {2904} (\bibinfo {year} {1989})}\BibitemShut {NoStop}%
\bibitem [{\citenamefont {Stephani}\ \emph {et~al.}(2003)\citenamefont
  {Stephani}, \citenamefont {Kramer}, \citenamefont {MacCallum}, \citenamefont
  {Hoenselaers},\ and\ \citenamefont {Herlt}}]{Stephani:Book:2003}%
  \BibitemOpen
  \bibfield  {author} {\bibinfo {author} {\bibfnamefont {H.}~\bibnamefont
  {Stephani}}, \bibinfo {author} {\bibfnamefont {D.}~\bibnamefont {Kramer}},
  \bibinfo {author} {\bibfnamefont {M.}~\bibnamefont {MacCallum}}, \bibinfo
  {author} {\bibfnamefont {C.}~\bibnamefont {Hoenselaers}}, \ and\ \bibinfo
  {author} {\bibfnamefont {E.}~\bibnamefont {Herlt}},\ }\href@noop {} {\emph
  {\bibinfo {title} {{Exact Solutions of Einstein's Field Equations}}}},\
  \bibinfo {edition} {2nd}\ ed.\ (\bibinfo  {publisher} {Cambridge University
  Press, Cambridge, England},\ \bibinfo {year} {2003})\BibitemShut {NoStop}%
\bibitem [{\citenamefont {Bateman}\ and\ \citenamefont
  {Erd{\'e}lyi}(1955)}]{Bateman:1955}%
  \BibitemOpen
  \bibfield  {author} {\bibinfo {author} {\bibfnamefont {H.}~\bibnamefont
  {Bateman}}\ and\ \bibinfo {author} {\bibfnamefont {A.}~\bibnamefont
  {Erd{\'e}lyi}},\ }\href@noop {} {\emph {\bibinfo {title} {{Higher
  Transcendental Functions}}}},\ \bibinfo {series} {Higher Transcendental
  Functions}, Vol.~\bibinfo {volume} {1}\ (\bibinfo  {publisher} {McGraw-Hill,
  New York},\ \bibinfo {year} {1955})\BibitemShut {NoStop}%
\bibitem [{\citenamefont {Geroch}(1970)}]{Geroch:1970b}%
  \BibitemOpen
  \bibfield  {author} {\bibinfo {author} {\bibfnamefont {R.~P.}\ \bibnamefont
  {Geroch}},\ }\href {\doibase 10.1063/1.1665427} {\bibfield  {journal}
  {\bibinfo  {journal} {J. Math. Phys.}\ }\textbf {\bibinfo {volume} {11}},\
  \bibinfo {pages} {2580} (\bibinfo {year} {1970})}\BibitemShut {NoStop}%
\bibitem [{\citenamefont {Hansen}(1974)}]{Hansen:1974}%
  \BibitemOpen
  \bibfield  {author} {\bibinfo {author} {\bibfnamefont {R.~O.}\ \bibnamefont
  {Hansen}},\ }\href {\doibase 10.1063/1.1666501} {\bibfield  {journal}
  {\bibinfo  {journal} {J. Math. Phys.}\ }\textbf {\bibinfo {volume} {15}},\
  \bibinfo {pages} {46} (\bibinfo {year} {1974})}\BibitemShut {NoStop}%
\bibitem [{\citenamefont {Ehlers}(1981)}]{Ehlers:1981}%
  \BibitemOpen
  \bibfield  {author} {\bibinfo {author} {\bibfnamefont {J.}~\bibnamefont
  {Ehlers}},\ }in\ \href@noop {} {\emph {\bibinfo {booktitle}
  {{Grundlagenprobleme der modernen Physik}}}},\ \bibinfo {editor} {edited by\
  \bibinfo {editor} {\bibfnamefont {J.}~\bibnamefont {Nitsch}}, \bibinfo
  {editor} {\bibfnamefont {J.}~\bibnamefont {Pfarr}}, \ and\ \bibinfo {editor}
  {\bibfnamefont {E.~W.}\ \bibnamefont {Stachov}}}\ (\bibinfo  {publisher}
  {BI-Verlag Mannheim},\ \bibinfo {year} {1981})\ pp.\ \bibinfo {pages}
  {65--84}\BibitemShut {NoStop}%
\bibitem [{\citenamefont {Quevedo}(1986)}]{Quevedo:1986}%
  \BibitemOpen
  \bibfield  {author} {\bibinfo {author} {\bibfnamefont {H.}~\bibnamefont
  {Quevedo}},\ }\href {\doibase 10.1103/PhysRevD.33.324} {\bibfield  {journal}
  {\bibinfo  {journal} {Phys. Rev. D}\ }\textbf {\bibinfo {volume} {33}},\
  \bibinfo {pages} {324} (\bibinfo {year} {1986})}\BibitemShut {NoStop}%
\bibitem [{\citenamefont {Young}\ and\ \citenamefont
  {Coulter}(1969)}]{Young1969}%
  \BibitemOpen
  \bibfield  {author} {\bibinfo {author} {\bibfnamefont {J.~H.}\ \bibnamefont
  {Young}}\ and\ \bibinfo {author} {\bibfnamefont {C.~A.}\ \bibnamefont
  {Coulter}},\ }\href {\doibase 10.1103/PhysRev.184.1313} {\bibfield  {journal}
  {\bibinfo  {journal} {Phys. Rev.}\ }\textbf {\bibinfo {volume} {184}},\
  \bibinfo {pages} {1313} (\bibinfo {year} {1969})}\BibitemShut {NoStop}%
\bibitem [{\citenamefont {Quevedo}(2011)}]{Quevedo:2011}%
  \BibitemOpen
  \bibfield  {author} {\bibinfo {author} {\bibfnamefont {H.}~\bibnamefont
  {Quevedo}},\ }\href {\doibase 10.1142/S0218271811019852} {\bibfield
  {journal} {\bibinfo  {journal} {Int. J. Modern Phys. D}\ }\textbf {\bibinfo
  {volume} {20}},\ \bibinfo {pages} {1779} (\bibinfo {year}
  {2011})}\BibitemShut {NoStop}%
\bibitem [{\citenamefont {Toktarbay}\ and\ \citenamefont
  {Quevedo}(2014)}]{Toktarbay:2014}%
  \BibitemOpen
  \bibfield  {author} {\bibinfo {author} {\bibfnamefont {S.}~\bibnamefont
  {Toktarbay}}\ and\ \bibinfo {author} {\bibfnamefont {H.}~\bibnamefont
  {Quevedo}},\ }\href {\doibase 10.1134/S0202289314040136} {\bibfield
  {journal} {\bibinfo  {journal} {Gravitation Cosmol.}\ }\textbf {\bibinfo
  {volume} {20}},\ \bibinfo {pages} {252} (\bibinfo {year} {2014})}\BibitemShut
  {NoStop}%
\bibitem [{\citenamefont {Quevedo}(2016{\natexlab{a}})}]{Quevedo:2016}%
  \BibitemOpen
  \bibfield  {author} {\bibinfo {author} {\bibfnamefont {H.}~\bibnamefont
  {Quevedo}},\ }\href@noop {} {\  (\bibinfo {year} {2016}{\natexlab{a}})},\
  \Eprint {http://arxiv.org/abs/1606.09361} {arXiv:1606.09361 [gr-qc]}
  \BibitemShut {NoStop}%
\bibitem [{\citenamefont {Quevedo}(2016{\natexlab{b}})}]{Quevedo:2016b}%
  \BibitemOpen
  \bibfield  {author} {\bibinfo {author} {\bibfnamefont {H.}~\bibnamefont
  {Quevedo}},\ }\href@noop {} {\  (\bibinfo {year} {2016}{\natexlab{b}})},\
  \Eprint {http://arxiv.org/abs/1606.05985} {arXiv:1606.05985 [gr-qc]}
  \BibitemShut {NoStop}%
\bibitem [{\citenamefont {Bach}(1922)}]{Bach:1922}%
  \BibitemOpen
  \bibfield  {author} {\bibinfo {author} {\bibfnamefont {B.}~\bibnamefont
  {Bach}},\ }\href {http://eudml.org/doc/167694} {\bibfield  {journal}
  {\bibinfo  {journal} {Math. Z.}\ }\textbf {\bibinfo {volume} {13}},\ \bibinfo
  {pages} {134} (\bibinfo {year} {1922})}\BibitemShut {NoStop}%
\bibitem [{\citenamefont {Griffiths}\ and\ \citenamefont
  {Podolsk\'{y}}(2009)}]{Griffiths:2009}%
  \BibitemOpen
  \bibfield  {author} {\bibinfo {author} {\bibfnamefont {J.~B.}\ \bibnamefont
  {Griffiths}}\ and\ \bibinfo {author} {\bibfnamefont {J.}~\bibnamefont
  {Podolsk\'{y}}},\ }\href@noop {} {\emph {\bibinfo {title} {{Exact
  {Space-Times} in Einstein's General Relativity}}}}\ (\bibinfo  {publisher}
  {Cambridge University Press, Cambridge, England},\ \bibinfo {year}
  {2009})\BibitemShut {NoStop}%
\bibitem [{\citenamefont {Philipp}\ and\ \citenamefont
  {Hackmann}(2017)}]{Philipp:2017b}%
  \BibitemOpen
  \bibfield  {author} {\bibinfo {author} {\bibfnamefont {D.}~\bibnamefont
  {Philipp}}\ and\ \bibinfo {author} {\bibfnamefont {E.}~\bibnamefont
  {Hackmann}},\ }\href@noop {} {\  (\bibinfo {year} {2017})},\ \bibinfo {note}
  {unpublished}\BibitemShut {NoStop}%
\bibitem [{\citenamefont {Sharp}(1981)}]{Sharp:1981}%
  \BibitemOpen
  \bibfield  {author} {\bibinfo {author} {\bibfnamefont {N.~A.}\ \bibnamefont
  {Sharp}},\ }\href {http://www.nrcresearchpress.com/doi/abs/10.1139/p81-086}
  {\bibfield  {journal} {\bibinfo  {journal} {Canad. J. Phys.}\ }\textbf
  {\bibinfo {volume} {59}},\ \bibinfo {pages} {688} (\bibinfo {year}
  {1981})}\BibitemShut {NoStop}%
\bibitem [{\citenamefont {Pelavas}\ \emph {et~al.}(2001)\citenamefont
  {Pelavas}, \citenamefont {Neary},\ and\ \citenamefont {Lake}}]{Pelavas:2001}%
  \BibitemOpen
  \bibfield  {author} {\bibinfo {author} {\bibfnamefont {N.}~\bibnamefont
  {Pelavas}}, \bibinfo {author} {\bibfnamefont {N.}~\bibnamefont {Neary}}, \
  and\ \bibinfo {author} {\bibfnamefont {K.}~\bibnamefont {Lake}},\ }\href
  {http://stacks.iop.org/0264-9381/18/i=7/a=314} {\bibfield  {journal}
  {\bibinfo  {journal} {Classical Quantum Gravity}\ }\textbf {\bibinfo {volume}
  {18}},\ \bibinfo {pages} {1319} (\bibinfo {year} {2001})}\BibitemShut
  {NoStop}%
\bibitem [{\citenamefont {Hackmann}\ and\ \citenamefont
  {L{\"a}mmerzahl}(2014)}]{Hackmann:2014}%
  \BibitemOpen
  \bibfield  {author} {\bibinfo {author} {\bibfnamefont {E.}~\bibnamefont
  {Hackmann}}\ and\ \bibinfo {author} {\bibfnamefont {C.}~\bibnamefont
  {L{\"a}mmerzahl}},\ }\href {\doibase 10.1103/PhysRevD.90.044059} {\bibfield
  {journal} {\bibinfo  {journal} {Phys. Rev. D}\ }\textbf {\bibinfo {volume}
  {90}},\ \bibinfo {pages} {044059} (\bibinfo {year} {2014})}\BibitemShut
  {NoStop}%
\bibitem [{\citenamefont {Soffel}\ \emph {et~al.}(2003)\citenamefont {Soffel},
  \citenamefont {Klioner}, \citenamefont {Petit}, \citenamefont {Wolf},
  \citenamefont {Kopeikin}, \citenamefont {Bretagnon}, \citenamefont
  {Brumberg}, \citenamefont {Capitaine}, \citenamefont {Damour}, \citenamefont
  {Fukushima}, \citenamefont {Guinot}, \citenamefont {Huang}, \citenamefont
  {Lindegren}, \citenamefont {Ma}, \citenamefont {Nordtvedt}, \citenamefont
  {Ries}, \citenamefont {Seidelmann}, \citenamefont {Vokrouhlick{\'y}},
  \citenamefont {Will},\ and\ \citenamefont {Xu}}]{Soffel:2003}%
  \BibitemOpen
  \bibfield  {author} {\bibinfo {author} {\bibfnamefont {M.}~\bibnamefont
  {Soffel}}, \bibinfo {author} {\bibfnamefont {S.~A.}\ \bibnamefont {Klioner}},
  \bibinfo {author} {\bibfnamefont {G.}~\bibnamefont {Petit}}, \bibinfo
  {author} {\bibfnamefont {P.}~\bibnamefont {Wolf}}, \bibinfo {author}
  {\bibfnamefont {S.~M.}\ \bibnamefont {Kopeikin}}, \bibinfo {author}
  {\bibfnamefont {P.}~\bibnamefont {Bretagnon}}, \bibinfo {author}
  {\bibfnamefont {V.~A.}\ \bibnamefont {Brumberg}}, \bibinfo {author}
  {\bibfnamefont {N.}~\bibnamefont {Capitaine}}, \bibinfo {author}
  {\bibfnamefont {T.}~\bibnamefont {Damour}}, \bibinfo {author} {\bibfnamefont
  {T.}~\bibnamefont {Fukushima}}, \bibinfo {author} {\bibfnamefont
  {B.}~\bibnamefont {Guinot}}, \bibinfo {author} {\bibfnamefont {T.-Y.}\
  \bibnamefont {Huang}}, \bibinfo {author} {\bibfnamefont {L.}~\bibnamefont
  {Lindegren}}, \bibinfo {author} {\bibfnamefont {C.}~\bibnamefont {Ma}},
  \bibinfo {author} {\bibfnamefont {K.}~\bibnamefont {Nordtvedt}}, \bibinfo
  {author} {\bibfnamefont {J.~C.}\ \bibnamefont {Ries}}, \bibinfo {author}
  {\bibfnamefont {P.~K.}\ \bibnamefont {Seidelmann}}, \bibinfo {author}
  {\bibfnamefont {D.}~\bibnamefont {Vokrouhlick{\'y}}}, \bibinfo {author}
  {\bibfnamefont {C.~M.}\ \bibnamefont {Will}}, \ and\ \bibinfo {author}
  {\bibfnamefont {C.}~\bibnamefont {Xu}},\ }\href
  {http://stacks.iop.org/1538-3881/126/i=6/a=2687} {\bibfield  {journal}
  {\bibinfo  {journal} {Astron. J.}\ }\textbf {\bibinfo {volume} {126}},\
  \bibinfo {pages} {2687} (\bibinfo {year} {2003})}\BibitemShut {NoStop}%
\bibitem [{\citenamefont {Kaplan}(2006)}]{Kaplan:2006}%
  \BibitemOpen
  \bibfield  {author} {\bibinfo {author} {\bibfnamefont {G.~H.}\ \bibnamefont
  {Kaplan}},\ }\href {https://arxiv.org/abs/astro-ph/0602086} {\  (\bibinfo
  {year} {2006})},\ \Eprint {http://arxiv.org/abs/astro-ph/0602086}
  {arXiv:astro-ph/0602086 [astro-ph]} \BibitemShut {NoStop}%
\end{thebibliography}%

\end{document}